\documentclass[structabstract]{aa}  

\usepackage{graphicx}
\usepackage{txfonts}  
\usepackage{epsfig}

\usepackage{natbib}
\bibpunct{(}{)}{;}{a}{}{,}
\usepackage{times}
\usepackage{units}

\newcommand{\spitzer}{\textit{Spitzer}}
\newcommand{\msun}{\hbox{$\hbox{M}_{\odot}$}}
\newcommand{\lsun}{\hbox{$\hbox{L}_{\odot}$}}
\newcommand{\herschel}{\textit{Herschel}}
\newcommand {\asp}{\mbox{$.\!\!^{\prime\prime}$}}

\begin{document}
   \title{The earliest phases of star formation - A Herschel\thanks{\herschel\ is an ESA space observatory with science instruments
   provided by European-led Principal Investigator consortia and with important participation from NASA.} key project}

   \subtitle{The thermal structure of low-mass molecular cloud cores\thanks{Partially based on observations carried out with the IRAM 30m Telescope, 
    with the Atacama Pathfinder Experiment (APEX), and with the James Clerk Maxwell Telescope (JCMT).
    IRAM is supported by INSU/CNRS (France), MPG (Germany) and IGN (Spain). 
    APEX is a collaboration between Max Planck Institut f\"ur Radioastronomie (MPIfR), 
    Onsala Space Observatory (OSO), and the European Southern Observatory (ESO).
    The JCMT is operated by the Joint Astronomy Centre on behalf of the Particle Physics and Astronomy Research Council of the United Kingdom, 
    the Netherlands Association for Scientific Research, and the National Research Council of Canada.}}

   \author{R. Launhardt\inst{1}
          \and
          A. M. Stutz\inst{1}
          \and
          A. Schmiedeke\inst{2,1}
          \and
          Th. Henning\inst{1}
          \and
          O. Krause\inst{1}
          \and
          Z. Balog\inst{1}
          \and
          H. Beuther\inst{1}
          \and
          S. Birkmann\inst{3,1}
          \and
          M. Hennemann\inst{4,1}
          \and
          J. Kainulainen\inst{1}
          \and
          T. Khanzadyan\inst{5}
          \and
          H. Linz\inst{1}
          \and
          N. Lippok\inst{1}
          \and
          M. Nielbock\inst{1}
          \and
          J. Pitann\inst{1}
          \and
          S. Ragan\inst{1}
          \and
          C. Risacher\inst{6,5} 
          \and
          M. Schmalzl\inst{7,1}
          \and
          Y. L. Shirley\inst{8}
          \and
          B. Stecklum\inst{9}
          \and
          J. Steinacker\inst{10,1}
          \and
          J. Tackenberg\inst{1}
           }

   \institute{Max-Planck-Institut f\"ur Astronomie (MPIA), 
              K\"onigstuhl 17, D-69117 Heidelberg, Germany\\
              \email{rl@mpia.de}
         \and
             Universit\"at zu K\"oln, Z\"ulpicher Strasse 77, D-50937 K\"oln, Germany
         \and
             ESA/ESTEC, Keplerlaan 1, Postbus 299, 2200 AG Noordwijk, The Netherlands 
         \and
             Laboratoire AIM Paris-Saclay, Service d'Astrophysique, CEA/IRFU -- CNRS/INSU -- Universit\'e Paris Diderot, 
             Orme des Merisiers Bat.~709, 91191 Gif-sur-Yvette Cedex, France
         \and
             Max-Planck-Institut f\"ur Radioastronomy (MPIfR), 
             Auf dem H\"ugel 69, D-53121 Bonn, Germany
         \and
             SRON Netherlands Institute for Space Research, PO Box 800, 9700 AV Groningen, 
             The Netherlands
         \and
             Leiden Observatory, Leiden University, PO Box 9513, 2300 RA, Leiden, The Netherlands
         \and
             Steward Observatory, 933 North Cherry Avenue, Tucson, AZ 85721, USA
         \and
             Th\"uringer Landessternwarte Tautenburg, Sternwarte 5, d-07778 Tautenburg, Germany
          \and
             Institut de Plan\'etologie et d'Astrophysique de Grenoble, Universit\'e de Grenoble,
             BP\,53, F-38041 Grenoble C\'edex 9, France
             }

   \date{Received October 1, 2012; accepted }


 
  \abstract
  {The temperature and density structure of molecular cloud cores are the most important physical 
  quantities that determine the course of the protostellar collapse and the properties of the stars they form. 
  Nevertheless, density profiles often rely either on the simplifying assumption of isothermality or on observationally 
  poorly constrained model temperature profiles. The instruments of the \herschel\ satellite provide us for the first 
  time with both the spectral coverage and the spatial resolution that is needed to directly measure the dust temperature 
  structure of nearby molecular cloud cores.}
   {With the aim of better constraining the initial physical conditions in molecular cloud cores 
   at the onset of protostellar collapse, in particular of measuring their temperature structure, 
   we initiated the Guaranteed Time Key Project (GTKP) ``The Earliest Phases of Star Formation'' (EPoS) with the 
   \herschel\ satellite. This paper gives an overview of the low-mass sources in the EPoS project, the \herschel\ and 
   complementary ground-based observations, our analysis method, and the initial results of the survey.}
   {We study the thermal dust emission of 12 previously well-characterized, isolated, nearby globules using FIR and submm 
   continuum maps at up to eight wavelengths between 100\,$\mu$m and 1.2\,mm. Our sample contains both globules with starless 
   cores and embedded protostars at different early evolutionary stages. The dust emission maps are used to extract spatially 
   resolved SEDs, which are then fit independently with modified blackbody curves to obtain line-of-sight-averaged dust 
   temperature and column density maps.}
   {We find that the thermal structure of all globules (mean mass 7\,\msun) is dominated by external heating from the interstellar 
   radiation field and moderate shielding by thin extended halos. All globules have warm outer envelopes (14-20\,K) and colder 
   dense interiors (8-12\,K) with column densities of a few $10^{22}$\,cm$^{-2}$. 
   The protostars embedded in some of the globules raise the local temperature of the dense cores only within radii out to about 5000\,AU, 
   but do not significantly affect the overall thermal balance of the globules. Five out of the six starless cores 
   in the sample are gravitationally bound and approximately thermally stabilized. The starless core in CB\,244 is found to be 
   supercritical and is speculated to be on the verge of collapse.
   For the first time, we can now also include externally heated starless cores in the $L_{\rm smm}\,/\,L_{\rm bol}$\ vs. $T_{\rm bol}$\ 
   diagram and find that $T_{\rm bol}<25$\,K seems to be a robust criterion to distinguish starless from  protostellar cores, including 
   those that only have an embedded very low-luminosity object.}
   {}


\titlerunning{The structure of low-mass cloud cores}
\authorrunning{R. Launhardt et al.}

   \maketitle


\section{Introduction and scientific goals}                  \label{sec-intro}

The formation of stars from diffuse interstellar matter is one of the
most fundamental and fascinating transformation processes in the
universe.  Stars form through the gravitational collapse of the
densest and coldest cores inside molecular clouds.  The initial
temperature and density structure of such cores are the most important
physical quantities that determine the course of the collapse and its
stellar end product
\citep[e.g.,][]{larson69,penston69,shu77,sal87,benoit2010}; however,
deriving the physical properties of such cores from observations and
thus constraining theoretical collapse models is a nontrivial
problem.  More than 98\% of the core mass is locked up in H$_2$\
molecules and in He atoms that do not have a permanent electric dipole
moment and that therefore do not radiate when cold. The excitation of
ro-vibrational and electronic states of H$_2$\ requires temperatures
much higher than are present in these cores \citep[e.g.,][]{burton92}.
This problem is usually overcome by observing radiation from heavier
asymmetric molecules, which are much less abundant, but have
rotational transitions that are easily excited via collisions with
hydrogen molecules \citep[e.g., CO, CS;][]{bergin07}.  However, at the
high densities ($n_{\rm H} \geq 10^6$\,cm$^{-3}$) and low temperatures
\citep[$T\leq 10$\,K; e.g.,][]{crapsi07} inside such cloud cores, most
of these heavier molecules freeze out from the gas phase and settle on
dust grains, where they remain 'invisible'
\citep[e.g.,][]{bl97,charnley97,caselli99,hilyblant2010}.  The few
remaining 'slow depleters' (e.g., NH$_3$\ or N$_2$H$^+$, $n_{\rm
  crit(1-0)}\sim 10^5$\,cm$^{-3}$) or deuterated molecular ions that
profit from the gas phase depletion of CO \citep[e.g., N$_2$D$^+$\ and
DCO$^+$;][]{caselli02} can be used to identify prestellar cores and
infer their kinematical and chemical properties.  However, the
respective observational data, in particular for the deuterated
species, are still sparse and their complex chemical evolution is not
yet understood sufficiently well to derive reliable density profiles.
A very promising and robust alternative tracer for the matter in such
cores is therefore the dust, which constitutes about 1\% of the total
mass of the interstellar matter (ISM) in the solar neighborhood.

Interstellar dust can be traced by its effect on the 
attenuation of background light \citep{lada94}, 
scattering of ambient starlight 
("cloudshine": \citealp{fg06}; "coreshine": \citealp{stein2010,pagani10}),
or by its thermal emission \citep[e.g.,][]{smith79}.  
In principle, it is preferable to use extinction measurements to 
trace the column density because they are to first order independent 
of the dust temperature (unlike emission).
However, extinction measurements require that at least some
measurable amount of light from background sources passes through the
cloud, which is no longer the case at the highest column densities
in the core centers ($N_{\rm H}>{\mathrm few} \times 10^{22}$\,cm$^{-2}$). 
Another limitation of this technique is the inability to exactly measure the 
extinction toward individual sources with with a priori unknown SEDs, and  
that it relies on statistically relevant ensembles of background stars, 
which restricts this method to fields with high background star density 
(i.e., close to the galactic plane) and limits the effective angular resolution.
Consequently, most attempts to derive the density structure of 
cloud cores from extinction maps have been done in the near-infrared (NIR) and have targeted 
less-opaque cloud cores and envelopes 
\citep[$A_V\le20$\,mag; e.g.,][]{alves01,lada04,kandori05,kai06}.
Going to wavelengths longer than optical or NIR, where 
the opacity is lower, also usually does not help much since the background stars 
also become dimmer at longer wavelengths.

In addition to the conventional NIR extinction mapping techniques,
mid and far-IR (MIR, FIR) shadows, or absorption features where the densest
portions of cores are observed in absorption against the background
interstellar radiation field (ISRF), can be used to map structure at high resolution.
Provided the absolute background level in the images can be accurately calibrated, 
they provide good measurements of the line-of-sight (LoS) projected structure and 
column density in very dense regions. Such features have been used to study cores at
8, 24, and 70\,$\mu$m \citep{Bacmann2000,stutz09a,stutz09b,tobin2010}. One of the main results from
these studies is that the LoS-projected geometry often drastically
departs from spherical geometry both in the prestellar and protostellar phases.  

Observations of scattered interstellar radiation in the form of cloudshine, 
although permitting certain diagnostics of cloud structure \citep{padoan06,juvela06}, 
also do not trace the interiors of dense cores. 
Scattered MIR interstellar light can penetrate most low-mass cores producing 
coreshine \citep{stein2010}. But, the effect requires the presence of larger 
dust grains for scattering to be efficient and is visible in only half of the cores \citep{pagani10}.
Beside the advantage of an only weak dependence on temperature effects,
scattered light is also sensitive to the 3D structure of the core and could 
allow to trace the actual density structure. But to use
this information, the outer radiation field and the scattering phase
function of the dust particles needs to be known precisely which poses
a problem for many cores in the complex environment of star-formation regions.

Hence, the thermal emission from dust grains remains as a superior tracer of matter 
in the coldest and densest molecular cloud cores where stars form.
Consequently, most of our current information
on the density structure of prestellar cloud cores and protostellar
envelopes comes from submm and mm dust continuum maps
\citep[e.g.,][]{dwt94,dwt99,lau97,henning98,evans01,motte01,shirley02,andre04,kirk05,kauf2008,lau10}.
Nevertheless, ultimately we will need to combine information from different tracers and methods
to calibrate the dust opacities as well as to derive quantities that cannot be derived from the 
dust emission alone (see discussion and outlook at the end of Sect.\,\ref{sec-sum}).

To relate the thermal dust emission at a given wavelength to the
mass of the emitting matter, three main pieces of information are
needed: the temperature of the dust grains, $T_{\rm d}$, their mass absorption coefficient,
$\kappa_{\nu}$, and the gas-to-dust mass ratio.
The mean opacity of the dust mixture is a function of
grain optical properties, wavelength, and temperature
\citep[e.g.,][]{henning96,agladze96,draine2003,boudet05} 
and may actually vary locally and in time as a function of the physical conditions. 
However, observational constraints of such opacity variations are very difficult 
to obtain due to various degeneracies and often insufficient data
\citep[e.g.,][]{shetty09a,shetty09b,shirley11,kelly2012}. Therefore, the conversion 
of flux into mass is usually done by adopting one or the other 'standard' dust model 
that is assumed to be constant throughout the region of interest. 

The definition of a common temperature for the dust
requires that the grains are in local thermal equilibrium (LTE), which
is fortunately the case, to first order, in the well-shielded dense interiors of
molecular cloud cores where collisions with H$_2$\ molecules occur
frequently and stochastic heating of individual grains 
(timescales $\tau_{\rm cool}<<\tau_{\rm heat}$), 
e.g., by UV photons, does not play a significant role \citep[e.g.,][]{sieben92,pav2012}.
However, with typical dust temperatures inside such dense cores 
being in the range $8-15$\,K \citep[e.g.,][]{difra07,bergin07},
the submm emission represents only the Rayleigh-Jeans tail of the 
Planck spectrum. Therefore, the dust temperature is only very poorly 
constrained by these data, which results in large uncertainties in 
the derived total masses and in the radial density profiles. 
E.g., at $\lambda=1$\,mm, where the thermal emission of low-mass cores 
is practically always optically thin, more than twice the amount of 8\,K dust is needed 
to emit the same flux density as 12\,K dust. 
Hence, a dust temperature estimate of $10\pm2$\,K still leaves the mass and 
the (local) column density uncertain to a factor of two. 
The effect of unaccounted temperature gradients on the derived density profile 
is even more significant and leads to large uncertainties 
in the observational constraints on protostellar collapse models.
Earlier attempts to reconstruct the dust temperature profiles of starless and star-forming
cores from, e.g., SCUBA 450/850\,$\mu$m flux ratio maps or self-consistent radiative transfer modeling 
are discussed in Sect.\,\ref{ssec-dis-comp}, but remain observationally poorly constrained and uncertain.

This situation is currently improving as new observing capabilities start to provide 
data that can potentially help to better constrain the dust temperatures.
The {\it Herschel} Space Observatory with its imaging photometers that cover the 
wavelengths range from 70 to 500\,$\mu$m  \citep{pil10} 
provides for the first time the opportunity to fully sample the 
SED of the thermal dust emission from these cold 
objects\footnote{The peak of the flux density of a blackbody 
with $T=10-20$\,K is at $\lambda\approx 290-145\,\mu$m. The peak of the energy density 
($\nu\,S_{\nu}$) of the optically thin globules at these temperatures is at only slightly 
shorter wavelengths ($230-130\,\mu$m).}
at a spatial resolution that is both appropriate to resolve these cores and 
comparable to that of the largest ground-based submm telescopes, which 
extend the wavelength coverage into the Rayleigh-Jeans regime of the SEDs.

To make use of {\it Herschel}'s new capabilities and improve our knowledge on the initial 
conditions of star formation by overcoming some of the above-mentioned limitations, 
we have initiated the GTKP "The Earliest Phases of Star Formation'' (EPoS).
The main goal of our {\it Herschel} observations of low-mass pre- and protostellar cores 
in the framework of this project is to derive the spatial temperature and density 
structure of these cores with the aim of constraining protostellar collapse models. 
For this purpose, we selected 12 individual, isolated, nearby, and previously 
well-characterized cores and obtained spatially resolved FIR dust emission 
maps at five wavelengths around the expected peak of the emission spectrum.
This paper gives an overview of the observations, data reduction, analysis methods, 
and initial results on the temperature and column density structure of the entire low-mass sample. 
Initial results on one of the sources (CB\,244) were already published by \citet{stutz10}.
A first detailed follow-up study of another source from this sample (B\,68) is published 
by \citet{nielbock12}.
Initial results on the high-mass cores observed in the framework of EPoS are published by 
\citet{beuther10,beuther12}, \citet{,henning10}, and \citet{linz10}. 
An overview on the high-mass part of EPoS is given in \citet{ragan12}. 
 
This paper is organized as follows.
Section\,\ref{sec-sources} describes the target selection criteria and the sample of target sources.
Section\,\ref{sec-obs} describes the {\it Herschel} and complementary ground-based observations and 
the corresponding data reduction.
In Sect.\,\ref{sec-mod} we introduce our method of deriving dust temperature and column density 
maps from the data.
Section\,\ref{sec-res} gives an overview of the initial results from the survey, and in 
Sect.\,\ref{sec-disc} we discuss the uncertainties and limitations of our approach and compare 
our results to earlier work by other authors. Section\,\ref{sec-sum} summarizes the paper.


\section{Sources}                  \label{sec-sources}


\begin{table*}[ht!]
\caption{Source list}             
\label{tab-sourcelist}      
\centering 
{\footnotesize       
\begin{tabular}{llclcllll}     
\hline\hline       
Source & Other & R.A., Dec. (J2000) & Region & Dist. & Ref. & Evol. & Ref. \\ 
       & names & [h:m:s, \degr:\arcmin:\arcsec]   &        & [pc]  &      & class &      \\ 
\hline  
CB\,4 	    & ...           & 00:39:03, $+$52:51:30 & Cas\,A, GB             & $350\pm 150$ & 18       & starless                  & 1      \\
CB\,6 	    & LBN\,613      & 00:49:29, $+$50:44:36 & Cas\,A, GB             & $350\pm 150$ & 18       & Cl.\,I                    & 2      \\
CB\,17 	    & L\,1389       & 04:04:38, $+$56:56:12 & Perseus, GB            & $250\pm 50$  & 2,18     & starless\tablefootmark{a} & 2      \\
CB\,26 	    & L\,1439       & 05:00:09, $+$52:04:54 & Taurus-Auriga          & $140\pm 20$  & 2,11     & starless\tablefootmark{b} & 2,3,4  \\
CB\,27 	    & L\,1512       & 05:04:09, $+$32:43:08 & Taurus-Auriga          & $140\pm 20$  & 1,11     & starless                  & 1,4    \\
BHR\,12     & CG\,30        & 08:09:33, $-$36:05:00 & Gum nebula             & $400\pm 50$  & 2,10     & Cl.\,0\tablefootmark{c}   & 2,10   \\
CB\,68 	    & L\,146        & 16:57:16, $-$16:09:18 & Ophiuchus              & $120\pm 20$  & 2,11     & Cl.\,0                    & 2      \\
B\,68 	    & L\,57, CB\,82 & 17:22:35, $-$23:49:30 & Ophiuchus, Pipe nebula & $135\pm 15$  & 15,16,17 &    starless               & 5      \\
CB\,130     & L\,507        & 18:16:15, $-$02:32:48 & Aquila rift, GB        & $250\pm 50$  & 12,14    & Cl.\,0\tablefootmark{d}   & 2      \\
B\,335 	    & CB\,199       & 19:36:55, $+$07:34:24 & Aquila, isolated       & $100\pm 20$  & 8        & Cl.\,0                    & 2,7    \\
CB\,230     & L\,1177       & 21:17:39, $+$68:17:32 & Cepheus flare, GB      & $400\pm 100$ & 2,13     & Cl.\,I                    & 2      \\
CB\,244     & L\,1262       & 23:25:47, $+$74:17:36 & Cepheus flare, GB      & $200\pm 30$  & 2,13     & Cl.\,0\tablefootmark{e}   & 2,9    \\

\hline                  
\end{tabular}
\tablefoot{
\tablefoottext{a}{CB\,17: additional low-luminosity Class\,I YSO 25\arcsec\ from starless core.}
\tablefoottext{b}{CB\,26: additional Class\,I YSO 3.6\arcmin\ south-west of the starless core.}
\tablefoottext{c}{BHR\,12: additional Class\,I core $\sim$20\arcsec\ south of Class\,0 source.}
\tablefoottext{d}{CB\,130: additional low-mass prestellar core $\sim$30\arcsec\ west and Class\,I YSO 
                  $\sim$15\arcsec\ east of Class\,0 core.}
\tablefoottext{e}{CB\,244: additional starless core $\sim$90\arcsec\ west of Class\,0 source.}
}
\tablebib{
(1)~this paper;
(2)~\citet{lau10};
(3)~\citet{lau01}; 
(4)~\citet{stutz09a};
(5)~\citet{alves01};
(7)~\citet{stutz08};
(8)~\citet{olof09};
(9)~\citet{stutz10};
(10)~\citet{bourke95};
(11)~\citet{loinard11};
(12)~\citet{lau97};
(13)~\citet{kun98};
(14)~\citet{straizys03};
(15)~\citet{degeus89};
(16)~\citet{lom06};
(17)~\citet{alves07};
(18)~\citet{perrot03}}
}
\end{table*}


\stepcounter{footnote}           
\footnotetext{http://lambda.gsfc.nasa.gov/product/cobe/dirbe\_overview.cfm}

\begin{figure}[htb]
 \centering
 \includegraphics[width=9cm]{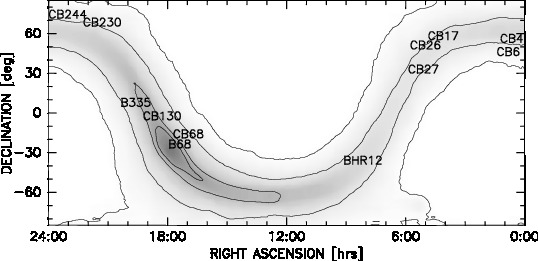}
 \caption{\label{fig-skydistr}
  All-sky map (gnomonic projection), showing the mean stellar K-band flux density distribution of the Milky Way 
  (grayscale, derived from COBE-DIRBE NIR all-sky maps${}^{\thefootnote}$) and the location of our target sources 
  (see Table\,\ref{tab-sourcelist}).}
\end{figure}

\begin{figure}[htb]
 \centering
 \includegraphics[width=9cm]{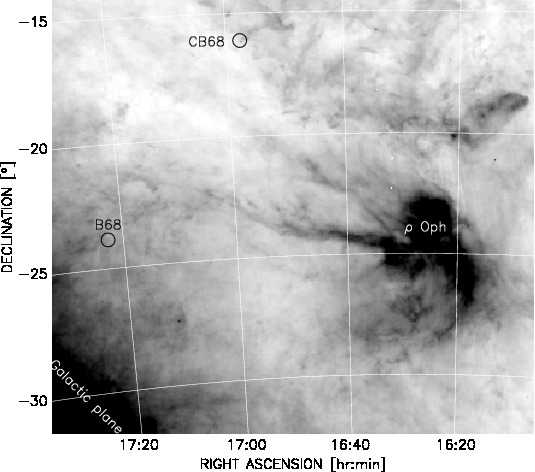}
 \caption{\label{fig-rhooph}
 IRAS 100\,$\mu$m dust continuum emission map, showing the $\rho$\,Oph region 
 and the loaction of two of our target sources (B\,68 and CB\,68). The plane of the Milky 
 Way becomes visible at the lower left corner of the map. Compare to Figs.\,\ref{fig-cb68}  
 and \ref{fig-cb68-morph}.}
\end{figure}


Based on the results of earlier studies \citep[e.g.,][]{lau97,laudwt97,henning98,stutz09a,lau10}, 
we have selected 12 well-isolated low-mass pre- and protostellar molecular cloud cores in regions 
of exceptionally low cirrus confusion noise. 
In order to obtain deep maps at 100\,$\mu$m, which is essential for a
precise estimate of the dust temperature, the absolute background levels and the 
point source confusion noise (PSCN) were important selection criteria. 
For this purpose, contributions from spatial fluctuations of the extragalactic background were derived from
\citet{negr04}. Estimates of the Galactic cirrus confusion noise are
based on the ISOPHOT confusion noise measurements, scaled down in the
power spectrum to the resolution of \herschel/PACS \citep{kiss05}.
Our selected globule sources have typical 100\,$\mu$m background levels of 
1\,mJy/$^{\square}$\arcsec\ or lower and PSCN\,$\le 1$\,mJy/beam. 
For comparison, typical 100\,$\mu$m background levels within the Taurus cloud are 
about $2-4$\,mJy/$^{\square}$\arcsec, within Ophiuchus about 
5\,mJy/$^{\square}$\arcsec, and reach values of $>100$\,mJy/$^{\square}$\arcsec\
in regions of high-mass star formation and infrared dark clouds (IRDCs).
This specific source selection strategy enabled us to obtain deeper 100\,$\mu$m maps than most 
large-area surveys and to derive robust dust temperature estimates.

Our target list contains only established and previously well-characterized 
sources and does not cover regions with unknown source content. 
All sources are nearby (100\,--\,400\,pc, with a mean distance of $240\pm100$\,pc), 
have angular diameters of 3\arcmin\ to 6\arcmin, linear sizes between 0.2 and 1.0\,pc, 
and total gas masses of 1 to 25\,M$_{\odot}$\ (see Table\,\ref{tab-res-physprop}).  
Coordinates, distances, spectral (evolutionary) classes, and references 
are listed in Table\,\ref{tab-sourcelist}.  Figure\,\ref{fig-skydistr} shows the
distribution of the target sources on the sky.
Figure\,\ref{fig-rhooph} illustrates that even those globules in our source
list that are loosely associated with larger dark cloud complexes
(like $\rho$\,Oph) are still isolated and located outside the regions
of high extinction and source confusion.

Seven out of the 12 target globules were known to contain starless
cores \citep{dwt94}. Of these, only CB\,17 was already shown before to be 
prestellar\footnote{We use the term ``prestellar'' only for those starless 
cores that have been shown to be gravitationally unstable or at the verge of collapse.} 
in nature \citep{pav2006}, while the star-forming potential of the
other sources is not known. CB\,17
contains, in addition to the prestellar core, a low-luminosity
Class\,I young stellar object (YSO) at a projected separation of
$\sim$25\arcsec\ \citep[$\sim$6000\,AU;][]{lau10}.  CB\,26 contains,
in addition to the starless core \citep{stutz09a}, a well-studied
Class\,I YSO at a projected separation of
$\sim$3.6\arcmin\ \citep[$\sim$0.15\,pc;][]{lau01,stecklum04,lau09,sauter09}.
Five out of the 12 target globules were known to host Class\,0
protostars.  Of these, three cores contain additional sources with
other evolutionary classifications.  CB\,130 contains an additional
low-mass starless core as well as a Class\,I YSO.
BHR\,12 contains an additional Class\,I core at a projected separation
of $\sim$20\arcsec\ ($\sim$8000\,AU).  CB\,244
contains an additional starless core at a projected separation of
$\sim$90\arcsec\ \citep[$\sim$18000\,AU;][]{lau10,stutz10}.
Two out of the 12 target globules (CB\,6 and CB\,230) are dominated by embedded Class\,I
YSOs, i.e., the extended emission in them supposedly arises from
remnant post-collapse envelopes.  Of these, CB\,230 is known to be a
binary source with a projected separation of
$\sim$10\arcsec\ \citep[$\sim$4000\,AU;][]{lau10}.

We have also re-evaluated the distance estimates toward all globules, 
confirming the earlier used distances for 7 sources \citep[e.g.,][]{lau10},
and adjusting the distances for 5 sources. 
CB\,4 and CB\,6, which we earlier associated with the so-called 
``$-$12\,km/s'' H\,I clouds at 600\,--\,800\,pc, are unlikely to be that far away,
as suggested by the lack of foreground stars within a 3\arcmin\ diameter 
area toward the cores. Despite their $v_{\rm LSR}\sim -12$\,km/s, 
they are more likely associated with the Cas\,A dark clouds 
in Gould's Belt at $\sim 350$\,pc \citep{perrot03}.
CB\,68 is associated with the $\rho$\,Oph dark clouds, for which we adopt 
the new precise trigonometric distance of 120\,pc from \citet{loinard11}.
B\,68 is located within the Pipe nebula and somewhat farther away from $\rho$\,Oph.
Various direct and indirect estimates suggest a distance of 135\,pc\,$\pm$15\,pc
\citep{degeus89,lom06,alves07}. For CB\,130, which is associated with the Aquila Rift clouds,
we adopt the distance estimate of 250\,pc\,$\pm$50\,pc from \citet{straizys03}.


\section{Observations and data reduction}                  \label{sec-obs}

The {\it Herschel} FIR continuum data are complemented by submm dust continuum emission 
maps at 450\,$\mu$m, 850\,$\mu$m, and 1.2\,mm, obtained with ground-based telescopes.
Observations and reduction of both the {\it Herschel} and complementary ground-based data 
are described in the following two subsections.
Representative general observing parameters are listed in Table\,\ref{tab-obs0},
and the observations of individual sources are summarized in Table\,\ref{tab-obs}.


\begin{table}[htb]
\caption{General observing parameters}             
\label{tab-obs0}      
\centering 
{\footnotesize       
\begin{tabular}{rcccc}     
\hline\hline       
$\lambda_0$~~   & Instr.    & $\langle$HPBW$\rangle$  &  $\langle$map size$\rangle$ &  $\langle$rms$\rangle$        \\
~[$\mu$m]       &           & [arcsec] & [arcmin] & [mJy/beam] \\
\hline  
100  & PACS    & 7.1\tablefootmark{a}       & 10    & 3    \\
160  & PACS    & 11.2\tablefootmark{a}      & 10    & 13   \\
250  & SPIRE   & 18.2\tablefootmark{a}      & 16-20 & 15   \\
350  & SPIRE   & 25.0 \tablefootmark{a}     & 16-20 & 17   \\
450  & SCUBA   & 8.5-9.5\tablefootmark{b}   & 2     & 200  \\
500  & SPIRE   & 36.4\tablefootmark{a}      & 16-20 & 14   \\
850  & SCUBA   & 14.4-15.1\tablefootmark{b} & 3     & 20   \\
870  & LABOCA  & 19.2\tablefootmark{c}      & 40    & 8    \\
1200 & MAMBO-2 & 11                         & 5-10  & 6    \\
\hline                  
\end{tabular}
\tablefoot{
\tablefoottext{a}{\citet{aniano11}}
\tablefoottext{b}{\citet{lau10}}
\tablefoottext{c}{\citet{nielbock12}}
}}
\end{table}


\begin{table*}[htb]
\caption{Summary of FIR and submm observations}             
\label{tab-obs}      
\centering 
{\footnotesize       
\begin{tabular}{llllll}     
\hline\hline       
Source &  PACS & SPIRE & 450\,$\mu$m & 850\,$\mu$m & 1.2\,mm \\
       &  Obs\,ID & Obs\,ID  &             &             &         \\
\hline  
CB\,4 	    & 134221614(5,6) & 1342188689 &  --                   & --                   & M11     \\
CB\,6 	    & 134222272(3,4) & 1342189703 &  L10                  & L10                  & L10/M11 \\
CB\,17 	    & 134219100(8,9) & 1342190663 &  --                   & L10                  & L10/M11 \\
CB\,26 	    & 134221739(7,8) & 1342191184 &  L10\tablefootmark{a} & L10\tablefootmark{a} & L10/M11 \\
CB\,27 	    & 134219197(5,6) & 1342191182 &  DF08                 & DF08                 & M11     \\
BHR\,12     & 13421985(39,40) & 1342193794 &  L10                 & L10                  & L10     \\
CB\,68 	    & 134220428(6,7) & 1342192060 &  L10                  & L10                  & L10     \\
B\,68 	    & 134219305(5,6) & 1342191191 &  DF08                 & DF08/N12             & B03     \\
CB\,130     & 134220598(2,3) & 1342191187 &  L10                  & L10                  & L10     \\
B\,335 	    & 134219603(0,1) & 1342192685 &  L10                  & L10                  & M11     \\
CB\,230     & 134218608(3,4) & 1342201447 &  L10                  & L10                  & L10/K08 \\
CB\,244     & 134218869(4,5) & 1342199366 &  L10                  & L10                  & L10     \\
\hline                  
\end{tabular}
\tablefoot{
\tablefoottext{a}{The SCUBA maps of CB\,26 cover only the
  western core with the embedded Class\,I YSO, but not the
  eastern starless core.}
}
\tablebib{
(L10)~\citet{lau10};
(M11)~dedicated MAMBO-2 observations, described in this paper (see Sect.\,\ref{ssec-obs-submm});
(DF08)~\citet[][SCUBA Legacy Catalogues]{difra08}; 
(B03)~\citet[][SEST-SIMBA observations]{bianchi03};
(N12)~\citet[][APEX-Laboca observations]{nielbock12};
(K08)~\citet[][IRAM-MAMBO-2 observations]{kauf2008}
}}
\end{table*}


\subsection{Herschel FIR Observations}      \label{ssec-obs-herschel}

FIR continuum maps at five wavelength bands were obtained with two different instruments 
on board the \herschel\ Space observatory: PACS at 100 and 160\,$\mu$m, 
and SPIRE at 250, 350, and 500\,$\mu$m.


\subsubsection{PACS data}      \label{sssec-obs-herschel_pacs}

All 12 sources were observed with the Photodetector Array Camera and
Spectrometer \citep[PACS; ][]{pog10} in the scan map mode 
at 100 and 160\,$\mu$m simultaneously. 
For each source, we obtained two scan directions oriented perpendicular to each 
other to eliminate striping in the final combined maps of effective size 
$\sim\!10\arcmin\times 10\arcmin$. The scan speed
was set to 20$\arcsec$~sec$^{-1}$, and a total of 30 repetitions were obtained in each 
scan direction, resulting in a total integration time of $\sim 2.6$\,hrs per map.
We emphasize that the accurate
recovery of extended emission is the main driver for the following
discussion and exploration of reduction schemes.  The PACS data at
100~$\mu$m and 160~$\mu$m were processed in an identical fashion.
They were reduced to level\,1 using HIPE v.\,6.0.1196 \citep{ott10}, except for CB4
(processed with HIPE v.\,6.0.2044), CB26 (processed with
v.\,6.0.2055), and CB6 (processed with v.\,7.0.1931).  
The rationale for using different versions of HIPE for different sources is that the latter three sources were 
observed much later (OD 660-770) than the other sources (OD 230-511) and the data were reduced with the respective latest 
HIPE version and calibration tree that was available at that time. Since processing of the PACS scanning data 
is very time-consuming, a re-reduction of all data with every new HIPE version with modifications 
relevant for our data is not feasible. Instead, we re-reduced the data for one source with the latest version 
of HIPE to verify how much the 
data reduction and post-processing modifications affect our results on the derived dust temperature and column density maps.
Since this comparison, which is presented and discussed in Sect.\,\ref{ssec-dis-data}, showed that 
the effects are well within the range of the other uncertainties, we do not re-reduce all data, but 
incorporate these effects into our uncertainty assessment in Sect.\,\ref{ssec-dis-data}.

Apart from the standard reduction steps, we applied {\it '2$^{nd}$ level deglitching'} 
to remove outliers in the time series data ({\it'time-ordered'} option) 
by applying a clipping algorithm (based on the median absolute deviation) to all
flux measurements in the data stream that will finally contribute to the
respective map pixel. For our data sets we applied an {\it 'nsigma'} value of 20.

After producing level\,1 data, we have generated final level\,2 maps
using two different methods: high-pass filtering with photproject (within HIPE) and
Scanamorphos \citep{roussel11}, which does not use straightforward high-pass filtering, 
but instead applies its own heuristic algorithms to remove artifacts caused by detector
flickering noise as well as spurious bolometer temperature drifts.
Based on the high-pass median-window subtraction method, the photproject images turned 
out to suffer from more missing flux and striping than the Scanamorphos images. 
Because the correct recovery of extended emission is critical for accurate 
temperature mapping, high-pass filtering was clearly disadvantageous for our science goals.
We also note that in a previous reduction, we found that the 
MADmap\footnote{http://herschel.esac.esa.int/twiki/pub/Pacs/PacsMADmap/\-UM\_MADmapDOC.txt} 
processing of our brighter sources (e.g., B\,335) used to produce large
artifacts\footnote{This problem has eventually been solved in the most recent
MADmap implementation.} in the
final maps due to the relatively bright central protostar; therefore
we do not utilize this reduction scheme in this work.

For these reasons we have decided to use Scanamorphos v.\,9 as our standard 
level 2 data reduction algorithm. A comparison of final maps processed with 
different Scanamorphos v.\,9 options showed that the ``galactic'' option 
recovered the highest levels of extended emission and was thus the best-suited 
for our data set and scientific goals.  We use a uniform final map pixel scale of
$1\farcs6$~pix$^{-1}$ at 100~$\mu$m and $3\farcs2$~pix$^{-1}$ at
160~$\mu$m for all objects.  Furthermore, these data were processed
including the non-zero-acceleration telescope turn-around data,
with no additional deglitching ({\it 'noglitch'} setting).


\subsubsection{SPIRE data}      \label{sssec-obs-herschel-spire}

All 12 sources were also observed at 250, 350, and 500\,$\mu$m with the Spectral and Photometric Imaging
Receiver \citep[SPIRE;][]{griffin10}.  The scan maps of each source were
obtained simultaneously at all three bands at the nominal scan
speed of 30\arcsec/sec and over a $(16\arcmin-20\arcmin)^2$\ area, resulting in a total integration time of 
$10-16$\,mins per map.
These data were processed up to level\,1 with HIPE v.\,5.0.1892 and calibration tree 5.1,
which was the most up-to-date version at the time all data became available.
In Sect.\,\ref{ssec-dis-data}, we show that using the newest HIPE v.\,9.1.0 only leads to  
nonsignificant minor changes in the SPIRE maps that are well below all other uncertainties.
For this reason, we considered it unnecessary to re-reduce the data.
Up to level\,1 (i.e., the level where the pointed photometer timelines are derived), the steps
of the official pipeline (POF5\_pipeline.py, dated 2.3.2010) provided
by the SPIRE Instrument Control Center (ICC) have been performed. 
The level\,1 frames were then processed with the Scanamorphos software
version 9 \citep{roussel11}. All maps were reduced using the
{\it 'galactic'} option.  The final map pixel sizes are
6, 10, and $14\arcsec$\ at 250, 350, and 500\,$\mu$m, respectively.

Photometric color corrections for both PACS and SPIRE data, that account for the difference 
in spectral slopes between flux calibration (flat spectrum) and actually observed source spectrum, 
are applied only in the data analysis and are described in Sect.\,\ref{ssec-mod-prep}.


\subsection{(Sub)mm continuum Observations}      \label{ssec-obs-submm}

Complementary submm dust continuum emission maps at 450\,$\mu$m,
850\,$\mu$m, and 1.2\,mm from ground-based telescopes are available or
were obtained through dedicated observations for all 12 sources.
References to the existing and published data used here are given in
Table\,\ref{tab-obs}.

Additional dedicated maps of the 1.2\,mm dust emission from 6
sources were obtained with the 117-pixel MAMBO-2 bolometer array of
the Max-Planck-Institut f\"ur Radioastronomie \citep{kreysa99} at the
IRAM 30\,m-telescope on Pico Veleta (Spain) during two pool observing
runs in October and November 2010.
The mean frequency (assuming a flat-spectrum source) was 250\,GHz
($\lambda_0\sim$\,1.2\,mm) with a half power (HP) bandwidth of $\sim$80\,GHz. 
The HP beam width (HPBW) on sky was 11\arcsec, and the FoV of the array
4\arcmin\ (20\arcsec\ pixel spacing).  Weather conditions were good,
with zenith optical depths between 0.1 and 0.35 for most of the time
and low sky noise.  Pointing, focus, and zenith optical depth (by
skydip) were measured before and after each map.  The pointing
stability was better than 3\arcsec\ rms.
Absolute flux calibration was obtained by observing Uranus several times during
the observing runs and monitored by regularly observing several
secondary calibrators. The flux calibration uncertainty is dominated
by the uncertainty in the knowledge of the planet fluxes and is
estimated to be $\sim$20\%.
The sources were observed with the standard on-the-fly dual-beam
technique, with the telescope scanning in azimuth direction at a speed
of 8\arcsec/s while chopping with the secondary mirror along the scan
direction at a rate of 2\,Hz.  Most sources were mapped twice, with
different projected scanning directions (ideally orthogonal) and
different chopper throws (46\arcsec\ and 60\arcsec) to minimize
scanning artifacts in the final maps.  For one source (CB\,27), 
we could obtain only one coverage.  Effective map sizes
(central pixel coverage) were adapted to the individual source sizes and
were in the range 5\arcmin\,--\,10\arcmin, resulting in total mapping
times per coverage of 40\,min to 1.5\,hrs.  

The raw data were reduced using the standard pipeline "mapCSF" provided with the MOPSIC 
software\footnote{Created and continuously updated by R. Zylka, IRAM, Grenoble; 
see http://www.iram.es/IRAMES/mainWiki/CookbookMopsic}. 
Basic reduction steps include
de-spiking from cosmic ray artifacts, 
correlated signal filtering to reduce the skynoise,
baseline fitting and subtraction,
restoring single-beam maps from the dual-beam scan data using a modified 
EKH algorithm \citep{emerson79}, 
and transformation and averaging of the individual horizontal maps into the equatorial system.
Correlated skynoise correction was performed iteratively, starting with no source model 
in the first run. The resulting map is smoothed and the region with source flux is 
selected manually to represent the first source model, which is then iteratively improved 
in 20 successive correlated skynoise subtraction runs.
Since our sources are faint, we follow the recommendation to apply this algorithm only up to level 1, 
except for the brightest source B335, where we go up to level 3. 
A third order polynomial baseline fit to the time-ordered data stream as well as 
first order fits to the individual scan legs are subtracted from the data, after masking out 
the region with source flux, to correct for slow atmospheric and instrumental drifts.
Despite some unsolved electronic problems during these observing runs, 
which limited the efficiency of the correlated 
skynoise subtraction, the resulting noise levels in the final maps are $\sim$6\,mJy/beam,
except for CB\,27, where it is $\sim$11\,mJy/beam because only one coverage could be obtained.

The assumption of a flat-spectrum source in the definition of the
nominal wavelength of the MAMBO detectors in combination with the
broad bandpass made it necessary to apply a "color correction". 
In contrast to the \herschel\ FIR observations, (sub)mm wavelengths are
sufficiently close to the Rayleigh-Jeans regime and the emission is optically 
thin for all our sources, such that the slope of the SED follows 
$S_{\nu}\propto\lambda^{-4}$\ (for the dust opacity submm spectral index $\beta=2$). 
Therefore, we follow \citet{schnee10} and 
correct the reference wavelength for all MAMBO data to 1100\,$\mu$m.  
Color corrections for the SCUBA data are negligibly small \citep[cf.][]{schnee10} 
and were not applied.

We also verified the pointing in the final maps by comparing the peak positions 
with interferometric positions where possible (CB\,26, CB\,68, CB\,230, CB\,244, B\,335, BHR\,12)
and found that deviations are in all cases smaller than 2\arcsec, 
hence confirming the pointing stability and making additional pointing corrections unnecessary.
The final maps of those sources for which we had both old and new 1.2\,mm data 
(CB\,6, CB\,17, CB\,26, and CB\,230) were generated by weighted averaging, 
after adapting beam sizes and correcting small pointing offsets.

Additional complementary NIR extinction maps as well as different molecular spectral line maps  
were also obtained with the aim of
studying the relation between density, temperature, and dust
properties on the one side, and gas phase abundances and chemistry on
the other side.  These latter data will be described and analyzed in
forthcoming papers and are not presented here.


\section{Modeling approach}                  \label{sec-mod}

\subsection{Strategy and data preparation}   \label{ssec-mod-prep}


\begin{table*}[htb]      
\centering 
\caption{DC--flux levels in the {\it Herschel} data}             
\label{tab-fdc}
{\footnotesize       
\begin{tabular}{lccccccc}     
\hline\hline       
Source & 
R.A.,Dec.\tablefootmark{a} & 
f${_{\rm DC}}$\ ($\sigma_f$) &
f${_{\rm DC}}$\ ($\sigma_f$) &
f${_{\rm DC}}$\ ($\sigma_f$) &
f${_{\rm DC}}$\ ($\sigma_f$) & 
f${_{\rm DC}}$\ ($\sigma_f$) \\
                           &
(J2000)                    & 
100\,$\mu$m                & 
160\,$\mu$m                & 
250\,$\mu$m                & 
350\,$\mu$m                & 
500\,$\mu$m                \\
                           &
[h:m:s, \degr:\arcmin:\arcsec]& 
[$\mu$Jy/$^{\square}$\arcsec]       &
[$\mu$Jy/$^{\square}$\arcsec]       &
[$\mu$Jy/$^{\square}$\arcsec]       &
[$\mu$Jy/$^{\square}$\arcsec]       &
[$\mu$Jy/$^{\square}$\arcsec]       \\
\hline  
CB\,4 	                     & 00:38:55, $+$52:56:06 & 8.4E2 (4.4E1) & 5.9E2 (1.0E2) & 2.8E2 (3.9E1) & 1.4E2 (2.1E1) & 6.9E1 (1.0E1) \\
CB\,6 	                     & 00:49:02, $+$50:42:34 & 1.1E3 (3.0E1) & 4.6E2 (3.7E1) & 1.8E2 (2.2E1) & 8.7E1 (1.1E1) & 3.8E1 (5.0E0) \\
CB\,17 	                     & 04:04:14, $+$56:53:07 & 1.3E3 (4.2E1) & 7.5E2 (4.5E1) & 2.6E2 (3.1E1) & 1.3E2 (1.6E1) & 6.5E1 (7.0E0) \\
CB\,26 	                     & 05:00:28, $+$52:00:36 & 7.8E2 (4.7E1) & 7.7E2 (1.1E2) & 1.4E2 (2.7E1) & 6.8E1 (1.0E1) & 3.3E1 (7.0E0) \\
CB\,27 	                     & 05:04:16, $+$32:38:19 & 1.0E3 (4.7E1) & 6.7E2 (1.3E2) & 1.9E2 (2.8E1) & 9.4E1 (1.7E1) & 5.0E1 (7.0E0) \\
BHR\,12                      & 08:09:10, $-$36:07:44 & 1.3E3 (5.7E1) & 6.7E2 (8.5E1) & 3.1E2 (2.9E1) & 1.7E2 (1.3E1) & 7.7E1 (5.0E0) \\
CB\,68 	                     & 16:57:07, $-$16:14:27 & 1.3E3 (5.4E1) & 5.1E2 (6.9E1) & 2.2E2 (3.5E1) & 1.3E2 (2.0E1) & 4.2E1 (1.1E2) \\
B\,68 	                     & 17:22:36, $-$23:44:11 & 9.1E2 (8.4E1) & 8.9E2 (7.2E1) & 1.2E2 (2.9E1) & 5.9E1 (1.4E1) & 1.3E1 (5.0E0) \\
CB\,130                      & 18:16:30, $-$02:29:23 & 1.4E3 (7.1E1) & 8.0E2 (1.5E2) & 3.0E2 (7.4E1) & 1.6E2 (3.9E1) & 7.7E1 (1.7E2) \\
B\,335 	                     & 19:36:48, $+$07:30:37 & 1.5E3 (4.3E1) & 7.2E2 (7.9E1) & 1.5E2 (2.1E1) & 8.7E1 (9.0E0) & 3.9E1 (6.0E0) \\
CB\,230                      & 21:17:00, $+$68:14:35 & 1.3E3 (4.8E1) & 5.1E2 (1.8E2) & 1.8E2 (4.3E1) & 1.4E2 (7.1E1) & 7.4E1 (2.8E1) \\
CB\,244                      & 23:24:55, $+$74:23:01 & 1.3E3 (4.0E1) & 5.1E2 (7.5E1) & 3.0E2 (2.7E1) & 1.7E2 (1.3E1) & 7.7E1 (7.0E0) \\
\hline
\end{tabular}
\tablefoot{ \tablefoottext{a}{Center of the DC reference field (see Sect.\,\ref{ssec-mod-prep}).} }}
\end{table*}


The calibrated dust emission maps at the various wavelengths were used 
to compile spatially and spectrally resolved data cubes that cover
the full extent of these relatively isolated sources on 
both sides of the peak of their thermal SEDs.
Ideally, one would want to compare such data to synthetic maps from radiative 
transfer models convolved with the respective beams (forward-modeling, virtual observations) 
in order to constrain the physical properties of the sources. However, 
to avoid circular averaging with its well-known caveats, one would need 3-D
modeling to account for the complex structure present even in these
relatively simple and isolated sources. Furthermore, not all of the observed features 
(like, e.g., the core-envelope temperature contrast, see Sects.\,\ref{ssec-res-temp} and 
\ref{ssec-res-isrf}) may be easily reproducible with existing self-consistent models, 
making the fully self-consistent forward-modeling approach very time-consuming.
This will be dealt with in forthcoming papers modeling individual cores. 

For this survey overview paper, we therefore take a simpler and more direct approach, 
giving up some of the spatial information in the short-wavelength maps and directly 
recovering and modeling beam and LoS optical-depth-averaged (hereafter LOS-averaged for short) 
SEDs, thus introducing as few as 
possible model-dependent assumptions into our analysis.  
For this purpose, the calibrated maps are prepared in the following way: 
\begin{enumerate}
\item all maps are registered to a common coordinate system,
\item pointing corrections are applied where necessary (see below),
\item all maps are converted to the same physical surface brightness units 
      (here Jy/$^{\square}$\arcsec) and extended emission calibration corrections are 
      applied where necessary (see below),
\item background levels are determined and subtracted from the maps (see below), and finally,
\item all maps are convolved to the SPIRE\,500\,$\mu$m beam (FWHM 36\asp4).
\end{enumerate}

\noindent {\it PACS pointing corrections:} 
Since the \herschel\ spacecraft absolute 1\,$\sigma$\ pointing error was $\sim$2\arcsec\ and 
the PACS and SPIRE maps were not obtained simultaneously, random relative pointing errors
between PACS and SPIRE maps can add up to $\sim$3\arcsec\,--\,4\arcsec, which is about 
half the PACS 100\,$\mu$m FWHM beam size \citep{aniano11}. This could severely hamper 
the usefulness of flux ratio (SED) maps, in particular around compact sources, 
if not corrected properly. For the long-wavelength SPIRE maps with beam sizes 
$\sim$18\arcsec\,--\,36\arcsec\ and no readily available astrometric reference, 
the pointing errors are fortunately negligible. For the PACS maps we utilize the fact 
that many 100\,$\mu$m point sources are also detected in the {\it Spitzer} MIPS 24\,$\mu$m images. 
We select in each field three stars detected in both the PACS 100\,$\mu$m and MIPS 24\,$\mu$m 
images to align the PACS astrometry in both the 100\,$\mu$m and 160\,$\mu$m images to the {\it Spitzer} images.  
The pointing corrections were found to be $\le 3$\arcsec\ in all cases, i.e., 
small compared to the SPIRE\,500\,$\mu$m beam. 

\noindent {\it Surface brightness calibration corrections:}
Since the PACS images are calibrated to Jy\,pix$^{-1}$, the
conversion to Jy/$^{\square}$\arcsec\ does not involve an assumption of the
beam sizes. The SPIRE data, on the other hand, are calibrated to
Jy\,beam$^{-1}$; to convert to surface brightness units we adopt the 
FWHM beam sizes from \citet{aniano11} listed in Tab.\,\ref{tab-obs0}. 
The (sub)mm data have also been converted to Jy/$^{\square}$\arcsec\
assuming the appropriate effective beam sizes for the respective observations 
(see references in Table\,\ref{tab-obs}).
Since the standard calibration is done on point sources, 
but we are using surface brightnesses to compile spatially 
resolved SEDs, we have applied the recommended extended emission calibration 
corrections to the SPIRE data
(see SPIRE Observer's Manual: HERSCHEL-DOC-0798, version 2.4, June 7, 2011).

\noindent {\it PACS and SPIRE map background subtraction:}  
The Scanamorphos ``galactic'' option preserves the large spatial frequency 
modes of the bolometer signal in the {\it Herschel} scan-maps. 
However, the absolute flux level of the background and spatial structures
more extended than the scan-maps cannot be recovered since the exact level of the strong thermal 
background from the only passively cooled mirrors M1 and M2 of the observatory is unknown. 
Therefore, it is not clear how the extended, large-scale emission levels 
in the final {\it Herschel} maps are related to the absolute flux level of 
the background, which is composed of emission features larger than the map size,
the general galactic background ISRF, the
cosmic microwave background (CMB), and other possible contributions,
all varying differently with wavelength.  
Since uncertainties in additive flux contributions affect flux ratio measurements 
and the respective derivation of physical parameters from the SEDs, in particular 
at low flux levels, we subtract the background levels from all {\it Herschel} maps,
thus making them consistent with the chopped (sub)mm data 
(although chopping is a much more aggressive filtering than that present in the
\herschel\ maps). The consequences and uncertainties of this background removal on the 
data analysis are discussed in Sects.\,\ref{ssec-mod-gb} and \ref{ssec-dis-data}.
We emphasize that this approach was feasible only because our sources 
were already initially selected to be relatively isolated on the sky (see Sect.\,\ref{sec-sources}).

To accomplish this re-zeroing, we adopt a method similar to that
applied to {\it Spitzer} MIPS images in
\citet{stutz09a}.  For each object we identify a
$4\arcmin\times4\arcmin$ region in the PACS and SPIRE images that is
relatively free from spatially varying emission and appears ``dark''
relative to the globule extended emission levels.  We impose the
additional requirement that this region is in or near a region in our
complementary molecular line  maps which is relatively free of \mbox{$^{12}$CO(2--1)} emission.  
We always use the same region in all five {\it Herschel} scan-maps.  
For each band, we then calculate the representative flux level, 
$f_{\rm DC}$, in the $4\arcmin\times4\arcmin$
region by implementing an iterative Gaussian fitting and
$\sigma$-clipping scheme to the pixel value distribution at each
wavelength.  
We do not consider pixels below 2$\sigma$\ from the mean and fit the
main peak of the pixel value distribution.  The $f_{\rm DC}$\ value is
then determined by iteratively fitting a Gaussian function to the
histogram of pixel values, where at each iteration we include one more
adjacent higher flux bin.  The final adopted value of $f_{\rm DC}$\ is
defined as the mean value of the best-fit Gaussian that has the
minimum $\sigma$\ value for all iterations. In this way we exclude
pixels with higher flux which effectively broaden the
distribution and would cause a bias in the $f_{\rm DC}$\ calculation. 
The $f_{\rm DC}$\ values are then subtracted from the corresponding 
\herschel\ maps at the respective wavelength 
(see Table\,\ref{tab-fdc} for the $f_{\rm DC}$ and $\sigma$ values and coordinates of the 
reference region).

\noindent {\it Image convolution:}  Finally, all maps are convolved to the beam
of the SPIRE\,500\,$\mu$m map, using the azimuthally averaged {\it Herschel} 
convolution kernels provided by \citet{aniano11}.
For the convolution of the (sub)mm maps, we use a Gaussian convolution kernel 
of width equal to the difference in quadrature between the effective FWHM of the (sub)mm 
observations and the SPIRE 500\,$\mu$m FWHM of 36\asp4.

After re-gridding to a common Nyquist-sampled pixel grid, 
we thus compile pixel-by-pixel SEDs for the wavelength
range from 100\,$\mu$m through 1.2\,mm.  
Photometric color corrections for each pixel and each band are iteratively derived 
from the 100--160\,$\mu$m and 250--500\,$\mu$m spectral indices and polynomial 
fits to color correction factors provided for PACS in Table\,2 of \citet{mueller11} and  
in Table\,5.3 of the SPIRE Observers’ Manual, and are applied in the subsequent SED modeling only.  

The nominal point-source calibration uncertainty of the PACS photometers listed in the latest calibration notes 
is 3\% at 100\,$\mu$m and 5\% at 160\,$\mu$m. However, additional uncertainites that are hard to quantify 
are introduced by, e.g., the conversion to surface brightness units for extended emission, 
the beam convolution (imperfect kernels), and the color corrections.  
Likewise, the final recommended calibration uncertainty of 7\% for SPIRE does not account for some of the 
uncertainties introduced in the post-processing. 
Recent systematic comparisons between the \herschel\ and \spitzer\ surface brightness calibrations 
showed that they agree to within about 10\%, with intrinsic uncertainties on both 
sides\footnote{see PICC-NHSC-TR-034 and PICC-NHSC-TN-029 under https://nhscsci.ipac.caltech.edu/.}.
Therefore, we adopt for all \herschel\ data the conservative value of 15\% for the relative calibration uncertainty
in the final maps.
This number is used for calculating the weights of the individual data points and with respect to the ground-based data 
in the subsequent fitting procedure (Sect.\,\ref{ssec-mod-gb}).


\subsection{Deriving dust temperature and column density maps}   \label{ssec-mod-gb}

The subtraction (or chopping out) of a flat background level from the emission maps 
implies that the remaining emission in the map at each image pixel is given by
\begin{equation}                      \label{eq-snu}
S_{\nu}(\nu) = \Omega\,\left(1-e^{-\tau(\nu)}\right)\,\left(B_{\nu}(\nu,T_{\rm d})-I_{\rm bg}(\nu)\right)
\end{equation}
with 
\begin{equation}                      \label{eq-tau1}
\tau(\nu) = N_{\rm H}\,m_{\rm H}\,\frac{M_{\rm d}}{M_{\rm H}}\,\kappa_{\rm d}(\nu)
\end{equation}
where
$S_{\nu}(\nu)$ is the observed flux density at frequency $\nu$, 
$\Omega$\ the solid angle from which the flux arises (here normalized to 1\,arcsec$^2$),
$\tau(\nu)$\ the optical depth through the cloud,
$B_{\nu}(\nu,T_{\rm d})$\ the Planck function,
$T_{\rm d}$\ the dust temperature,
$I_{\rm bg}(\nu)$\ the background flux level, 
$N_{\rm H} = 2\times N{\rm (H_2)} + N{\rm (H)}$\ the total hydrogen column density,
$m_{\rm H}$\ the proton mass, 
$M_{\rm d}/M_{\rm H}$\ the dust-to-hydrogen mass ratio, and 
$\kappa_{\rm d}(\nu)$\ the dust mass absorption coefficient.
$T_{\rm d}$\ and $\kappa_{\rm d}$\ may vary along the LoS,
in which case Eq.\,\ref{eq-snu} has to be written in differential form 
and integrated along the LoS. However, for this overview paper we assume 
both parameters to be constant along the LoS and discuss the uncertainties and 
limitations of this approach in Sect.\,\ref{ssec-dis-mod}.

For $\kappa_{\rm d}(\nu)$, we assume for all sources in this paper the tabulated values 
listed by \citet{oh94} for mildly coagulated ($10^5$\,yrs coagulation time at gas density $10^6$\,cm$^{-3}$)
composite dust grains with thin ice mantles (column\,5 in their Table\,1, usually called ``OH5''),
logarithmically interpolated to the respective wavelength where necessary.
For the hydrogen-to-dust mass ratio in the Solar neighborhood, we adopt \mbox{$M_{\rm H}/M_{\rm d}=110$} 
\citep[e.g.,][]{sod1997}. 
Note that the total gas-to-dust mass ratio, accounting for helium and heavy elements, 
is about 1.36 times higher, i.e., \mbox{$M_{\rm g}/M_{\rm d}\approx150$}.

The exact background flux levels at the location of the individual clouds, $I_{\rm bg}(\nu)$,  
are a priori unknown as explained in Sect.\,\ref{ssec-mod-prep}.
At the frequencies relevant for this paper, this background radiation is composed mainly of the CMB, 
the extragalactic cosmic infrared background (CIB), and the diffuse galactic background (DGB).
While the first contribution (CMB) is well-known and dominates at wavelength $> 500\,\mu$m, 
the mean levels of the CIB have been compiled by, e.g., \citet{hauserdwek2001}
from flux measurements in the "Lockman Hole" by various space observatories 
(e.g., {\it COBE} and {\it ISO}) and ground-based (sub)mm instruments 
(e.g., SCUBA). However, at wavelengths $<350\,\mu$m, the DGB, 
which strongly varies with position, dominates over CMB and CIB. 
Here we follow the approach of \citet{stutz10} and use the \citet{schlegel98} 100\,$\mu$m IRAS maps 
and the ISO Serendipity Survey observations at 170\,$\mu$m to extrapolate approximate mean 
flux levels at the typical galactic locations of our sources. 
Since the uncertainty in the resulting temperature and column density estimates introduced 
by the uncertainty in the exact knowledge of the $I_{\rm bg}$\ levels is negligible,
as discussed quantitatively in Sect.\,\ref{ssec-dis-data}, we do not attempt to derive  
the local values of the DGB for each source separately.
Instead, we adopt the following mean values for all sources: 
0.3, 0.8, 0.5, 0.3, 0.2, 2.9, and 6.5\,mJy/arcsec$^2$\ 
at $\lambda$\ 100, 160, 250, 350, 500, 850, and 1100\,$\mu$m, respectively.

\begin{figure*}
 \centering
 \includegraphics[width=15cm]{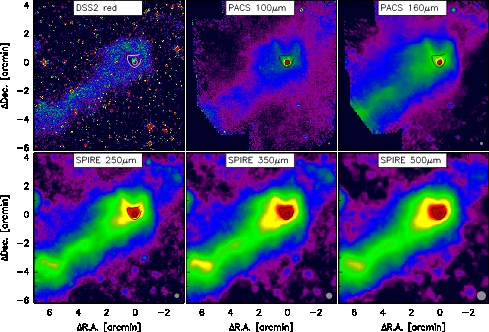}
 \caption{\label{fig-cb6}
          CB\,6 (Class\,I YSO): Visual (red) DSS2 image and Herschel FIR maps (color; log scale) at 100, 160, 250, 350,
          and 500\,$\mu$m, with 850\,$\mu$m continuum contours overlaid 
          (180 and 300\,mJy/20$^{\prime\prime}$beam). Herschel beam sizes are indicated 
          as gray circles in the lower right corners.}
\end{figure*}

\begin{figure*}
 \centering
 \includegraphics[width=15cm]{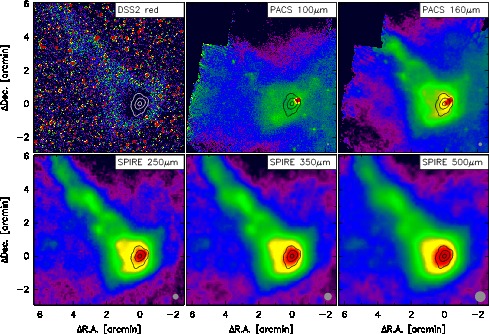}
 \caption{\label{fig-cb17}
          CB\,17 (starless core and embedded Class\,I YSO): 
          Visual (red) DSS2 image and Herschel FIR maps (color; log scale) at 100, 160, 250, 350,
          and 500\,$\mu$m, with 1.2\,mm continuum contours overlaid 
          (20, 40, and 65\,mJy/20$^{\prime\prime}$beam). Herschel beam sizes are indicated 
          as gray circles in the lower right corners.}
\end{figure*}

\begin{figure*}
 \centering
 \includegraphics[width=15cm]{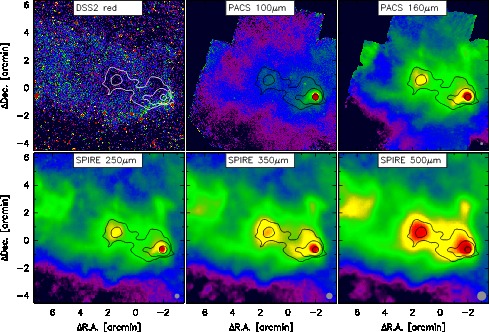}
 \caption{\label{fig-cb26}
          CB\,26 (starless core and embedded Class\,I YSO): 
          Visual (red) DSS2 image and Herschel FIR maps (color; log scale) at 100, 160, 250, 350,
          and 500\,$\mu$m, with 1.2\,mm continuum contours overlaid 
          (10, 26, and 150\,mJy/30$^{\prime\prime}$beam). 
          Herschel beam sizes are indicated 
          as gray circles in the lower right corners.}
\end{figure*}

\begin{figure*}
 \centering
 \includegraphics[width=15cm]{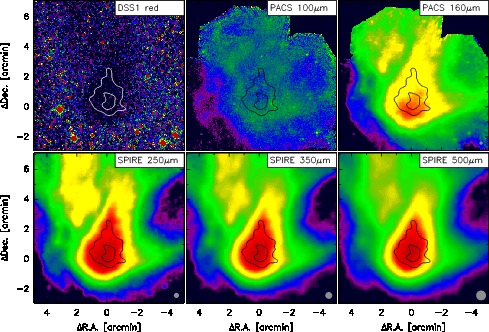}
 \caption{ \label{fig-cb27}
          CB\,27 (starless core): Visual (red) DSS1 image and Herschel FIR maps (color; log scale) at 100, 160, 250, 350,
          and 500\,$\mu$m, with 1.2\,mm continuum contours overlaid 
          (40 and 70\,mJy/30$^{\prime\prime}$beam). 
          Herschel beam sizes are indicated as gray circles in the lower right corners.}
\end{figure*}

\begin{figure*}
 \centering
 \includegraphics[width=15cm]{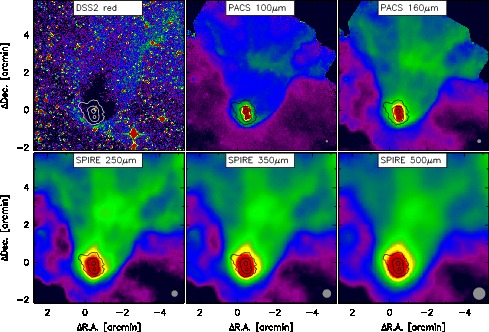}
 \caption{\label{fig-bhr12}
          BHR\,12 (Class\,0/I protostars): 
          Visual (red) DSS2 image and Herschel FIR maps (color; log scale) at 100, 160, 250, 350,
          and 500\,$\mu$m, with 850\,$\mu$m continuum contours overlaid (SCUBA; 120, 
          300, and 900\,mJy/15$^{\prime\prime}$beam). Herschel beam sizes are indicated 
          as gray circles in the lower right corners.}
\end{figure*}

\begin{figure*}
 \centering
 \includegraphics[width=15cm]{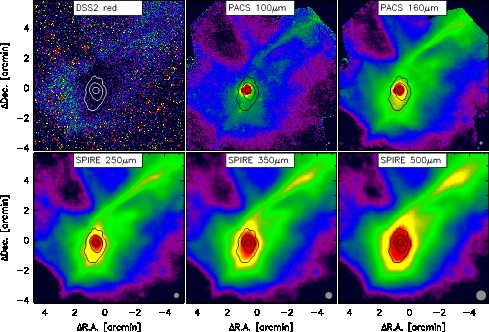}
 \caption{\label{fig-cb68}
          CB\,68 (Class\,0 protostar): 
          Visual (red) DSS2 image and Herschel FIR maps (color; log scale) at 100, 160, 250, 350,
          and 500\,$\mu$m, with 850\,$\mu$m continuum contours overlaid 
          (150, 300, and 600\,mJy/20$^{\prime\prime}$beam). Herschel beam sizes are indicated 
          as gray circles in the lower right corners.}
\end{figure*}

\begin{figure*}
 \centering
 \includegraphics[width=15cm]{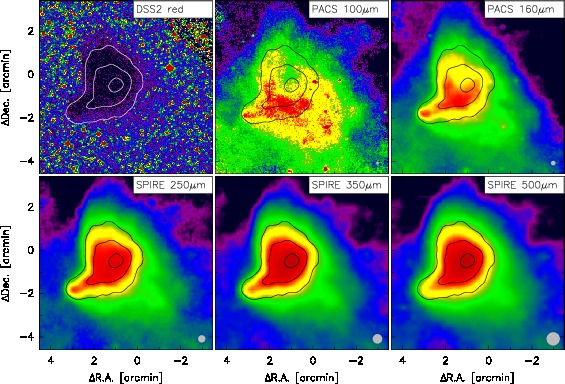}
 \caption{\label{fig-b68}
          B\,68 (starless core):  Visual (red) DSS2 image and Herschel FIR maps (color; log scale) at 100, 160, 250, 350,
          and 500\,$\mu$m, with 870\,$\mu$m continuum contours overlaid 
          (40, 120, and 220\,mJy/27$^{\prime\prime}$beam). Herschel beam sizes are indicated 
          as gray circles in the lower right corners.}
\end{figure*}

\begin{figure*}
 \centering
 \includegraphics[width=15cm]{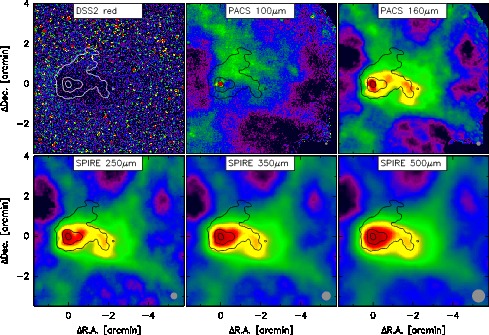}
 \caption{\label{fig-cb130}
          CB\,130 (Class\,0 protostar): 
          Visual (red) DSS2 image and Herschel FIR maps (color; log scale) at 100, 160, 250, 350,
          and 500\,$\mu$m, with 1.2\,mm continuum contours overlaid 
          (20, 60, and 100\,mJy/20$^{\prime\prime}$beam). Herschel beam sizes are indicated 
          as gray circles in the lower right corners.}
\end{figure*}

\begin{figure*}
 \centering
 \includegraphics[width=15cm]{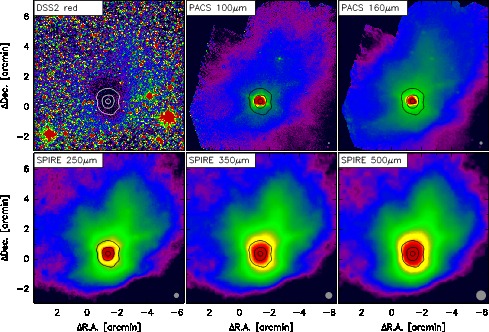}
 \caption{\label{fig-b335}
          B\,335 (Class\,0 protostar): 
          Visual (red) DSS2 image and Herschel FIR maps (color; log scale) at 100, 160, 250, 350,
          and 500\,$\mu$m, with 850\,$\mu$m continuum contours overlaid 
          (100, 300, and 900\,mJy/20$^{\prime\prime}$beam). Herschel beam sizes are indicated 
          as gray circles in the lower right corners.}
\end{figure*}

\begin{figure*}
 \centering
 \includegraphics[width=15cm]{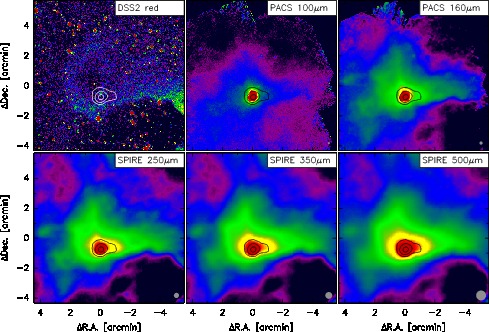}
 \caption{\label{fig-cb230}
          CB\,230 (Class\,I YSO): 
          Visual (red) DSS2 image and Herschel FIR maps (color; log scale) at 100, 160, 250, 350,
          and 500\,$\mu$m, with 850\,$\mu$m continuum contours overlaid 
          (70, 200, and 700\,mJy/20$^{\prime\prime}$beam). Herschel beam sizes are indicated 
          as gray circles in the lower right corners.}
\end{figure*}

\begin{figure*}
 \centering
 \includegraphics[width=15cm]{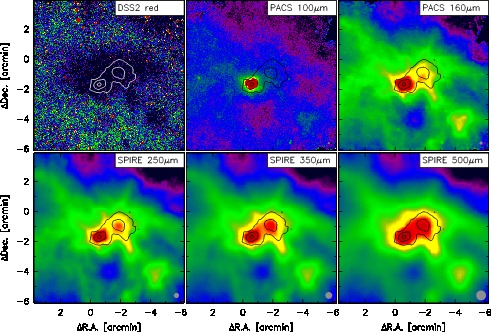}
 \caption{\label{fig-cb244}
          CB\,244 (Class\,0 protostar and starless core): 
          Visual (red) DSS2 image and Herschel FIR maps (color; log scale) at 100, 160, 250, 350,
          and 500\,$\mu$m, with 1.2\,mm continuum contours overlaid 
          (30, 80, and 180\,mJy/20$^{\prime\prime}$beam). Herschel beam sizes are indicated 
          as gray circles in the lower right corners.}
\end{figure*}

\begin{figure}
 \centering
 \includegraphics[width=9cm]{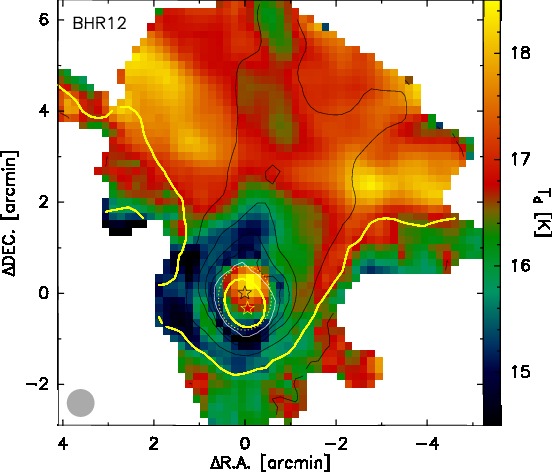}
 \caption{\label{fig-tmap-bhr12}
  Dust temperature map of BHR\,12 (color scale), overlaid with contours of the 
  hydrogen column density (white: $10^{21}$\ - masked by yellow $N_{\rm out}$\ contour - 
  and $10^{22}$\,cm$^{-2}$, black: 0.5, 2, 4, 6, 8, and 20\,$10^{21}$\,cm$^{-2}$).
  Yellow contours mark $N_{\rm out}$\ (Eq.\,\ref{eq-densprof}) and $N_{1^{\prime}}$\ (Eq.\,\ref{eq-dens1a}).
  The dotted yellow ellipse marks the $N_{1^{\prime}}$\ approximation described in Sect.\,\ref{ssec-res-seds}.
  The two asterisks mark the positions of embedded sources SMM1 and SMM2 \citep[see][]{lau10}.}
\end{figure}

\begin{figure}
 \centering
 \includegraphics[width=9cm]{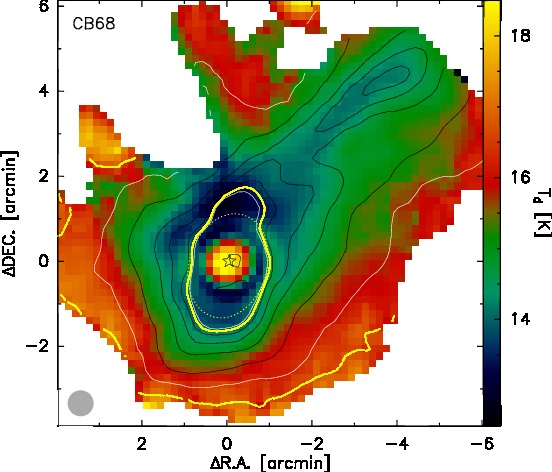}
 \caption{\label{fig-tmap-cb68}
  Dust temperature map of CB\,68 (color scale), overlaid with contours of the 
  hydrogen column density (white: $10^{21}$\ and $10^{22}$\,cm$^{-2}$, black: 0.5 - masked by yellow $N_{\rm out}$\ contour - 
  2, 4, 6, 8, and 20\,$10^{21}$\,cm$^{-2}$).
  Yellow contours mark $N_{\rm out}$\ (Eq.\,\ref{eq-densprof}) and $N_{1^{\prime}}$\ (Eq.\,\ref{eq-dens1a}).
  The dotted yellow ellipse marks the $N_{1^{\prime}}$\ approximation described in Sect.\,\ref{ssec-res-seds}.
  The asterisk marks the position of the protostar and the reference position listed in Table\,\ref{tab-res-prof}.}
\end{figure}

From the calibrated and background-subtracted dust emission maps, we  
extract for each image pixel the SED with up to 8 data points between 
100\,$\mu$m and 1.2\,mm. These individual SEDs are independently fit 
($\chi^2$\ minimization) with a 
single-temperature modified blackbody of the form of Eqs.\,\ref{eq-snu} and \ref{eq-tau1}
with $T_{\rm d}$\ and $N_{\rm H}$\ being the free parameters.
The individual flux data points are weighted with 
$\sigma^{-2}$, 
where $\sigma$\ is the quadratic sum of flux times the relative calibration 
uncertainty (see Sect.\,\ref{ssec-mod-prep}) and the mean rms noise 
in the respective map measured in regions of zero or lowest emission outside 
the sources (see Table\,\ref{tab-fdc}). 
In order to minimize the effect of $T\,-\,N_{\rm H}$\ degeneracies 
in the fitting, because at the low temperatures in these cores, $T\leq10$\,K, 
even the 1.2\,mm emission does not represent the Rayleigh-Jeans regime where the 
SED slope is independent of $T$, the weight of data points with $S_{\nu}<\sigma$\ 
was set to zero, i.e., these points were not considered in the fitting.
To further avoid erroneous extrapolations of the SED fit to unconstrained 
shorter wavelengths, only pixels with valid fluxes in all 5 \herschel\ bands 
were fitted. This latter criterion was usually constrained by the 
PACS\,100\,$\mu$m maps.
The best-fitting parameters were derived by employing a least squares fit 
to all flux values between 100\,$\mu$m and 1.2\,mm at a given image pixel, 
using a ``robust'' combination of the classical simplex amoeba search and a modified 
Levenberg Marquardt method with adaptive  steps,  as implemented in the 
{\it 'mfit'} tool of GILDAS{\footnote{http://www.iram.fr/IRAMFR/GILDAS}.

This procedure yields LoS-averaged dust tem\-pe\-ra\-ture and column density 
maps of the sources, which are presented in Sect.\,\ref{ssec-res-tmaps}.
The temperature maps provide a robust estimate of the actual dust 
temperature of the envelope in the projected outer regions, where the emission is optically 
thin at all wavelengths and LoS temperature gradients are negligible. Toward the source centers,
where cooling and shielding or embedded heating sources can produce significant 
LoS temperature gradients and the observed SEDs are therefore broader than 
single-temperature SEDs, the central dust temperatures will be overestimated 
(in the case of a positive gradient in cold sources) or underestimated 
(negative gradient in internally heated sources). 
For these reasons, the column density maps, which come out usually very smooth, 
are corrupted and exhibit artifacts in regions of $\sim$20\arcsec\,--\,40\arcsec\ radius 
around the warm protostars (see, e.g., Figs.\,\ref{fig-tmap-cb68} or \ref{fig-tmap-bhr12}).
These caveats are discussed and quantified in Sect.\,\ref{ssec-dis-mod}, including some already 
worked-on solutions.


\section{Results}           \label{sec-res}

\subsection{Herschel FIR maps}              \label{ssec-res-maps}

The resulting calibrated dust emission maps at $\lambda$\,100, 160, 250, 350, 
and 500\,$\mu$m and at original angular resolution are presented in Figs.\,\ref{fig-cb4} through 
\ref{fig-cb244}. 
The \herschel\ maps are accompanied by optical (red) images 
from the second Digitized Sky Survey (DSS2\footnote{The Digitized Sky Surveys were produced at the Space Telescope Science 
Institute under U.S. Government grant NAG W-2166. The images of these surveys are based on photographic data obtained 
using the Oschin Schmidt Telescope on Palomar Mountain and the UK Schmidt Telescope. The plates were processed into 
the present compressed digital form with the permission of these institutions.}), which were obtained 
through the {\it SkyView}\footnote{{\it SkyView} has been developed with generous support from the NASA AISR and 
ADP programs (P.I. Thomas A. McGlynn) under the auspices of the High Energy Astrophysics Science Archive Research 
Center (HEASARC) at the NASA/ GSFC Astrophysics Science Division. } interface.
These optical images clearly show the regions of highest extinction as well as, in several cases, extended cloudshine structures. 
All maps are overlaid with contours of the (sub)mm dust continuum emission observed with 
ground-based telescopes.

The 160\,$\mu$m through 500\,$\mu$m maps are usually very similar to each other in appearance, 
outlining the thermal dust emission from the dense cores along with the often 
filamentary or tail-like, more tenuous envelopes. 
Except for some filamentary or tail-like extensions like, e.g., in CB\,17 (Fig.\,\ref{fig-cb17}),
the $\sim$18\arcmin\ SPIRE maps usually cover the entire extent of the detectable 
FIR dust emission down to the more or less flat background levels.
The slightly smaller PACS maps ($\sim$10\arcmin) also usually cover the entire extent of 
the clouds, but in some cases cut off a bit more of some filamentary or tail-like extensions 
(e.g., Fig.\,\ref{fig-cb17}).

The PACS 100\,$\mu$m maps often show significantly fainter or less extended emission than 
the maps at longer wavelengths, owing to the fact that 100\,$\mu$m samples the steep 
short-wavelength side of the SED at the low temperature of the dust in these clouds 
(Fig.\,\ref{fig-sedall}).
In some cases, the diffuse 100\,$\mu$m emission even exhibits a ``hole'' or ``shadow'' 
at the location of the column density peak (e.g., CB\,27; Fig.\,\ref{fig-cb27}), 
hinting at extremely cold and dense (high column density) cores with thin warmer envelopes,
which are thus good candidates of cores at the verge of protostellar collapse
\citep[see][for a discussion of 24\,$\mu$m and 70\,$\mu$m shadows in dense cores]{stutz09a}. 

On the other hand, the 100\,$\mu$m maps, which also have the highest angular resolution 
of all our \herschel\ maps ($\approx7$\arcsec), are most sensitive to embedded heating 
sources like, e.g., protostars. Although even the previously known very low luminosity object (VeLLO) in CB\,130 
(\citealp{kim2011}; see also \citealp{dunham08}) is well-detected (Fig.\,\ref{fig-cb130}), 
our maps do not reveal any obvious previously unknown warm compact source in any of the globules,
confirming that all our presumably starless cores are indeed starless. 
The only possible exception might be CB\,17 (see Fig.\,\ref{fig-cb17}), 
where we find hints of an extremely low-luminosity ($L_{\rm bol}<0.04$\,\lsun) 
cold embedded source very close (in projection) to CB17\,-\,IRS 
\citep[][]{chen12,schmalzl13}.


\subsection{Dust temperature and column density maps and radial profiles}  \label{ssec-res-tmaps}


\begin{figure}[ht!]
 \centering
 \includegraphics[width=9cm]{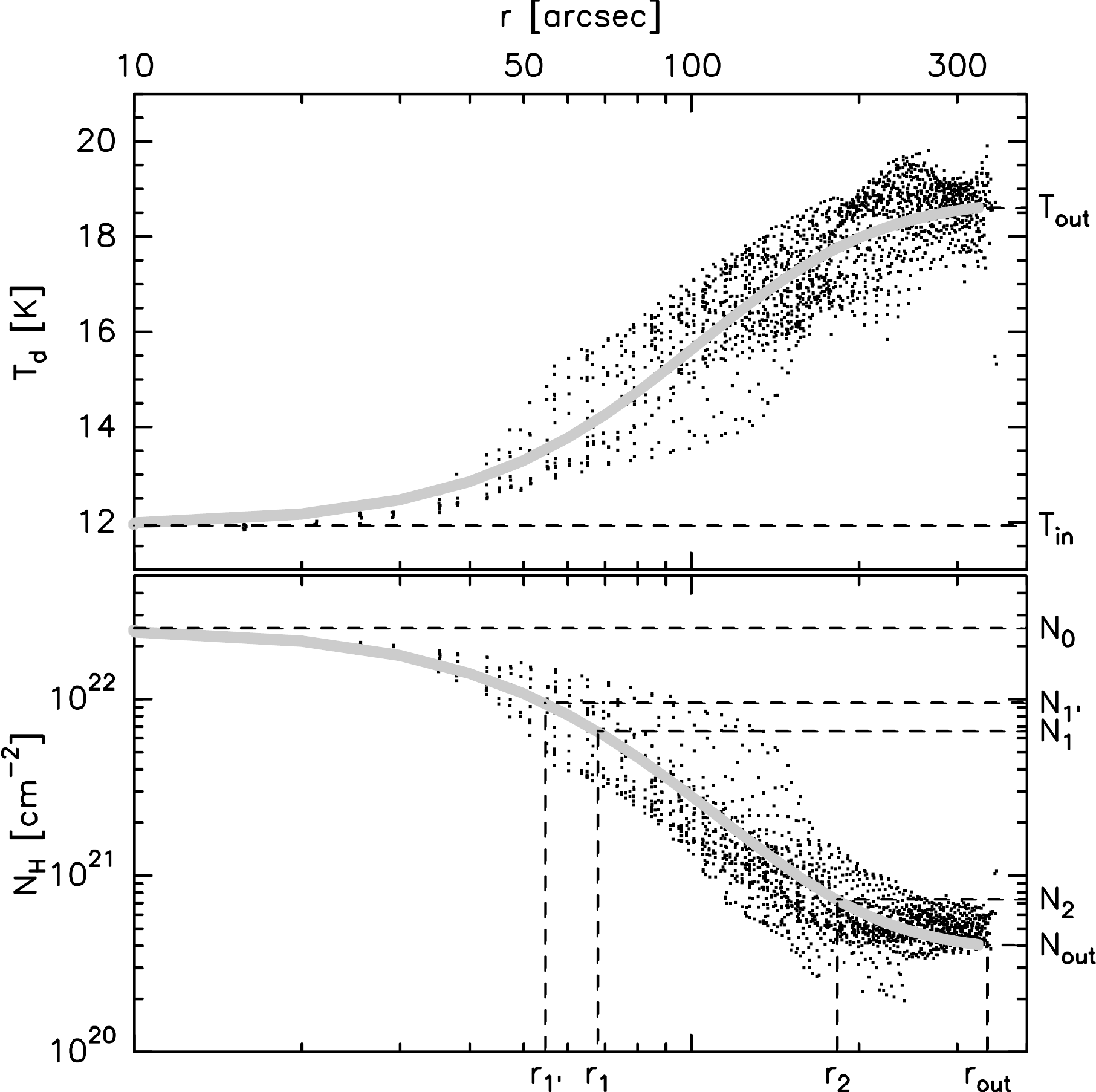}
 \caption{\label{fig-b68-profiles1}
 Radial profiles of column density and LoS-averaged dust temperature of B\,68 
 (cf. Fig.\,9 of \citealp{nielbock12} for the ray-tracing results of the same data).
 Small black dots show the data pixel values over radial distance from the column density peak 
 (see Fig.\,\ref{fig-tmap-b68}). Solid gray lines show best fits to these data with Eqs.\,\ref{eq-densprof} 
 and \ref{eq-tempprof} and the parameters listed in Table\,\ref{tab-res-prof}. 
 The parameters are also labeled on the diagram axes to illustrate their meaning, 
 with the following characteristic radii: 
 $r_1$\,=\,flat core profile radius (Eq.\,\ref{eq-densprof}),
 $r_{1^{\prime}}$\,=\,radius to define consistent dense core boundary (Eqs.\,\ref{eq-dens1a} and \ref{eq-r1a}),
 $r_2$\,=\,cloud radius at transition to halo (Eq.\,\ref{eq-r1}), and 
 $r_{\rm out}$\,=\,outer cloud boundary (Eq.\,\ref{eq-densprof}).
}
\end{figure}


\begin{table*}[htb]
\caption{Column density and dust temperature profiles}             
\label{tab-res-prof}      
\centering 
{\footnotesize       
\begin{tabular}{lcccccccccc}     
\hline\hline       
Source\tablefootmark{a}                 &
$N_0$\tablefootmark{b}                  &
$N_{\rm out}$\tablefootmark{b}          &
$r_1$\tablefootmark{b}                  &
$r_2$\tablefootmark{b}                  &
$r_{\rm out}$\tablefootmark{b}          &
$p$\tablefootmark{b}                    &
$T_{\rm in}$\tablefootmark{b}           & 
$T_{\rm peak}$\tablefootmark{c}         & 
$T_{\rm out}$\tablefootmark{b}          \\
                                        &
[cm$^{-2}$]                             &
[cm$^{-2}$]                             &
[AU]                                    &
[AU]                                    &
[AU]                                    &
                                        &
[K]                                     &
[K]                                     &
[K]                                     \\
\hline
CB\,4 	                           & 7.5E21 & 6.0E19 & 2.5E4 & 6.2E4 & 9.5E4 & 5.0 & 14.0 & --   & 19.5 \\
CB\,17\,-\,SMM                     & 2.5E22 & 1.0E21 & 1.0E4 & 3.1E4 & 3.5E4 & 3.0 & 11.7 & --   & 13.1 \\
CB\,26\,-\,SMM2                    & 1.4E22 & 1.2E21 & 8.4E3 & 2.2E4 & 3.4E4 & 2.4 & 12.8 & --   & 14.0 \\
CB\,27                             & 1.9E22 & 2.1E21 & 1.6E4 & 2.1E4 & 2.8E4 & 4.0 & 11.9 & --   & 14.2 \\
B\,68                              & 2.5E22 & 4.0E20 & 1.0E4 & 2.1E4 & 4.6E4 & 5.0 & 11.9 & --   & 18.5 \\
CB\,244\,-\,SMM2                   & 4.6E22 & 1.1E21 & 7.0E3 & 6.0E4 & 9.0E4 & 1.8 & 11.3 & --   & 14.5 \\
CB\,130\,-\,SMM\tablefootmark{d}   & 2.6E22 & 2.0E21 & 7.2E3 & 2.8E4 & 4.7E4 & 1.8 & 12.1 & --   & 13.5 \\[1mm]
CB\,6\tablefootmark{e,f}           & 5.0E21 & 5.0E20 &...    &...    & 1.4E5 &...  & 14.5 & 18.4 & 15.7 \\ 
CB\,26\,-\,SMM1\tablefootmark{e}   & 1.0E22 & 9.0E20 &...    &...    &...    &...  & ...  & 19.0 & 14.6 \\
BHR\,12\tablefootmark{f}           & 5.4E22 & 1.0E21 & 1.4E4 & 3.5E4 & 1.1E5 & 4.0 & 14.8 & 18.5 & 17.3 \\
CB\,68                             & 2.4E22 & 5.0E20 & 1.1E4 & 2.7E4 & 2.9E4 & 4.0 & 12.7 & 19.0 & 16.5 \\
B\,335                             & 3.1E22 & 6.2E20 & 3.6E3 & 1.8E4 & 3.8E4 & 2.3 & 14.7 & 18.8 & 16.2 \\
CB\,230\,-\,SMM\tablefootmark{e,f} & 2.7E22 & 5.0E20 &...    &...    & 1.4E5 &...  & 14.5 & 19.6 & 15.6 \\
CB\,244\,-\,SMM1                   & 1.5E22 & 1.0E21 & 1.2E4 & 5.5E4 & 9.0E4 & 1.7 & 11.5 & 19.6 & 14.2 \\[1mm]
\hline
starless cores     & 2.3$\pm$1.0E22 & 1.1$\pm$0.6E21 & 1.2$\pm$0.6E4 & 3.5$\pm$1.6E4 & 5$\pm$2E4 & 3.3$\pm$1.2 & 12.2$\pm$0.8 & ...          & 15.3$\pm$2.2 \\
protostellar cores & 2.4$\pm$1.4E22 & 0.7$\pm$0.2E21 & 0.9$\pm$0.5E4 & 3.4$\pm$1.2E4 & 9$\pm$4E4 & 3.0$\pm$0.9 & 13.6$\pm$1.2 & 19.0$\pm$0.4 & 15.7$\pm$0.9 \\
all                & 2.3$\pm$1.3E22 & 0.9$\pm$0.5E21 & 1.1$\pm$0.5E4 & 3.5$\pm$1.5E4 & 7$\pm$4E4 & 3.2$\pm$1.2 & 13.0$\pm$1.2 & ...          & 15.5$\pm$1.8 \\
\hline                  
\end{tabular}
\tablefoot{{In contrast to the other tables, sources in this table have been
grouped into starless and protostellar.}
\tablefoottext{a}{Coordinates are listed in Table\,\ref{tab-res-physprop}.}
\tablefoottext{b}{Derived from circular fits to the dust temperature and column density maps; 
                  see Sect.\,\ref{ssec-res-tmaps}, Eqs.\,\ref{eq-densprof} through \ref{eq-tin}.}
\tablefoottext{c}{Peak temperature at the position of the embedded heatig source (protostar), severely underestimates 
                  the true central dust temperature because of beam smoothing and LoS averaging.}
\tablefoottext{d}{CB\,130: the embeded VeLLO causes only a very small temperature increase in the smoothed dust temperature map.}
\tablefoottext{e}{Bad fit due to large unresolved LoS gradients, no meaningful values for some of the parameters.}
\tablefoottext{f}{The embedded double sources in BHR\,12 and CB\,230 are not resolved in the smoothed dust temperature maps.}
}}
\end{table*}


The LoS-averaged dust temperature and hydrogen column density maps derived 
with the fitting procedure described in Sect.\,\ref{ssec-mod-gb} are presented 
in Figs.\,\ref{fig-tmap-cb4} through \ref{fig-tmap-cb244}.
All sources show systematic dust temperature gradients with cold 
interiors (11\,--\,14\,K) and significantly warmer outer rims (14\,--\,20\,K). Embedded heating sources 
like protostars, including the VeLLO in CB\,130 \citep[Fig.\,\ref{fig-tmap-cb130};][]{kim2011},
show up very clearly in the temperature maps.

To characterize the column density profiles quantitatively, we fit circular profiles 
around the column density peak to the maps, using a 
``Plummer-like'' profile (\citealp{plummer11}; see also \citealp{ww01}),
modified by a constant term to account for the observed outer column density ``plateau'' 
(see discussion in Sect.\,\ref{ssec-res-isrf}):
\begin{equation}                      \label{eq-densprof}
N_{\rm H}(r) =  
\left\{ \begin{array}{ll}
\frac{\Delta N}{\left(1+\left(r/r_1\right)^2\right)^{p/2}} + N_{\rm out} &\mbox{ if $r\le r_{\rm out}$} \\
0            &\mbox{ if $r > r_{\rm out}$}\quad .
       \end{array} \right.
\end{equation}
The peak column density is then 
\begin{equation}                      \label{eq-dens0}
N_0 = N_{\rm H}(r=0) = \Delta N + N_{\rm out}\quad .
\end{equation}
This profile accounts for an inner flat (column) density core
at $r< r_1$\ where
\begin{equation}                      \label{eq-dens1}
N_1 = N_{\rm H}(r_1) = \frac{\Delta N}{2^{p/2}} + N_{\rm out}\quad ,
\end{equation}
approaches a power-law with index $p$\ at $r\gg r_1$, 
turns over into a flat outer column density ``plateau'' outside
\begin{equation}                      \label{eq-r1}
r_2 = r_1\,\sqrt{\,\left(\frac{\Delta N}{N_{\rm out}}\right)^{2/p} - 1}
\end{equation}
where 
\begin{equation}                      \label{eq-dens2}
N_2 = N_{\rm H}(r_2) = 2\times N_{\rm out}\quad ,
\end{equation}
and is cut off at $r_{\rm out}$.
This outer boundary of the thin envelope or halo is not well-recovered in most cases because 
filamentary extensions emphasize deviations from circular\,/\,spherical geometry 
with increasing distance from the core center and the flux levels gradually decline 
below the noise cut-off (see Sect.\,\ref{ssec-mod-gb}).
For the same reason, the flatness of the outer column density profile at level $N_{\rm out}$\
might be partially an artifact of circularly averaging noncircular structure 
at a level just above the noise. 

To empirically fit and parameterize the radial temperature profiles of purely externally heated cores, 
we use a similar profile of the form: 
\begin{equation}                      \label{eq-tempprof}
T(r) =  T_{\rm out} - \frac{\Delta T}{\left(1+\left(\frac{r}{r_{\rm T}}\right)^2\right)^{q/2}}\quad .
\end{equation}
The central temperature minimum is then given by
\begin{equation}                      \label{eq-tin}
T_{\rm in} = T_{\rm out} - \Delta T\quad .
\end{equation}

Before fitting radial profiles to the maps, we mask some of the spurious low-SNR edge 
features and tail-like (asymmetric) extensions, always applying the same spatial mask in both 
the column density and dust temperature maps.
For cores with embedded heating sources (protostars), we also mask the region with local temperature 
increase before fitting the radial profiles with Eqs.\,\ref{eq-densprof} and \ref{eq-tempprof} 
and also list in Table\,\ref{tab-res-prof} the value of the central temperature maximum $T_{\rm peak}$.
The radius of the masked region varied between 20\arcsec\ and 40\arcsec.
This extrapolation for $N_0$\ and $T_{\rm in}$\ of the outer profile into the core center, 
where both the local temperature and the column density estimates are very uncertain 
due to large and unresolved LoS temperature gradients, leads to an increased uncertainty 
in particular of the value for $N_0$\ in the protostellar cores.
This and other shortcomings of this model and the interpretation of its parameters are discussed in 
Sect.\,\ref{ssec-dis-mod}.

To illustrate the quality of these radial profile fits and the meaning of the various parameters, 
we show in Fig.\,\ref{fig-b68-profiles1} the radial distribution of $N_{\rm H}$\ and $T_{\rm d}$\ 
values in the resulting maps for B\,68 along with the best-fit profiles and the parameters
labeled on the diagram axes.
The best-fit parameter values for 
$N_{\rm 0}$, $N_{\rm out}$, $r_1$, $r_2$, $r_{\rm out}$, $p$, $T_{\rm in}$, 
$T_{\rm peak}$\ (for cores with embedded heating sources), and $T_{\rm out}$\ 
of all sources are listed in Table\,\ref{tab-res-prof}. 
We do not list parameters $r_{\rm T}$\ and $q$, but note that from the core centers outward, 
the LoS-averaged dust temperature typically raises by 1\,K at a radius of $(1\pm 0.3)\,10^4$\,AU.

Note that in the next section, we define core and cloud sizes via the mean extent of certain 
column density contours in the maps, and not via the radial profile fit parameters 
$r_1$\ and $r_{\rm out}$. 
In particular for very elliptical sources and sources with extended halos, the derived mean diameters 
are therefore not necessarily identical to twice the corresponding radial profile-fit radii.

 
\subsection{Source morphologies, integral properties, and SEDs}  \label{ssec-res-seds}


\begin{table*}[htb]
\caption{Physical parameters of globules and embedded cores}             
\label{tab-res-physprop}
\centering 
{\footnotesize       
\begin{tabular}{lllllcccll}     
\hline\hline       
Source                                  &
R.A.,Dec. (J2000)\tablefootmark{a}      &
Size\tablefootmark{b}                   & 
b:a \tablefootmark{b}                   & 
$M_{\rm H}$\tablefootmark{c}            &  
$L_{\rm bol}$\tablefootmark{c}          & 
$T_{\rm bol}$\tablefootmark{c}          &
$\frac{L_{\rm smm}}{L_{\rm bol}}$\tablefootmark{c} &
Morphology\tablefootmark{d}             &
Figures                                 \\
                                        &
[h:m:s, \degr:\arcmin:\arcsec]          & 
[AU]                                    &
                                        &
[M$_{\odot}$]                           &
[L$_{\odot}$]                           &
[K]                                     &
[\%]                                    &
Evol. class                             &
                                        \\
\hline  
CB\,4              & ...                     & 1.2E5 & 0.74 & 1.6  & 2.8  & ...  & ...  & R, T(0.5)                 & \ref{fig-cb4}, \ref{fig-cb4-morph}, \ref{fig-tmap-cb4} \\
~~CB\,4 \,-\,SMM   & 00:39:05.2, $+$52:51:47 & 5.2E4 & 0.62 & 0.7  & 0.77 &   23 & 11.7 & starless, unbound         & ... \\[1mm]
CB\,6              & ...                     & 1.2E5 & 0.40 & 5.9  & 3.5  & ...  & ...  & C, T(2.6)                 & \ref{fig-cb6}, \ref{fig-cb6-morph}, \ref{fig-tmap-cb6}\\
~~CB\,6\,-\,SMM    & 00:49:24.3, $+$50:44:36 & 3.9E4 & 1.0  & 1.0  & 2.5  &  158 &~~5.1 & Cl.\,I                    & ... \\[1mm]
CB\,17             & ...                     & 8.6E4 & 0.67 & 3.0  & 1.3  & ...  & ...  & C, T($>$1.0)              & \ref{fig-cb17}, \ref{fig-cb17-morph}, \ref{fig-tmap-cb17}\\
~~CB\,17\,-\,SMM   & 04:04:37.7, $+$56:55:59 & 2.1E4 & 0.86 & 1.0  & 0.37 &   18 & 22.3 & starless, stable          & ... \\
~~CB\,17\,-\,IRS   & 04:04:34.0, $+$56:56:16 & ...   & ...  & ...  & 0.12 &   80 &~~2.2 & Cl.\,I\tablefootmark{e}   & ... \\[1mm]
CB\,26             & ...                     & 7.3E4 & 0.75 & 2.4  & 1.9  & ...  & ...  & C, M, T(0.6)              & \ref{fig-cb26}, \ref{fig-cb26-morph}, \ref{fig-tmap-cb26}\\
~~CB\,26\,-\,SMM1  & 04:59:49.3, $+$52:04:39 & 2.0E4\tablefootmark{f} & 1.0\tablefootmark{f} & 0.3\tablefootmark{f} & 0.47\tablefootmark{f} & 81\tablefootmark{f} &~~6.5\tablefootmark{f} & Cl.\,I     \\
~~CB\,26\,-\,SMM2  & 05:00:14.5, $+$52:05:59 & 2.2E4\tablefootmark{f} & 1.0\tablefootmark{f} & 0.6\tablefootmark{f} & 0.30\tablefootmark{f} & 20\tablefootmark{f} & 17.3\tablefootmark{f} & starless, stable  \\[1mm]
CB\,27             & ...                     & 6.7E4 & 0.77 & 3.7  & 2.2  & ...  & ...  &  C, T(0.3)                & \ref{fig-cb27}, \ref{fig-cb27-morph}, \ref{fig-tmap-cb27}\\
~~CB\,27\,-\,SMM   & 05:04:08.1, $+$32:43:30 & 2.6E4 & 0.61 & 1.2  & 0.44 &   18 & 20.5 & starless, stable          & ... \\[1mm]
BHR\,12            & ...                     & 1.2E5 & 0.39 & 11   & 29   & ...  & ...  & C, M, T(0.9)              & \ref{fig-bhr12}, \ref{fig-bhr12-morph}, \ref{fig-tmap-bhr12}\\
~~BHR\,12\,-\,SMM  & 08:09:32.7, $-$36:05:19 & 3.1E4\tablefootmark{g} & 0.82\tablefootmark{g} & 3.7\tablefootmark{g} & 14.9\tablefootmark{g} & 104\tablefootmark{g} &~~3.2\tablefootmark{g} & Cl.\,0\,/\,I\tablefootmark{h} \\[1mm]
CB\,68             & ...                     & 6.5E4 & 0.55 & 2.1  & 2.1  & ...  & ...  & C, T(0.7)                 & \ref{fig-cb68}, \ref{fig-cb68-morph}, \ref{fig-tmap-cb68}\\
~~CB\,68\,-\,SMM   & 16:57:19.7, $-$16:09:23 & 1.5E4 & 0.79 & 0.5  & 0.86 &   41 &~~6.3 & Cl.\,0                    & ... \\[1mm]
B\,68              & ...                     & 5.6E4 & 0.83 & 1.3  & 1.3  & ...  & ...  & C                         & \ref{fig-b68}, \ref{fig-b68-morph}, \ref{fig-tmap-b68}\\
~~B\,68\,-\,SMM    & 17:22:38.3, $-$23:49:51 & 1.6E4 & 0.88 & 0.6  & 0.23 &   19 & 17.4 & starless, stable          & ... \\[1mm]
CB\,130            & ...                     & 4.9E4 & 0.52 & 2.7  & 1.6  & ...  & ...  & M                         & \ref{fig-cb130}, \ref{fig-cb130-morph}, \ref{fig-tmap-cb130}\\
~~CB\,130\,-\,SMM  & 18:16:15.6, $-$02:32:45 & 2.0E4 & 0.63 & 1.0  & 0.37 &   38 & 20.8 & VeLLO\tablefootmark{i}    & ... \\[1mm]
B\,335             & ...                     & 4.8E4 & 0.98 & 0.8  & 1.3  & ...  & ...  & R                         & \ref{fig-b335}, \ref{fig-b335-morph}, \ref{fig-tmap-b335}\\
~~B\,335\,-\,SMM   & 19:37:00.7, $+$07:34:08 & 8.5E3 & 0.80 & 0.2  & 0.57 &   37 &~~4.4 & Cl.\,0                    & ... \\[1mm]
CB\,230            & ...                     & 2.1E5 & 0.69 & 18   & 25   & ...  & ...  & C                         & \ref{fig-cb230}, \ref{fig-cb230-morph}, \ref{fig-tmap-cb230}\\
~~CB\,230\,-\,SMM  & 21:17:39.9, $+$68:17:36 & 3.5E4 & 0.55 & 3.0  & 10.4 &  189 &~~3.7 & Cl.\,I\tablefootmark{k}   & ... \\[1mm]
CB\,244            & ...                     & 1.4E5 & 0.61 & 12   & 8.2  & ...  & ...  & M                         & \ref{fig-cb244}, \ref{fig-cb244-morph}, \ref{fig-tmap-cb244}\\
~~CB\,244\,-\,SMM1 & 23:25:47.3, $+$74:17:44 & 2.0E4\tablefootmark{l} & 0.87\tablefootmark{l} & 0.8\tablefootmark{l} & 1.82\tablefootmark{l} & 56\tablefootmark{l} &~~4.3\tablefootmark{l} & Cl.\,0  \\
~~CB\,244\,-\,SMM2 & 23:25:26.8, $+$74:18:22 & 2.3E4 & 0.87 & 2.1  & 0.53 &   17 & 24.2 & prestellar                & ... \\
\hline                  
\end{tabular}
\tablefoot{
\tablefoottext{a}{Coordinates of column density peak (starless cores) or dust temperature peak (embedded source).}
\tablefoottext{b}{Mean diameter ($\sqrt{a\times b}$) and aspect ratio of the $N_{\rm out}$\ column density contour listed in Table\,\ref{tab-res-prof} 
     for the globules (i.e., $\approx 2\times r_{\rm out}$, Eq.\,\ref{eq-densprof} and Fig.\,\ref{fig-b68-profiles1}), 
     and of the $N_{1^{\prime}}$\ contour (Eq.\,\ref{eq-dens1a}) for the embedded cores (i.e., $2\times r_{1^{\prime}}$, Eq.\,\ref{eq-r1a}).}
\tablefoottext{c}{Measured inside the $N_{\rm out}$\ contour (globules) or the $N_{1^{\prime}}$\ contour (cores), except where marked otherwise.}
\tablefoottext{d}{Morphology description based on visual inspection of DSS2-red optical images and \herschel\ dust emission maps:
                  R: round or slightly elliptical; C: cometary-shaped; M: multiple cores; T($x$): tail of length $x$\,pc.}
\tablefoottext{e}{CB\,17\,-\,IRS is identified in the PACS 100\,$\mu$m map and as a local temperature maximum in the $T$-map, 
                  but cannot be separated from SMM in the column density map. }
\tablefoottext{f}{CB\,26\,-\,SMM1 and SMM2: ellipses, marked by dotted lines in Fig.\,\ref{fig-tmap-cb26}, were used instead 
                  of the $N_{1^{\prime}}$\ contour (cf. Sect.\,\ref{ssec-res-seds}).}
\tablefoottext{g}{BHR\,12: An ellipse, marked by dotted lines in Fig.\,\ref{fig-tmap-bhr12}, was used instead of the $N_{1^{\prime}}$\ contour (cf. Sect.\,\ref{ssec-res-seds}).}
\tablefoottext{h}{BHR\,12: The two embedded sources SMM1 (Class\,I) and SMM2 \citep[Class\,0; sep. 21\arcsec; see][]{lau10} are not resolved here.}
\tablefoottext{i}{CB\,130: additional low-mass prestellar core $\sim$30\arcsec\ west and Class\,I YSO 
                  $\sim$15\arcsec\ east of Class\,0 core.}
\tablefoottext{k}{CB\,230: the two embedded sources IRS1 and IRS2 \citep[sep. 10\arcsec; see][]{lau10} are not resolved here.}
\tablefoottext{l}{CB\,244\,-\,SMM1: an ellipse, marked by dotted lines in Fig.\,\ref{fig-tmap-cb244}, was used instead 
                  of the $N_{1^{\prime}}$\ contour (cf. Sect.\,\ref{ssec-res-seds}).}
}}
\end{table*}



\begin{figure*}
 \centering
 \includegraphics[width=15cm]{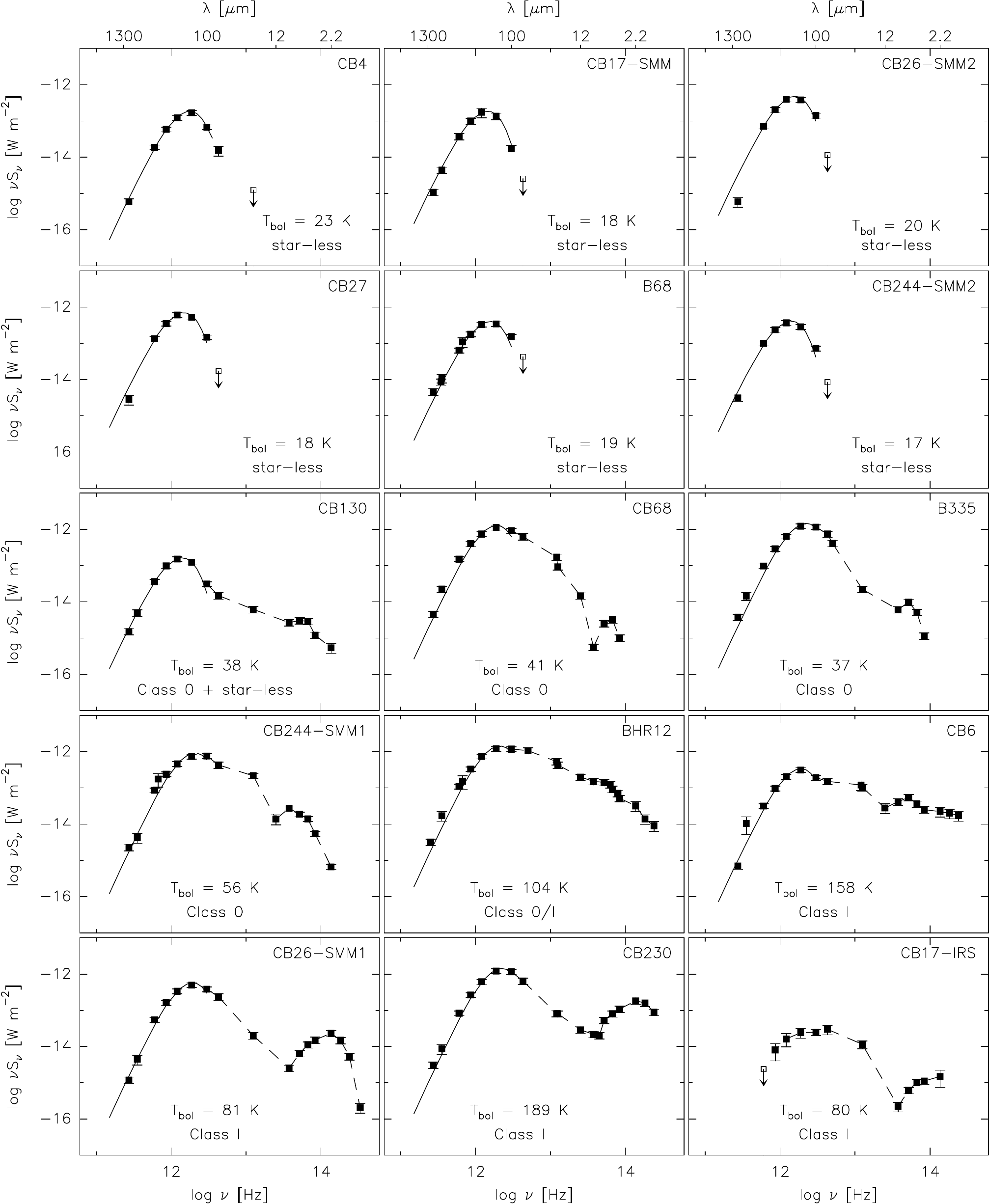}
 \caption{\label{fig-sedall}
   Integrated spectral energy distributions of the dense cores and embedded sources 
   listed in Table\,\ref{tab-res-physprop}. 
   Fluxes are integrated within the $N_{1^{\prime}}$\ contour (Eq.\,\ref{eq-dens1a}), as are the core 
   masses and luminosities listed in Table\,\ref{tab-res-physprop}. 
   Solid lines between 3\,mm and 100\,$\mu$m mark the modified blackbody fits to the data points, 
   while dashed lines only represent a logarithmic interpolation of the data points shortward of 100\,$\mu$m
  (see Sect.\,\ref{ssec-res-seds}).
}
\end{figure*}


In addition to the pixel-by-pixel SEDs and the column density and dust temperature maps, 
we also derive integrated source pro\-per\-ties using certain column density levels to define the 
integration \mbox{areas}.
Total cloud sizes are derived by measuring the major and minor extent of the $N_{\rm out}$\ 
contour in the column density maps, which is marked as the outer yellow contour in 
Figs.\,\ref{fig-tmap-cb4} through \ref{fig-tmap-cb244}.
The resulting linear diameters (invoking the distances listed in Tab.\,\ref{tab-sourcelist})
are listed in Tab.\,\ref{tab-res-physprop}.
The mean radii of the globules in our sample range from 0.1\,pc (CB\,130) 
to 0.5\,pc (CB\,230), with a mean of $0.22\pm0.1$\,pc ($4.8\pm 2.2\,10^4$\,AU),
which is about twice the typical Jeans lengths of these clouds 
($\sim 0.13$\,pc, assuming the mean radius and mass and a mean temperature of 13\,K; 
see also stability discussion in Sect.\,\ref{ssec-res-stability}).
The mean projected aspect ratio of the clouds is $0.66\pm 0.16$, 
with the most extreme cases CB\,6 and BHR\,12 (0.4) and B\,335 (0.98).
These measures do not account for the long tails, e.g., in CB\,17 or CB\,68, nor do they accurately reflect the 
actual aspect ratios of the individual sources, since the 3-D projection of the elliptical 
clouds onto the plane of sky remains unknown.}
Note that these and the following uncertainty values only represent the variance of the mean and do not include 
systematic uncertainties, which are discussed in Sect.\,\ref{sec-disc}.

Since cometary tails in globules appear to be failry ubiquitous 
(see Tab.\,\ref{tab-res-physprop} and Sect.\,\ref{ssec-res-env}), 
we have also assessed 
the possibility that some of the apparently tail-less globules (e.g., B\,335 or CB\,130) 
could possibly also have a tail pointing at us or away from us. 
However, given the length of the tails observed, e.g., in CB\,6 (Fig.\,\ref{fig-cb6-morph}),
CB\,17 (Fig.\,\ref{fig-cb17-morph}), or CB\,68 (Fig.\,\ref{fig-cb68-morph}),
it is extremely unlikely that a globule like B\,335 has a tail that is directly aligned 
with the line of sight. Another argument against this possibility comes from the fact that, 
at least in the starless cores, the (LoS-averaged) temperature minimum always agrees with the 
column density maximum. Should a long tail, like that seen in e.g., CB\,17, be oriented along the line of sight, 
then its total column density would sum up to a value similar to or even larger than that observed toward 
the core centers. However, since the tails are not as well-shielded from the IRSF as spherical cores and 
therefore have approximately the same average temperature as the envelopes, such a projection could 
not result in a clear temperature minimum at the position of the column density maximum, 
as observed in all starless cores.

Total hydrogen masses are derived by integrating the column density maps 
within the $N_{\rm out}$\ contour and are also listed in Tab.\,\ref{tab-res-physprop}.
These masses range from 0.8\,\msun\ (B\,335) to 18\,\msun\ (CB\,230), 
with a mean value of $\langle M_{\rm cloud}\rangle= 5.4\pm5.0$\,\msun. 
Using the somewhat more conservative column density threshold $N_2$\ (Eq.\,\ref{eq-dens2}) 
would have the benefit of being less susceptible to inhomogeneities of the background level 
and having closed contours in all sources, but would cut off some of the extended emission 
that most likely originates from the globule. Using $N_2$\ instead of $N_{\rm out}$\ would lead 
to $\sim10-15$\% smaller masses and sizes for the globules compared to the values listed in Tab.\,\ref{tab-res-physprop}.

Integral cloud SEDs (not shown) were obtained by integrating the various emission maps 
within the $N_{\rm out}$\ contour. Where available and appropriate, we also included 
ground-based optical and NIR data, \spitzer\ IRAC and MIPS maps, as well as {\it ISO} and {\it IRAS}
fluxes \citep[see][]{lau10}. The resulting SEDs where fitted between $\lambda$\,100\,$\mu$m 
and 1.2\,mm in the same way as the individual image pixel SEDs (Sect.\,\ref{ssec-mod-gb}). 
To facilitate integration of the SEDs at wavelengths shortward of the SED peak, fluxes at $\lambda \le 100\,\mu$m 
were logarithmically interpolated as described in \citet{lau10}.
Bolometric luminosities ($L_{\rm bol}$) were derived by integrating the SEDs over the complete wavelength range.
Submillimeter luminosities ($L_{\rm smm}$) were derived by integrating the SEDs at wavelengths longward of 350\,$\mu$m.
The resulting total cloud luminosities range from 1.3\,\lsun\ (CB\,17, B\,68, B\,335) to 29\,\lsun\ (BHR\,12), 
with a mean of $\langle L_{\rm cloud}\rangle \sim 6.7$\lsun. See Table\,\ref{tab-res-physprop} 
and Fig.\,\ref{fig-mh-lbol} for the distribution of masses and luminosities.

In four out of the twelve globules, the {\it Herschel} and submm maps reveal multiple cores 
(CB\,26, BHR\,12, CB\,130, CB\,244 - all known before to be multiple), while the other 
eight globules appear single-cored at the {\it Herschel} resolution.
In the temperature and column density maps, the double cores are only resolved clearly in CB\,26 and CB\,244.

The identification of a common criterion for characterizing the extent and measuring masses and luminosities 
of the dense cores was less straight-forward.
Since the (column) density profile of a single-cored globule continuously decreases from the center 
toward the outer boundary of the globule, there is no simple way to observationally distinguish between a 
``core'' and an ``envelope'', like in larger molecular clouds where the outer boundary of a core is usually 
judged to be where the (column) density profile merges with the surrounding cloud level.
The flat core radius, $r_1$, of the column density profile is also not a good criterion 
since its value depends on the fitted power-law index $p$\ (Eq.\,\ref{eq-densprof}),
which is not well-constrained in most cases since $r_2\ngg r_1$, i.e., the range where 
the column density profile approaches a power-law and $p$\ can be derived is too small in many cases.
Therefore, we define an empirical, but well-reproducible column density threshold for 
all dense cores by
\begin{equation}                      \label{eq-dens1a}
N_{1^{\prime}} = N_{\rm H}(r_{1^{\prime}}) = \frac{\Delta N}{e} + N_{\rm out}\quad ,
\end{equation}
where $e$\ is Euler's number and
\begin{equation}                      \label{eq-r1a}
r_{1^{\prime}} = r_1\,\sqrt{e^{2/p}-1}\quad .
\end{equation}
For $p = 2/ln(2)$\ ($\sim$2.9, which is actually very close to the mean $\langle p\rangle$\ of our sampe), 
we obtain $N_{1^{\prime}} = N_1$\ and $r_{1^{\prime}} = r_1$\ (Eq.\,\ref{eq-dens1}).
This column density threshold is high enough to permit a reasonable separation of the subcores in CB\,26 
(Fig.\,\ref{fig-tmap-cb26}) and CB\,244 (Fig.\,\ref{fig-tmap-cb244}), while at the same time being 
low enough to fully include the regions of corrupted column density values around embedded protostars 
(due to strong unresolved LoS temperature gradients; cf. Sect.\,\ref{ssec-mod-gb}).
Note that this is an observationally driven definition of ``core'' that is not based on a well-characterized 
physical transition between core and envelope (see also discussion in Sect.\,\ref{ssec-res-evol}).
In a few cases, we had to use ellipses to define the core areas, which were however guided by 
the $N_{1^{\prime}}$ contour.
In CB\,26 (Fig.\,\ref{fig-tmap-cb26}), the $N_{1^{\prime}}$ contour still includes a bridge between cores 
SMM1 and SMM2, such that an artifical cut-off by an ellipse was necessary.
In addition, the corruption of the column density map around SMM1 by unresolved LoS temperature gradients 
leads to an offset between the column density peak (artifact) and the temperature peak 
(actual location of the embedded YSO).
This latter effect required to also use ellipses for CB\,244-SMM1 (Fig.\,\ref{fig-tmap-cb244}) 
and BHR\,12 (Fig.\,\ref{fig-tmap-bhr12}). 

The resulting integral SEDs of the core regions enclosed by the $N_{1^{\prime}}$\ contours or the respective 
ellipses (where marked in Figs.\,\ref{fig-tmap-cb4} through \ref{fig-tmap-cb244}) are shown in 
Fig.\,\ref{fig-sedall}. These SEDs where fit in the same way as 
described above for the integral cloud SEDs. 
The integral properties derived from these subregions are much more representative 
of the cold starless cores or warm protostellar cores than the total globule quantities which include 
the extended envelopes that are dominated by external heating from the ISRF. 
Resulting sizes, aspect ratios, masses, and bolometric luminosities of the cores are listed 
in Table\,\ref{tab-res-physprop}. The relation between SEDs and evolutionary stages 
is discussed in Sect.\,\ref{ssec-res-evol}.

The mean core diameters, as defined by the $N_{1^{\prime}}$ contour and listed in Table\,\ref{tab-res-physprop},
range from $8.5\times10^3$\,AU (0.04\,pc; B\,335) 
to $5.2\times10^4$\,AU (0.25\,pc; CB\,4), with a mean of $(2.5\pm1)\times10^4$\,AU ($0.12\pm 0.05$\,pc).
The mean aspect ratio of the cores is $\langle b:a\rangle = 0.8\pm 0.14$,
i.e., they are somewhat rounder than the globule envelopes.
However, this latter difference is not significant since beam-smoothing does affect the cores more than the envelopes.
The mean angular diameter of the cores derived this way is 
for all sources more than twice larger than the beam (HPBW 36\asp4), and for half of the sources 
more than three times larger.
Hence, the effect of the beam size on the core definition is small, although not 
completely negligible in some sources. 
A more severe limitation on the accuracy of core mass estimates may come from the fact that in the presence 
of unresolved warm protostars and large LoS temperature gradients, the column density maps are corrupted 
at the location of the protostar and the peak column density can be underestimated (see Sect.\,\ref{ssec-mod-gb}).
In these cases, the core masses are underestimated and the core sizes have larger uncertainties.

The mean total hydrogen mass of all dense cores in our sample is $1.2\pm 0.9$\,\msun, with the lowest-mass 
core being in B\,335 (0.2\,\msun), the highest-mass core in BHR\,12 (3.7\,\msun),
and no significant systematic difference between starless and protostellar cores.
The cores in single-core globules contain on average 25$\pm$6\,\% of the total mass of the globule,
irrespective of whether they already contain a protostar or not.
In globules with two cores (CB\,26 and CB\,244), the total core mass fraction is comparable (30$\pm$5\,\%),
but the individual cores constitute only 15$\pm$6\,\% of the total globule mass.
In terms of their luminosity, starless and protostellar cores are more distinct from each other.
The mean bolometric luminosity of the seven starless cores (Tab.\,\ref{tab-res-physprop}, including CB\,130) 
is $0.43\pm 0.15$\,\lsun, while that of the cores with embedded protostars is 4.5\,\lsun\ 
with a large scatter ranging from 0.5\,\lsun\ to 15\,\lsun\ (excluding CB\,17\,-\,IRS).
The dense starless cores emit on average 20$\pm$5\,\% of the total luminosity of their hosting globules,
while protostellar cores emit on average 40$\pm$10\,\%.
These relations are discussed in more detail and in the context of the thermal structure of the globules 
and the relative effects of external and internal heating in Sect.\,\ref{ssec-res-temp}.

 
\subsection{Evolutionary stage tracers}  \label{ssec-res-evol}

\begin{figure}[htb]
 \centering
 \includegraphics[width=8cm]{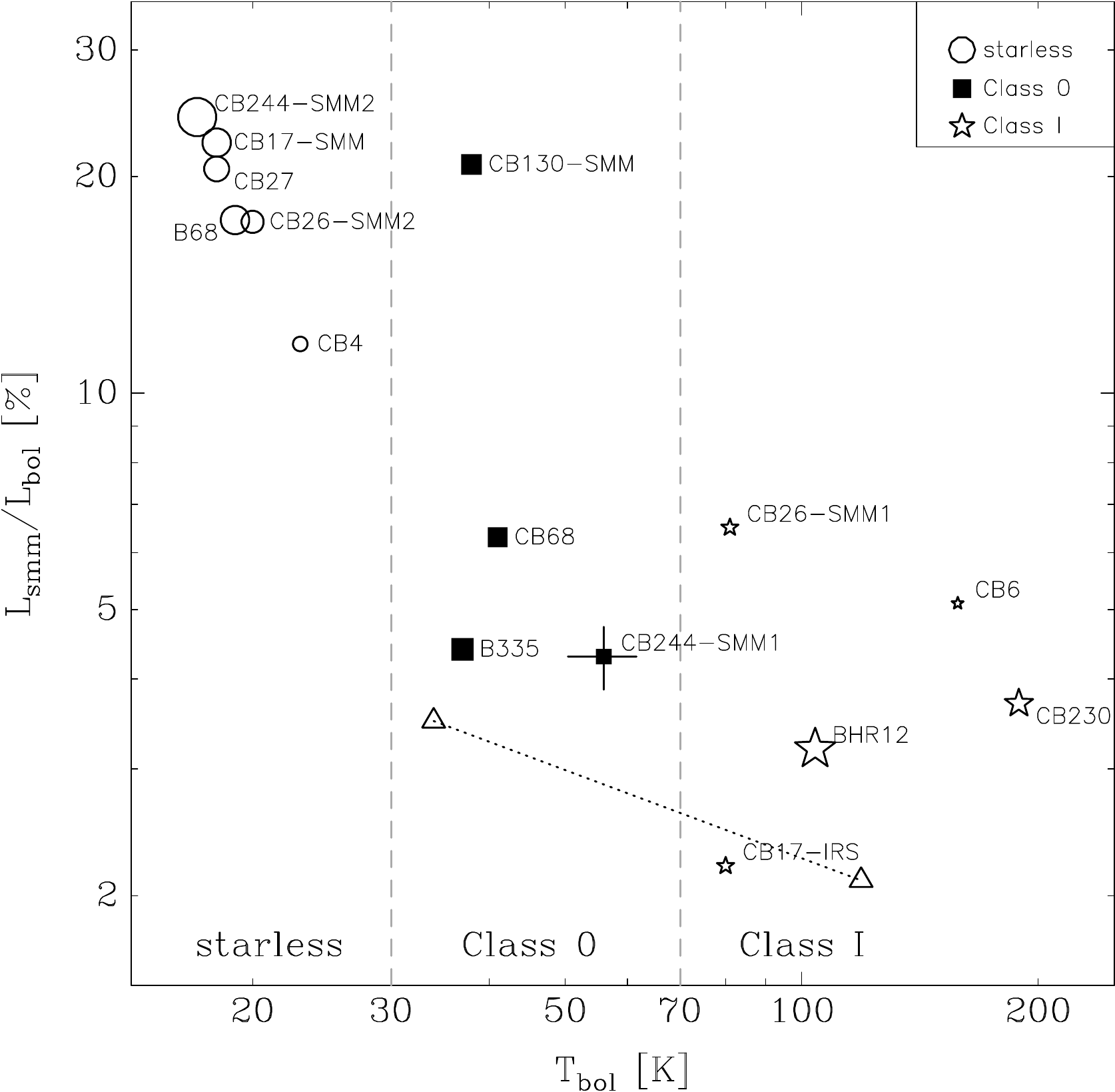}
 \caption{\label{fig-tbol-lratio}
  $L_{\rm bol}/L_{\rm submm}$\ ratio vs. bolometric temperature of the globule cores 
  (see Table\,\ref{tab-res-physprop} and Sect.\,\ref{ssec-res-evol}). 
  The size of the symbols scales with the peak column density of the respective sources. 
  Error bars on CB\,244\,-\,SMM1 illustrate the 10\% relative uncertainity on both values (Sect.\,\ref{ssec-res-evol}).
  The ``classical'' $T_{\rm bol}$\ boundaries for Class\,0 protostars are indicated by the 
  vertical dashed lines. The two triangles, connected by a dotted line, represent one synthetic 
  protostar\,-\,envelope system taken from \citep{robit2006}, once seen close to edge-on ($T_{\rm bol}=34$\,K) 
  and once seen nearly pole-on into the outflow cavity ($T_{\rm bol}=119$\,K; 
  see discussion in Sect.\,\ref{ssec-res-evol}).}
\end{figure}

The spectral classes and the approximate evolutionary stages of the sources in our sample 
were already known from earlier observations and modeling \citep[see Tab.\,\ref{tab-sourcelist} and][]{lau10}.
With the Herschel data in hand, which provide a much more complete 
coverage of the thermal SEDs and allow for more robust 
mass, luminosity, and $T_{\rm bol}$\ estimates than before, 
we re-evaluate the classical evolutionary tracers
$L_{\rm smm}\,/\,L_{\rm bol}$\ \citep{awb93}
and $T_{\rm bol}$\ \citep{ml93}.
Figure\,\ref{fig-tbol-lratio} shows the $L_{\rm smm}\,/\,L_{\rm bol}$\ ratio 
vs. $T_{\rm bol}$\ for all dense cores in the 12 globules of our sample. 
Uncertainties of these values are not straigt-forward to assess because the formal error bars 
are much smaller than the effect of exactly how a core and the respective integration area are defined. 
Based on tests with slightly different core definitions we estimate the relative uncertainties to be about 10\%.
In contrast to earlier ``pre-\herschel'' papers,
the new $L_{\rm smm}\,/\,L_{\rm bol}$\ vs. $T_{\rm bol}$\ diagram 
now also includes purely externally heated starless cores.
At the end of this section, we also use complementary information to relate the 
empirical ``Class'' of the individual sources to their actual evolutionary stage.


All starless cores in our sample, i.e., cores that do not have a compact 100\,$\mu$m 
source or any other signs of central heating or star formation, have 
$T_{\rm bol}<25$\,K and \mbox{$10\%<L_{\rm smm}\,/\,L_{\rm bol}<30\%$}.
We also find a correlation between $T_{\rm bol}$\ and $L_{\rm smm}\,/\,L_{\rm bol}$\ ratios 
and both mass and peak column density of starless cores in the sense 
that colder cores (lower $T_{\rm bol}$) with higher $L_{\rm smm}\,/\,L_{\rm bol}$\ ratios 
tend to be more massive (see Fig.\,\ref{fig-tbol-lratio}).
Since these cores have no significant internal heating sources, 
this correlation is likely to reflect purely the degree of external heating by the ISRF and shielding 
rather than an evolutionary effect (see also discussion in Sect.\,\ref{ssec-res-isrf}).
Massive cores with higher column densities are 
better shielded and can cool down more than less-massive and less-shielded cores. 
Even in the unlikely case that some of these cores would already have undetected first hydrostactic 
cores (FHSCs), their existence would have no measurable effect on $T_{\rm bol}$\ or the  
$L_{\rm smm}\,/\,L_{\rm bol}$\ ratio. 
Furthermore, an evolution of these integral quantities through the lifetmine of the FHSC is 
for the same reasons also not expected theoretically \citep[e.g.,][]{benoit2012}.
On the other hand, cores in larger molecular clouds with active star formation may be expected to 
have both more shielding and a stronger local ISRF, which may lead to different temperature profiles 
and different location in the $L_{\rm smm}\,/\,L_{\rm bol}$\ vs. $T_{\rm bol}$\ diagram as compared to isolated cores. 

All cores with embedded sources previously classified as Class\,0 protostars 
are confirmed to have \mbox{30\,K\,$<T_{\rm bol}<70$\,K} and and have 
\mbox{$3\%<L_{\rm smm}\,/\,L_{\rm bol}<7\%$}. 
The only exception is the core of CB\,130 with the embedded VeLLO, 
which has $L_{\rm smm}\,/\,L_{\rm bol}\sim 21\%$. 
All cores with embedded sources previously classified as Class\,I YSOs 
are confirmed to have $T_{\rm bol}>70$\,K and have \mbox{$2\%<L_{\rm smm}\,/\,L_{\rm bol}<7\%$}.

While the $L_{\rm smm}\,/\,L_{\rm bol}$\ ratios we derive for the isolated Class\,0 sources 
agree well with the upper half of range of the values listed by, e.g., \citet{awb2000} for Class\,0 sources in 
larger molecular clouds, we find significantly higher (up to a factor of 10) 
$L_{\rm smm}\,/\,L_{\rm bol}$\ ratios for the more evolved Class\,I YSOs than derived and 
proposed as threshold by \citet{awb93} and \citet[][$L_{\rm smm}\,/\,L_{\rm bol}^{\rm (Class\,I)}<0.5$\%]{awb2000}.
In fact, we do not find a significant systematic difference in the $L_{\rm smm}\,/\,L_{\rm bol}$\ ratios 
between Class\,0 protostars and Class\,I YSOs, but see at most a slight trend.
As in \citet{lau10}, we argue here that this systematic difference in $L_{\rm smm}\,/\,L_{\rm bol}$\ ratios 
between isolated and embedded Class\,I YSOs is unlikely to be an artifact of distance bias, 
angular resolution, or SED coverage, but rather reflects a combination of 
different methods of identifying the core boundaries and possibly real environmental differences. 
Luminosities for these earlier embedded samples were derived by \citet{am94}, \citet{ms94}, 
and others based on IRAS point source fluxes and ground-based 800 and 1100\,$\mu$m or 1.3\,mm maps only.
The compact, warm embedded sources dominate the small submm maps completely and the remaining low-level emission 
from a possible envelope was most likely chopped out to a large degree against the extended emission 
from the surrounding cloud (inter-core) material. Hence, these older observations could not correctly account 
for the envelope emission. 
In contrast, the column density profile of a globule core continuously decreases toward the "edge" of the small cloud, 
the isolated envelopes are heated more by the ISRF, and we do have spatially resolved emission maps 
at wavelengths between 100\,$\mu$m and 1.3\,mm at hand. The net effect is that we recover more envelope emission 
than those earlier observations and the envelopes of the isolated sources are more luminous than those 
of the embedded sources, thus elevating the $L_{\rm smm}\,/\,L_{\rm bol}$\ ratios.

When trying to interpret the quantities $T_{\rm bol}$\ and $L_{\rm smm}\,/\,L_{\rm bol}$\ 
as evolutionary tracers, one has to keep in mind that, 
as soon as an outflow starts to open a cavity in the envelope,
i.e., already early in the Class\,0 collapse phase, the observed SED at wavelengths 
shortward of the thermal peak at $\lambda\approx 160$\,$\mu$m becomes highly dependent on the 
viewing angle toward that cavity \citep[e.g.,][]{whitney2003,enoch2009,offner2012}.
Correspondingly, as long as one does not take 
into account the full source geometry, but instead simply integrates the observed fluxes 
over 4\,$\pi$, quantities derived from the SED also become projection-dependent.
To illustrate this, we have taken one specific synthetic protostar\,-\,envelope system 
from \citet[][model ID: 3008984]{robit2006} with properties similar to our ``Class\,0'' sources 
(stellar and envelope mass, luminosity, envelope radius, etc.), and analyzed its synthetic SED in the same 
way as our real sources. This synthetic source was once ``observed'' close to edge-on ($i=87$\degr, 
w.r.t. the outflow axis), and once close to pole-on ($i=18$\degr). 
Although the absolute $L_{\rm smm}\,/\,L_{\rm bol}$\ ratios derived for this model are systematically somewhat 
lower than those of our real sources, possibly due to a combination of the simplicity of the specific model 
and the same physical reasons as discussed above, the $T_{\rm bol}$\ values for the two projections (34\,K and 119\,K) 
and the $L_{\rm smm}\,/\,L_{\rm bol}$\ trend cover the full range of ``Class\,0'' and ``Class\,I''
sources (see Fig.\,\ref{fig-tbol-lratio}).
Hence, without involving more specific information on the opening angle and orientation 
of the outflow cavities, e.g., from NIR images and high-resolution molecular line maps, 
together with radiative transfer modeling, a robust evolutionary classification 
of sources with $T_{\rm bol}>30$\,K just from the SEDs is not possible. Protostars seen pole-on 
have SEDs that can be very similar to those of more evolved YSOs (``Class\,I'' stage) seen edge-on,
i.e., for individual sources there is no clear one-to-one correspondence between spectral Class and 
evolutionary stage.

For CB\,26\,-\,SMM1, we know that it is seen nearly edge-on \citep{lau09}, i.e., 
its classification as YSO in the Class\,I stage seems to be robust. 
BHR\,12 is harder to judge since the two sub-cores with nonaligned outlows are not resolved in our SED.
However, the NIR images also seem to indicate that we are viewing them from a more edge-on orientation
\citep[e.g.,][]{hodapp95}. Hence, its classification as YSOs in the Class\,I stage seems also robust. 
For CB\,6 and CB\,230, the NIR images suggest that we are at least partially looking into the outflow 
cones \citep{lau10}, i.e., they could well be protostars seen pole-on.

 
\subsection{Stability of starless cores}  \label{ssec-res-stability}


\begin{figure}[htb]
 \centering
 \includegraphics[width=8cm]{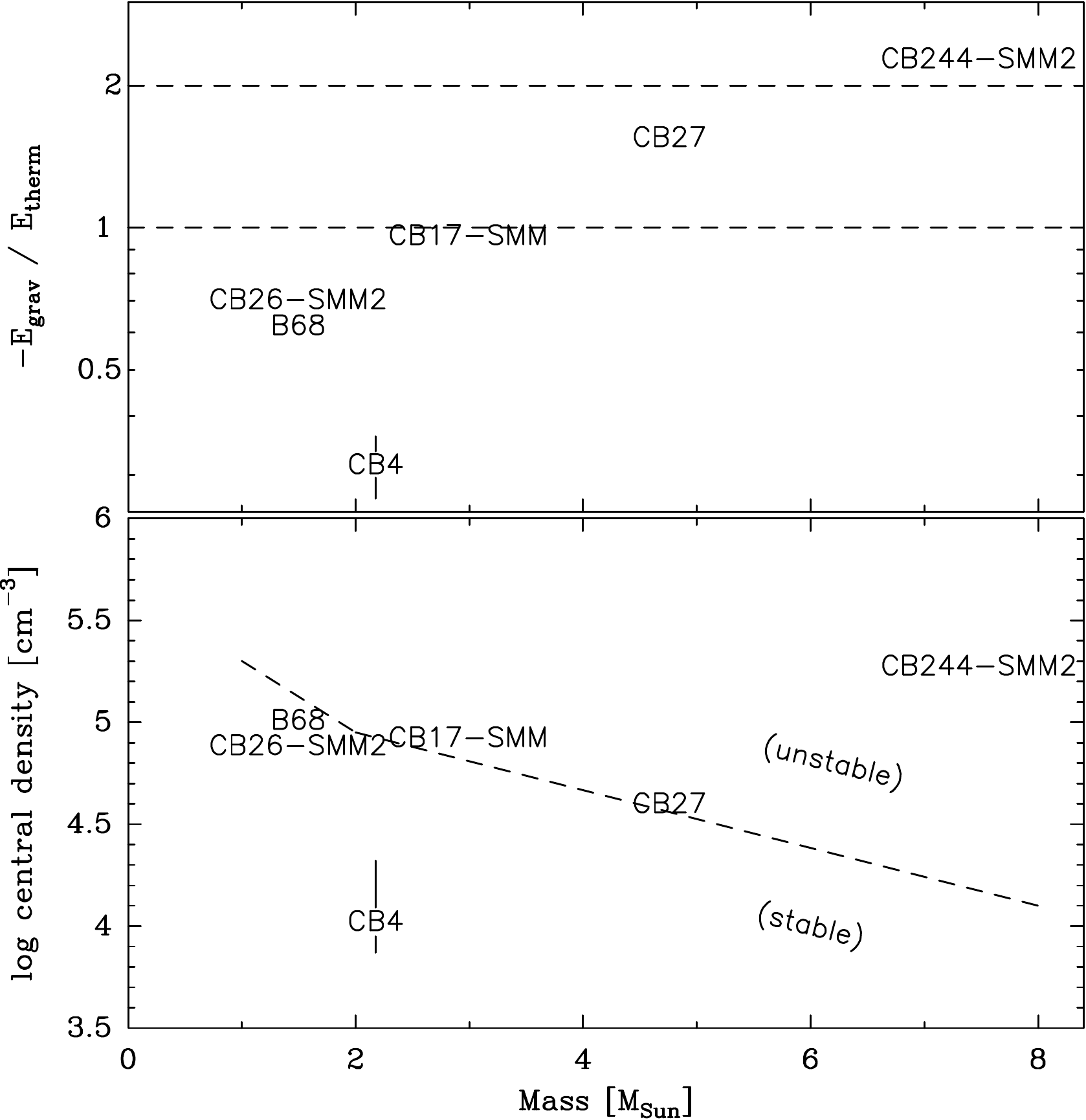}
 \caption{\label{fig-stability}
 Stability of starless cores.
 Top: Ratio of gravitational potential to thermal kinetic energy vs. total gas mass ($1.36\times M_{\rm H}$), 
 both integrated within the $N_2$\ column density contours (Table\,\ref{tab-res-prof}). 
 The lower dashed horizontal line marks the bounding limit of $-E_{\rm grav}/E_{\rm therm}=1$, 
 while the upper line marks the state of virialization at $-E_{\rm grav}/E_{\rm therm}=2$.`
 Bottom: Estimated central density (see Sect.\,\ref{ssec-res-stability}) vs. total gas mass for the same starless cores.
  The dashed line marks the maximum stable density of a pressure-supported, self-gravitating 
  modified (nonisothermal) BES (with photelectric heating at the core boundary taken into account) as calculated by 
  \citet[][their Fig.\,14]{keto08}. Typical estimated uncertainties on the source parameters are indicated as 
  Y error bars on CB\,4. The uncertainty in mass corresponds to the size of the label.}
\end{figure}


In order to assess which of the starless cores are unstable against gravitational collapse und must therefore be 
prestellar in nature, and which ones are stable, we evaluate their total gravitational potential and 
thermal kinetic energies and compare their central densities with the theoretically predicted maximum stable central density 
of self-gravitating starless cores.  

In a first step, we calculated the gravitational potential energy of the cores as
\begin{equation}                      \label{eq-egrav}
E_{\rm grav} = - \frac{G M^2}{R} \alpha_{\rm vir}\quad ,
\end{equation}
with $G$\ being the gravitational constant, and using 
$R=r_2$\ (Eq.\,\ref{eq-r1} and Table\,\ref{tab-res-prof}),
$M=M_2$\ the total gas mass enclosed in the $N_2$\ contour (Eq.\,\ref{eq-dens2}, mass not explicitely listed here), and 
the profile shape correction factor $\alpha_{\rm vir} = 0.75(\pm 0.15)$\ \citep{bertoldi92,sipila2011}.
The thermal kinetic energy is calculated as follows: 
\begin{equation}                      \label{eq-etherm}
E_{\rm therm} = \frac{3}{2} \frac{M k T}{\mu m_{\rm H}} \approx \frac{3 k d^2}{2\mu}\,\int N_{\rm H} T d\Omega \quad ,
\end{equation}
with $k$\ being the Boltzmann constant,
$\mu=2.33$\ the mean molecular weight,
$m_{\rm H}$\ the proton mass, 
and $N_{\rm H}\times T$\ being integrated within the $N_2$\ contour.
Figure\,\ref{fig-stability}\,(top) shows for the starless cores in our sample the ratio of gravitational to thermal energy 
vs. total gas mass $M_2$. The turbulent energy content of the starless cores 
(derived from C$^{18}$O(2--1) line widths; Lippok et al., in prep.)
is by a factor $20-50$\ smaller 
than the thermal energy for all cores and is thus negligible in this balance.
Given the uncertainties involved in this estimate, this energy ratio indicates that 
CB17\,-\,SMM, CB\,26\,-\,SMM2, CB\,27, and B\,68 are all gravitationally bound and approximately stable.
CB244\,-\,SMM2 must be supercritical, which is consistent with the conclusion derived by \citet{stutz10}, 
and CB\,4 does not seem to be gravitationally bound.

We also estimate the central volume number density of the starless cores by 
assuming spherical symmetry and dividing the 
peak column density $N_0$\ by the flat core radius $r_1$\ 
(Eqs.\,\ref{eq-densprof}, \ref{eq-dens0} and Table\,\ref{tab-res-prof}) 
and plot it in Fig.\,\ref{fig-stability}\,(bottom) against the total gas mass $M_2$.
Also indicated in this plot is the maximum stable central density of a pressure-supported, self-gravitating 
modified (nonisothermal) Bonnor-Ebert sphere \citep[BES;][]{bonnor56,ebert55}, 
calculated by \citet[][their Fig.\,14]{keto08}. For comparison with the isolated globule cores, 
we adopt their results for the model with photoelectric heating at the core boundary.
This evaluation supports the previous conclusions even more clearly, i.e., 
CB17\,-\,SMM, CB\,26\,-\,SMM2, CB\,27, and B\,68 are all gravitationally bound but seem to 
be thermally subcritical and just at the limit of stability. 

This complies well with the notion that the mean radii of the globules are similar to or only slightly larger 
than their Jeans lengths (Sect.\ref{ssec-res-seds}).
Interestingly, the projected separations between the two sub-cores in CB\,26 (0.15\,pc) and CB\,244 (0.09\,pc) 
agrees quite well with the typical Jeans lengths in these clouds 
\mbox{($\sim 0.13$\,pc; see Sect.\,\ref{ssec-res-seds})}, indicating that large globules 
tend to fragment gravitationally on the Jeans scale. Also interesting in this respect is that in both 
globules, one of the two cores has already formed a protostar, while the other one is still starless/prestellar.

CB\,4 is clearly a subcritical globule and must be purely pressure-confined.
CB\,244\,-\,SMM2 must be a supercritical prestellar core at the verge of collapse,
unless it has significant other nonthermal support like, e.g., strong magnetic fields.
The CB\,244 region has been studied by sub-mm polarization observations reported 
in \citet{wolf03}, and the SMM1 core has indeed been found to exhibit a relatively strong 
magnetic field (257\,$\mu$G), but the signal-to-noise for the SMM2 core turned out to be too 
low to derive meaningful field strengths.
Hence, CB\,244\,-\,SMM2 is a good candidate to look for kinematic infall signatures or 
signs of a possibly already existing FHSC.

Finding the majority of starless cores are bound and approximately stable 
is an expected result of a natural selection effect.
Due to the short collapse time scales \citep[at most $10^4$\,yrs, e.g.,][]{st2011},
unstable cores should be much rarer than stable cores which can have lifetimes of up to 
$10^6$\,yrs \citep[e.g.,][]{enoch08}.


\subsection{Thermal structure of the cores}  \label{ssec-res-temp}


\begin{figure}[htb]
 \centering
 \includegraphics[width=9cm]{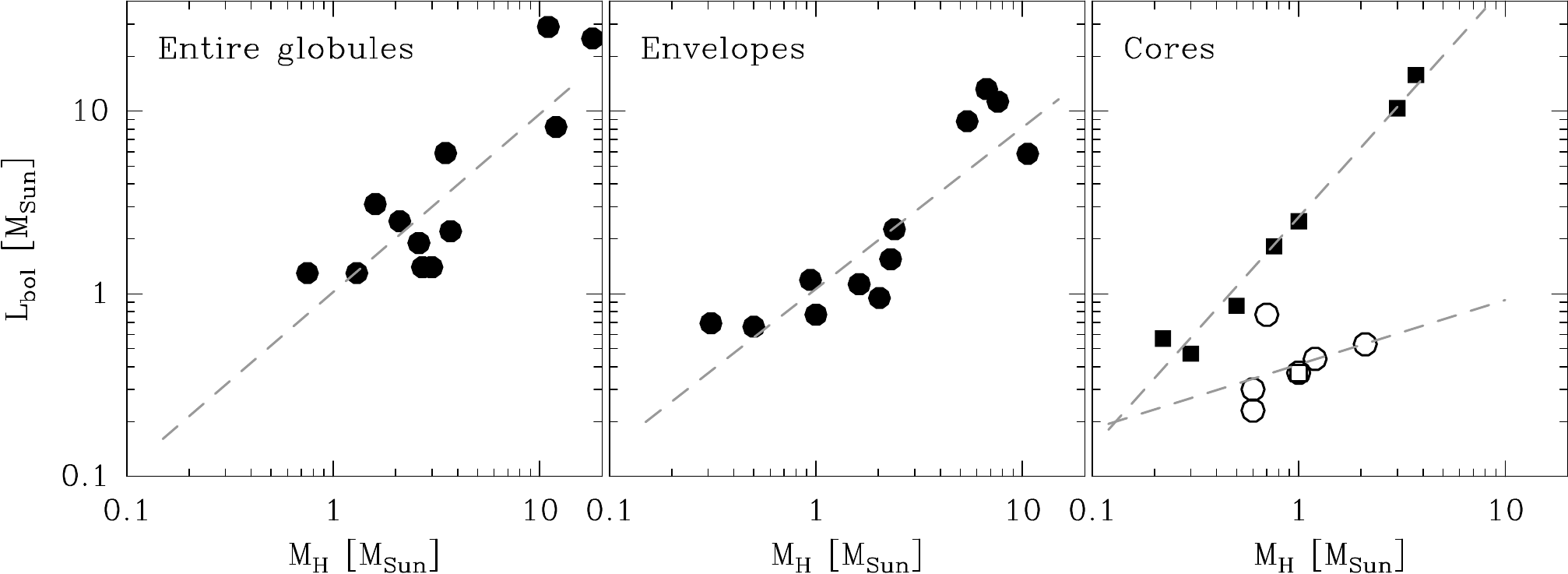}
 \caption{\label{fig-mh-lbol}
  Bolometric luminosity vs. total hydrogen mass of the entire globules (left),
  the envelopes only (without the dense cores; center) and the dense cores only (right;  
  see Table\,\ref{tab-res-physprop} and Sect.\,\ref{ssec-res-temp}). 
  Filled quares in the right-most panel mark protostellar cores, empty circels mark starless cores, 
  and the open square marks the core of CB\,130 with the embedded VeLLO.}
\end{figure}

\begin{figure}[htb]
 \centering
 \includegraphics[width=9cm]{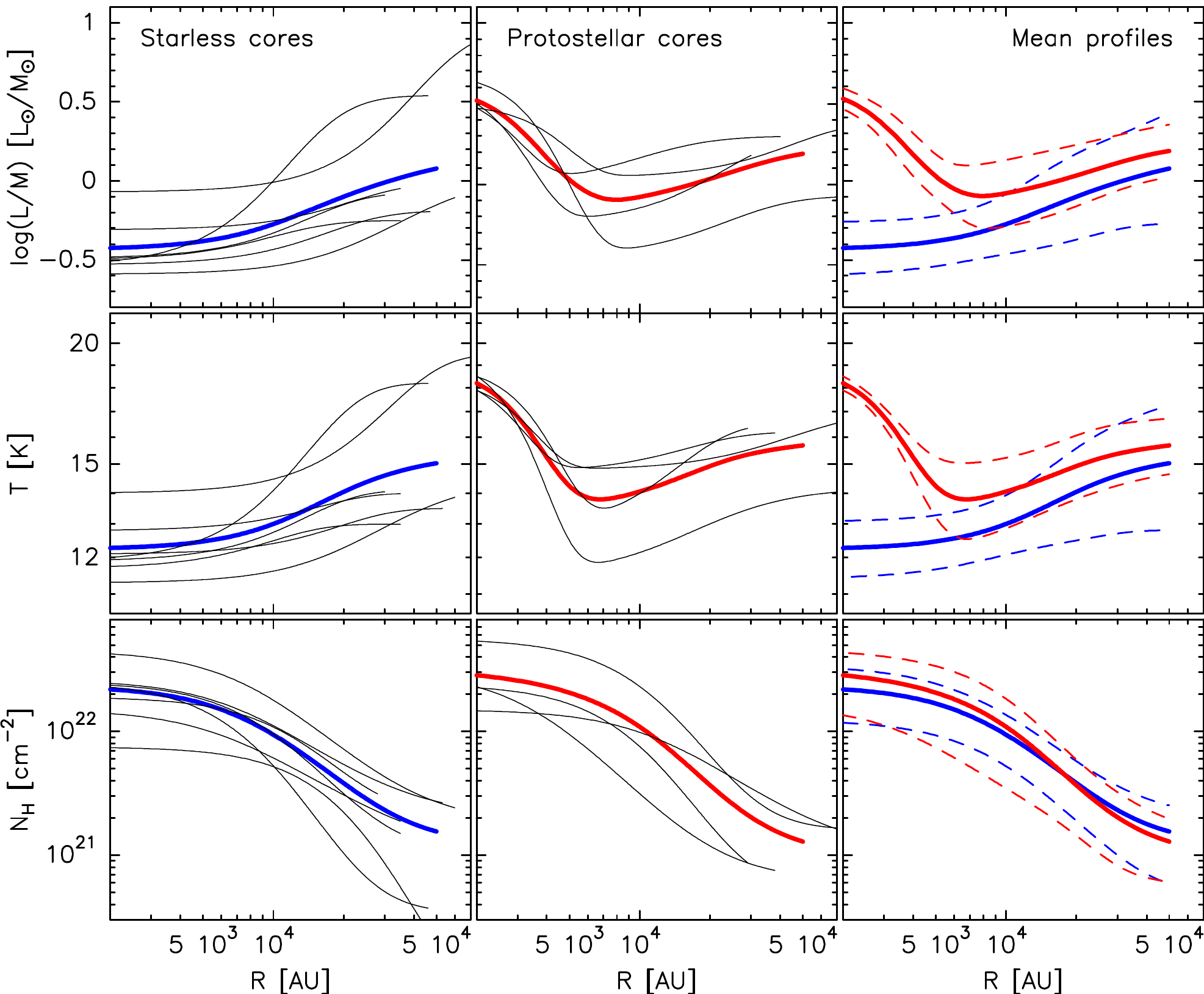}
 \caption{\label{fig-profiles-all}
  Radially averaged column density profiles (bottom), LoS-averaged dust temperature (center), 
  and luminosity-to-mass ratio profiles (top), for individual sources (thin black solid lines) 
  as well as mean profiles (thick solid lines) of starless and protostellar cores. 
  Note that the latter ones are fits to the mean profiles, while we list the means of the indivudual 
  profile parameters in Table\,\ref{tab-res-prof}, which is not exactly the same.
  Dashed lines in the rightmost panels indicate the 1\,$\sigma$\ uncertainty ranges of the 
  mean profiles.}
\end{figure}


Figure\,\ref{fig-mh-lbol} shows the bolometric luminosity vs. total hydrogen mass 
of the entire globules, the dense cores only, and the envelopes only (globule minus core).
All globules turn out to have about the same luminosity-to-mass ratio of $1.1\pm 0.6$\,\lsun/\msun,
irrespective of whether they harbor starless cores or protostars.
The mean luminosity-to-mass ratio of the envelopes (cores subracted) is very similar, 
with somewhat smaller scatter \mbox{($1\pm 0.5$\,\lsun/\msun)}. 
A significant difference in the luminosity-to-mass ratios between star-less and 
protostar-harboring sources exists only for the central dense cores,
with starless cores having $0.5\pm 0.25$\,\lsun/\msun\ and 
protostellar cores having $2.5\pm 1$\,\lsun/\msun.
This alone already points toward the dominating effect of the externally heated envelopes 
for the globules and the very local effect of internal heating 
by embedded (low-mass) protostars. The embedded protostars have very little effect on 
the overall energy balance of their host globule.

This dominating effect of external heating and shielding becomes even more evident 
in the radial profiles of the dust temperature and the local luminosity-to-mass ratios.
Figure\,\ref{fig-profiles-all} shows these radial profiles along with the column density 
profiles for all resolved starless and protostellar cores as well as the 
mean profiles for these two groups. 
Both starless globules and globules harboring protostars approach at their outer boundaries,  
at radii of $\sim 5\times 10^4$\,AU and column densities of $\sim 10^{21}$\,cm$^{-2}$, 
dust temperatures of $15.5\pm 1.5$\,K and local luminosity-to-mass ratios of $2\pm 1$\,\lsun/\msun.
From the outer boundaries inward, both the dust temperatures and local luminosity-to-mass ratios 
decrease to reach mean values of 12\,--\,13\,K and $\sim 0.4$\,\lsun/\msun\ at radii of 
$\sim 5000$\,AU. Hence, all of these cores are far from being isothermal. They are heated externally by 
the IRSF and their innermost dense centers are well-shielded from the energetic short-wavelength radiation 
of this heating source. 
Local heating by the embedded protostars raises the dust temperature 
and local luminosity-to-mass ratio only inside radii of $\sim 5000$\,AU, where the derived local temperature
is only a lower limit due to the effects of beam-convolution and LoS-averaging.
However, there are also some systematic differences between certain globules, 
which we attribute to differences in the shielding by extended halos and difference in the local ISRF. 
This is being discussed in the next section.
Table\,\ref{tab-res-prof} also lists the column density and temperature profile fit parameters 
of the mean profiles of protostellar and starless cores in our sample.
Note that these mean profile parameters are averages of the individual source profile parameters,
while the mean profiles shown in Fig.\,\ref{fig-profiles-all} are derived by fitting the averaged profiles.
These observationally derived mean profile parameters can be used to construct representative 
cores for radiative transfer models. Such models must be able to explain not only the outer 
and inner dust temperatures, but also the temperature contrast between core and envelope.


\subsection{Constraints on the interstellar radiation field}  \label{ssec-res-isrf}


\begin{figure}[htb]
 \centering
 \includegraphics[width=8cm]{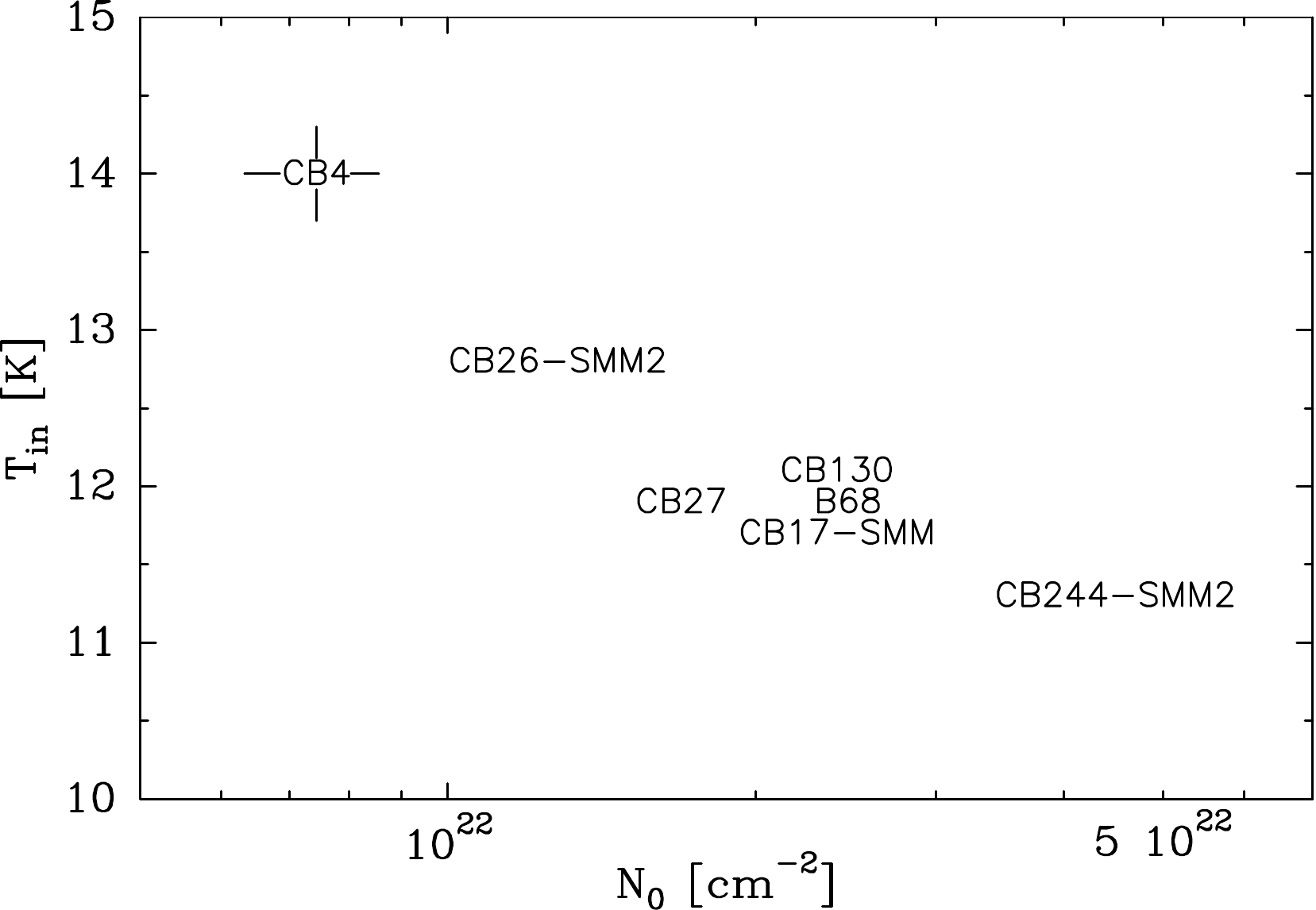}
 \caption{\label{fig-tin-nh}
  Central minimum (LoS-averaged) dust temperature, $T_{\rm in}$, of starless cores vs. peak 
  column density $N_0$. Representative error bars are indicated only for CB\,4.}
\end{figure}


Figure\,\ref{fig-tin-nh} shows the central minimum (LoS-averaged) dust temperature, 
$T_{\rm in}$, of all starless cores in our sample vs. their peak column density $N_0$\ (see Table\,\ref{tab-res-prof}).
The very tight anti-correlation between total column density and central dust temperature 
indicates very clearly that the inner temperatures of passive molecular cloud cores (without 
internal heating sources) are controled by the shielding against the ISRF. It also indicates 
that these nearby and isolated globules are all exposed to about the same ISRF.
The correlated errors between $T_{\rm in}$\ and $N_0$\ are much smaller than the range of 
values for these sources and therefore do not affect this conclusion.

We also use the observed outer dust temperatures and halo column densities ($N_{\rm out}$) of our sources 
to derive constraints on the mean ISRF.
For this purpose, we compare the observed temperature profiles of the cores in this paper
to a grid of 1D radiative transfer models (Shirley et al., in prep.).  
Dust temperatures are self-consistently
calculated using the CSDUST3 code \citep{egan88}
for an input density profile, input dust opacities, and input
heating conditions.  
The density profiles are isothermal BES with the BES temperature derived iteratively at the half-central
density point of the profile from the radiative transfer calculation. 
The central densities of the BESs vary from $9 \times 10^3$\,cm$^{-3}$\
to $6 \times 10^6$\,cm$^{-3}$\ with an outer
radius set to 30,000\,AU.

The dust opacities are taken from OH5 and are augmented by scattering coefficients according
to the prescription given in \citet{ye2005}.  The heating is derived from the ISRF
plotted in Figure\,3 of \citet{shirley05} and
is modified in two ways: the overall strength of the UV to FIR
ISRF is multiplied by a factor ($S_{\rm ISRF}$), and the model core is assumed to
be embedded within a medium with extinction 
given by $A_{V}(R_{\rm out})$\ affecting only the shortest wavelengths ($\lambda <$ few $\mu$m) of the ISRF.  
The final grid consists of 3600 models with self-consistently calculated
dust temperature profiles.  

\begin{figure}
 \centering
 \includegraphics[width=8cm]{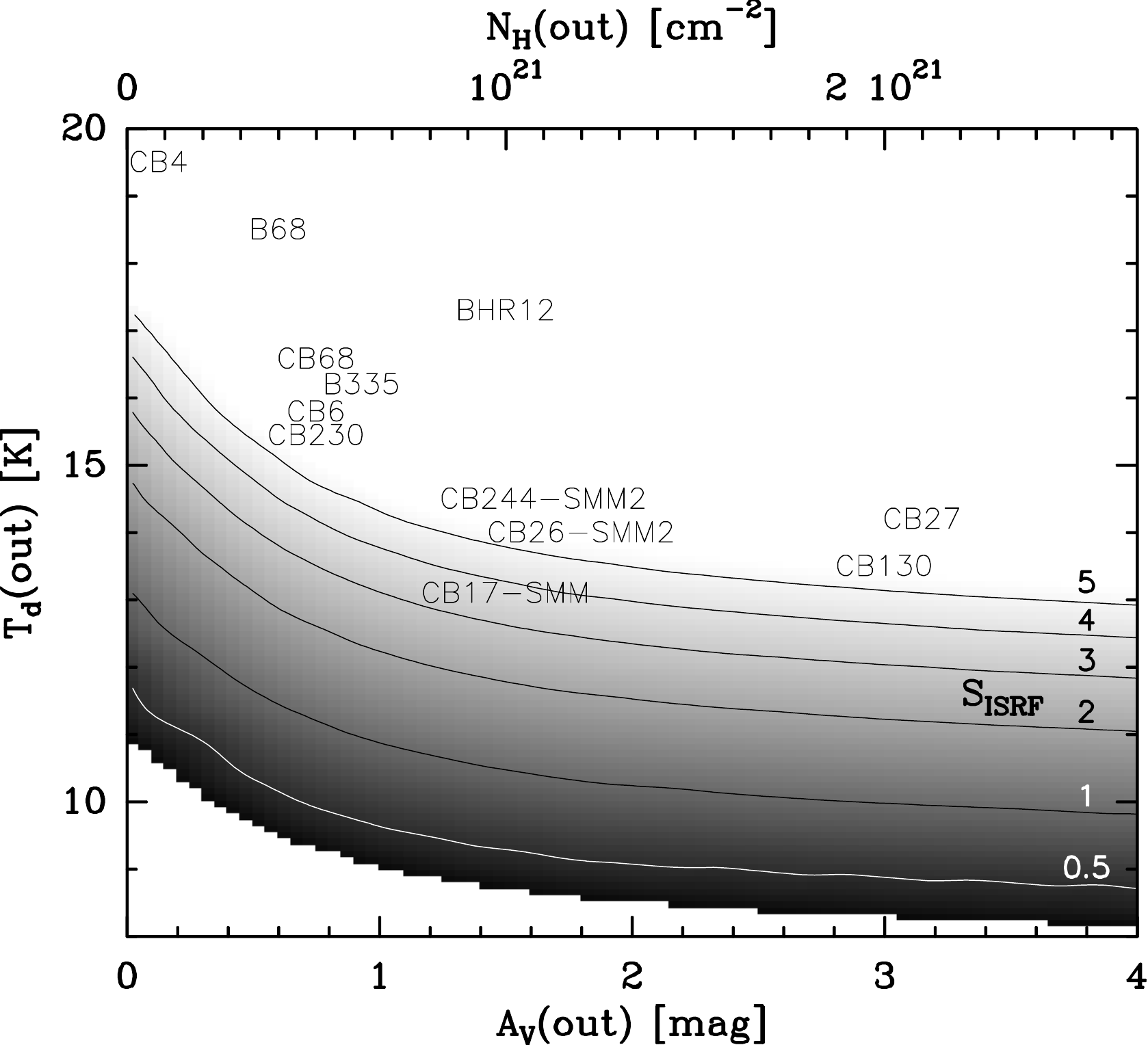}
 \caption{\label{fig-isrf}
  Observed mean outer dust temperatures of the globules 
  vs. column density of the outer halo (see Table\,\ref{tab-res-prof}). 
  The conversion to $A_V$\ 
  was done with the mean $A_V$-to-$N_{\rm H}$\ ratio of $(1.5\pm0.5)\times10^{-21}$\ 
  derived from the comparison of NIR extinction maps (not used further in this paper) 
  of 9 globules and the $N_{\rm H}$\ maps derived in this paper and thus remains somewhat uncertain.
  Errorbars mark uncertainties of $\pm 0.5$\,K for $T_{\rm d}$\ and 20\% for $N_{\rm H}$(out), repectively.
  Plotted as grayscale and contours is the dust temperature at the outer radius (30,000\,AU) of a model cloud, 
  derived with the radiative transfer modeling descibed in Sect.\,\ref{ssec-res-isrf}, 
  versus shielding of the cloud against the ISRF by a halo and as function 
  of the relative strengths of the ISRF.}
\end{figure}


Figure\,\ref{fig-isrf} compares the modeled relation between the outer temperature (at 30,000\,AU), 
the relative strengths of the ISRF, and the additional shielding of the cloud against the ISRF by a halo 
for a model core that matches the mean column density profile of the observed starless globules 
(Fig.\,\ref{fig-profiles-all}) with the observed outer dust temperature 
and column densities of the outer halos (see Table\,\ref{tab-res-prof}).
The observed outer temperatures range from 13.1\,K to 19.5\,K and occupy the upper 
half of the model outer temperatures which range from 8.1\,K to 17.8\,K. 
The general shape of the observed dependency of the outer temperature on the shielding 
against the ISRF agrees well with the model calculations, suggesting that the outer temperature 
is indeed controled to first order by the shielding against an approximately uniform 
(not necessarily isotropic) ISRF. The only source that deviates significantly from the observed 
relation toward higher outer temperature or stronger ISRF is BHR\,12 (CG\,30). 
This cometary globule in the Gum nebula region is exposed to the 
Vela pulsar and supernova remnant \citep{large68} and to the O4 supergiant star $\zeta$\,Puppis, 
and might thus indeed experience a stronger local ISRF than the other globules (Fig.\,\ref{fig-bhr12-morph}).

However, the comparison between observations and model calculations also shows that the isolated globules 
have on average higher outer dust temperatures than predicted by \citet{shirley05}. 
This is probably due to a combination of a stronger ISRF than adopted and lower submm opacities for grains 
with a smaller size distribution than OH5. For this calculation, we used the same dust opacity model for 
both the dense cold cores and for the thin extended envelopes. OH5 opacities, which are calculated for coagulated 
dust grains with ice mantles, are certainly not appropriate for the thin extended halos around the globules. 
Though opacity models for smaller grains will not significantly affect the observational derivation of dust 
temperatures from the SEDs, they would lead to smaller values of the derived outer column densities (by a factor of a few).
At the same time, smaller grains would lead to higher outer dust temperatures in the self-consistently 
calculated models. These issues require further modeling before robust conclusions on the strength of the ISRF can be drawn 
and will be addressed in a forthcoming paper that will also include information from NIR extinction maps.

Nevertheless, the observed relation between outer dust temperatures and halo column densities complies very well 
with the general picture of external heating by the ISRF and, at the same time, suggests that 
our background flux level treatment and extended emission recovery (Sect.\,\ref{ssec-mod-prep}) worked well 
and that the detected ``halos'' indeed belong to the globules and do not represent unrelated background material.


\subsection{Environmental effects}           \label{ssec-res-env}

Although all globules selected for this study appear at first glance 
compact and not very filamentary, eight out of the 12 globules 
are cometary-shaped with one sharp rim and either a long cometary tail seen 
both in the optical cloudshine as well as in the FIR dust emission (e.g. CB\,6, CB\,17, CB\,68) 
or at least a very diffuse side opposite the sharp rim (e.g., CB\,26, CB\,230; 
see Table\,\ref{tab-res-physprop} and Figs.\,\ref{fig-cb4-morph} through \ref{fig-cb244-morph}).
Most of these globules where not classified originally as cometary globules like, e.g., 
BHR\,12 \citep[CG\,30;][]{zealey83}, but appear at first glance rather round and isolated when seen on 
less deep optical images or older dust emission maps. The diffuse tails have projected physical lengths 
between 0.3\,pc (CB\,27; Fig.\,\ref{fig-cb27-morph}) and 2.6\,pc (CB\,6; Fig.\,\ref{fig-cb6-morph}).
In some cases, like CB\,68, the long diffuse tail connects the globule to larger (diffuse) clouds 
(Figs.\,\ref{fig-rhooph} and \ref{fig-cb68-morph}). In other cases, like BHR\,12, 
the cometary tail is pointing away from an energetic irradia\-ting source (Fig.\,\ref{fig-bhr12-morph}).
In some sources, like, e.g., CB\,17, neither a connection of the long tail to another structure 
nor an illuminating source could be clearly identified.  
This diversity suggests that cometary shape and tails may not have the same physical origin in all sources.

\begin{figure*}
 \centering
 \includegraphics[width=15cm]{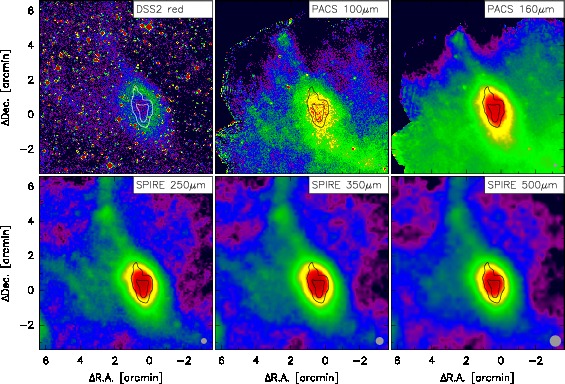}
 \caption{\label{fig-cb4}
          CB\,4 (starless core): Visual (red) DSS2 image and Herschel FIR maps (color; log scale) at 100, 160, 250, 350,
          and 500\,$\mu$m, with 1.2\,mm continuum contours overlaid 
          (20 and 40\,mJy/30$^{\prime\prime}$beam). Herschel beam sizes are indicated 
          as gray circles in the lower right corners.}
\end{figure*}

We have also evaluated the location and orientation of warm rims on the outer edges of the globules 
(like, e.g., CB\,26: Fig.\,\ref{fig-tmap-cb26}) and their relation to (potential) external heating sources, 
like nearby stars or the galactic plane. 
Even though the one-sided warm rim appears at first glance questionable in some cases because 
we have recorded no information for the temperature of the opposite rim (e.g., CB\,4 and CB\,17:
Figs.\,\ref{fig-tmap-cb4} and \ref{fig-tmap-cb17}), the existence and location of the one-sided warm rims
is indeed significant. The warm rims, were present, are on average $2\pm 0.4$\,K warmer than the opposite 
cold rims. Their presence can be understood by inspecting the PACS and SPIRE maps of, e.g., 
CB\,4 (Fig.\,\ref{fig-cb4}). While the long-wavelength SPIRE maps, which are to first order sensitive 
to column density, show an approximate symmetry between northern and southern side of the globule 
(apart from the cometary tail), the shorter-wavelength PACS maps, which are more sensitive to 
temperature, show a significant asymmetry with much stronger emission on the southern side.
Indeed, after fitting the data, the column density contours indicate an approximate symmetry between 
northern and southern side down to a few times $10^{20}$\,cm$^{-2}$\ (Fig.\,\ref{fig-tmap-cb4}). 
At lower column densities, 
we can still derive the temperature at the warm southern rim where the short-wavelength PACS emission is 
sufficiently strong. But on the northern side, the short-wavelength PACS flux densities drop below 
the detection threshold. Hence, we can no longer reliably derive a temperature for the 
cold material which is still traced by the SPIRE maps and may have the same column density as 
the warmer material on the southern side. When allowing the fitting routine to derive temperatures 
from only three (SPIRE) wavelengths (cf. Sect.\,\ref{ssec-mod-gb}), we obtain a temperature map that 
fully confirms the above result and rules out a misinterpretation due to a detection bias 
driven the data reduction.
However, these SPIRE-only derived temperature maps are too unreliable and noisy to be used in the analysis 
of the source structure presented in this paper. Furthermore, the comparison of the results for 
CB\,244 obtained with different 
versions of the data reduction and calibration pipeline (Sect.\,\ref{ssec-dis-data}) also show 
that the extent and location of the warm south-eastern rim in CB\,244 (Fig.\,\ref{fig-tmap-cb244}) 
remains unaffected by the specific data reduction version or background subtraction level.

Figures\,\ref{fig-cb4-morph} through \ref{fig-cb244-morph} show the 
large-scale morphology and surroundings 
of all regions studied in this paper, including the schematic indication of the warm rims and the location 
of potential external illumination sources.
We find no clear indication for a general link between the loaction of the warm rim side 
in a globule and the location of identified potential external heating sources. 
In some globules, the warm rim points toward the galactic plane
(CB\,26, B\,68, CB\,68, CB\,230, CB\,244: 
Figs.\,\ref{fig-cb26-morph}, \ref{fig-b68-morph}, \ref{fig-cb68-morph}, \ref{fig-cb230-morph}, and \ref{fig-cb244-morph}), 
in others it points in the opposite direction 
(CB\,4, CB\,27, CB\,130: Figs.\,\ref{fig-cb4-morph}, \ref{fig-cb27-morph}, and \ref{fig-cb130-morph}). 
In CB\,6 (Fig.\,\ref{fig-cb6-morph}) and CB\,17 (Fig.\,\ref{fig-cb17-morph}), 
there are bright nearby stars identified that could be located at the same distance 
and thus illuminate the globules directly. However, the warm rims do not point 
toward the potential illuminators.
Except for BHR\,12, no other nearby heating or illuminating sources could be uniquely identified.
However, even in this globule, the tail pointing away from the illuminating source is significantly 
warmer than the side facing it (Figs.\,\ref{fig-tmap-bhr12} and \ref{fig-bhr12-morph}). 
Hence, we have to conclude that, as with the cometary tails, the link between warm rims and illuminating 
sources or the physical reason for the asymmetric warm rims remains inconclusive and needs to be addressed 
by future studies.


\section{Discussion}           \label{sec-disc}
 

\subsection{Calibration and data reduction uncertainties}  \label{ssec-dis-data}

The total flux scale uncertainty of the \herschel\ data is assumed to be 15\% (Sect.\,\ref{ssec-mod-prep}),
while that for the ground-based submm bolometer data is about 20\%  (Sect.\,\ref{ssec-obs-submm}).
These flux calibration uncertainties propagate directly into the column density and mass estimates.
However, the effect on the dust temperature estimates should be much smaller for the following reason.
Since all bands within one of the \herschel\ instruments were photometrically calibrated in a consistent way,
any systematic uncertainties in the effective temperature of the calibrators should be correlated between the 
bands and thus to first order not affect the flux ratios that determine the temperature. 
There could still be a systematic offset between PACS and SPIRE due to different calibration procedures.
However, if such an offset exists, it would have become immediately evident in the combined 
pixel-by-pixels SEDs since we have more than one wavelength band from each instrument. 
Since, after the background offset subtraction (Sect.\,\ref{ssec-mod-prep}), 
we have not noticed any such offset, it must be very small (see also the integrated SEDs in Fig.\,\ref{fig-sedall}).
Similarly, any systematic errors that we could have possibly introduced in the various post-processing 
steps described in Sect.\,\ref{ssec-mod-prep} (e.g., spatial filtering) should also be correlated to 
first order between the \herschel\ bands.
Here we assess these effects quantitatively and estimate the uncertainties introduced by using different and older 
versions of HIPE and Scanamorphos for different sources (Sect.\,\ref{ssec-obs-herschel}). 
I particular, we have reprocessed 
all \herschel\ data from one source (CB\,244, originally reduced with HIPE v.\,5/6 and Scanamorphos v.\,9) 
with the newest versions of HIPE (v.\,9) and Scanamorphos (v.\,18), with all additional post-processing steps 
(e.g., background subtraction) unchanged. The new reduction resulted in up to 20\% lower PACS fluxes
(as compared to the old reduction), but left the SPIRE fluxes unchanged within $\approx 1$\%. As a result, 
the LoS-averaged dust temperatures derived with the newly processed \herschel\ maps are on average 
2.2\% lower than the old values, and the resulting column densities are on average 7.5\% higher. We also interpret 
these numbers here as a general assessment of the possible changes that new data reduction versions might 
bring, since  HIPE v.\,9.1.0 may not be the ``final'' version.

The complementary ground-based submm data were obtained independently with different instruments and 
processed with the respective instrument pipelines. Not only are the 
calibration uncertainties between the instruments and with respect to \herschel\ no longer correlated,
these data are also much more and in a different way affected by spatial filtering. 
The sum of these effects would make any results based on simple submm flux ratio maps very susceptible 
to a number of calibration uncertainites.
However, we use all data in a combined analysis in which the dust temperature estimate is to first order 
determined by the flux ratios of the \herschel\ bands, while the column density is to first order determined
by the long-wavelength SPIRE bands and the submm data.
In order to quantify the effect of the calibration uncertainties on our temperature estimates, 
we have carried out a Monte Carlo test on one pixel in the center of B\,68, 
artificially increasing all \herschel\ fluxes by 15\% and randomly modifying the three individual 
submm fluxes with a rms of 20\% and fitting the SEDs in the same way as we did for the normal analysis. 
As a result, the mean temperature of that pixel was elevated by 0.2\,K from 11.8\,K to 12.0\,K with a rms scatter of 0.2\,K. 
The mean column density was elevated by 6\% with a rms scatter of 10\%. 

The background subtraction described in Sect.\,\ref{ssec-mod-gb} could have potentially introduced 
additional uncertainties. However, since $I_{\rm bg} \ll B_{\nu}(10\,{\rm K})$\ at all wavelengths for all 
our sources, thanks to the initial source selection in regions of low background, 
the difference in the final temperature estimates between considering $I_{\rm bg}$\
or setting $I_{\rm bg}=0$\ is $\Delta T \le 0.15$\,K. The uncertainty introduced by 
not knowing the exact level of the background, but still subtracting a mean guess of $I_{\rm bg}$\ is 
correspondingly smaller, i.e., negligible.
We therefore estimate the overall relative 1\,$\sigma$\ uncertainty introduced by flux calibration and data processing 
uncertainties to be $\sigma_T\approx 3$\% and $\sigma_N\approx 15$\% (quadratic sums of individual 
contributions described above).


\subsection{Uncertainties and limitations of the modeling approach}  \label{ssec-dis-mod}


\begin{figure}
 \centering
 \includegraphics[width=8cm]{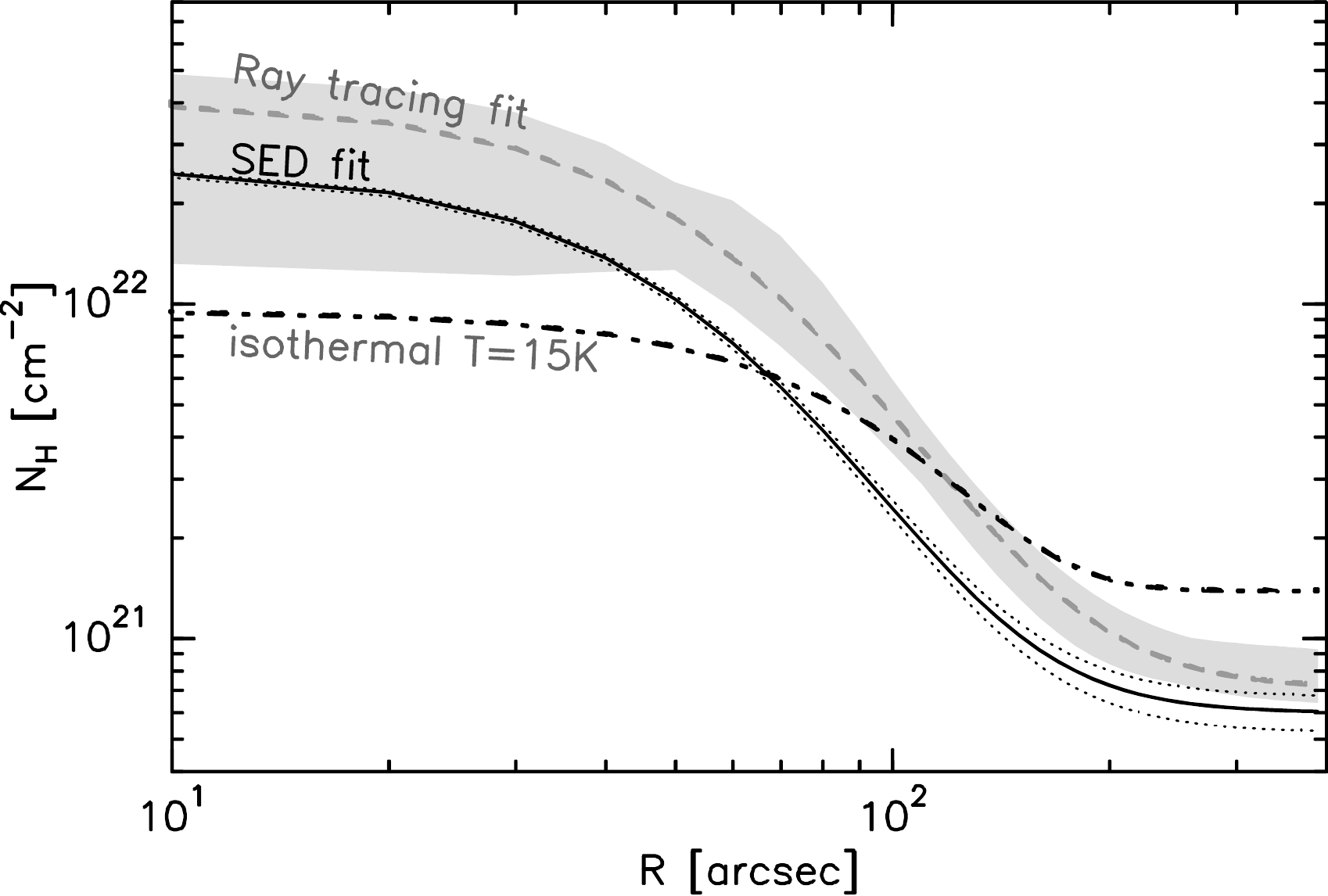}
 \caption{\label{fig-b68-profiles2}
 Comparison of radial column density profiles of B\,68,
 derived with the LoS-averaged temperature recovery method described in this paper (black solid line,
 uncertainty range marked by dotted lines), 
 with the assumption of isothermal dust emission (dash-dotted line), 
 and with a ray-tracing approach that takes LoS temperature gradients into account 
 \citep[gray dashed line, uncertainty range marked by gray-shaded area,][corrected for OH5 opacities]{nielbock12}.
}
\end{figure}

Throughout this paper, we used OH5 opacities \citep{oh94} for all sources and regions. 
Although this opacity model seems to be a reasonable assumption for this kind of sources, 
it is not well-constrained observationally and may not accurately reflect the actual dust properties 
of the individual objects. Furthermore, a single opacity model may not apply to both the dust in the 
cold dense cores, where coagulation and ice mantle growth may have affected the effective optical properties of the grains,
and that in the thin and warm outer envelopes.
Since the data do not provide a direct and unambiguous 
constraint on the dust opacities (this will be subject of a subsequent paper), we tested 
the dependency of the temperature and column density estimation on different opacity models and report the results 
here in the form of an uncertainty discussion.
In particular, we test for one source (B\,68) 
how the temperature and column density estimates derived with the following five dust opacity models 
differ with respect to the use of OH5 opacities: 
{\it (1)} OH1 \citep[MRN size distribution,][]{mrn1977},
{\it (2)} OH8 (coagulated grains like OH5, but with thick ice mantles),
{\it (3)} Weingartner \& Draine opacities for $R_V=5.5$\ 
    \citep{wd01,draine2003}\footnote{See http://www.astro.princeton.edu/$\sim$draine/dust/dustmix.html}, and 
{\it (4)} two pure beta models with $\beta=1.5$\ and $\beta=2.5$, normalized to 
   OH5 (which has $\beta\approx 1.9$) at $\lambda = 500\,\mu$m.
The first three of these models lead to deviations in the derived temperature w.r.t. OH5 
of about 4\% (0.4\,K at 10\,K), with the maximum deviation occuring for WD01 in the warm 
envelope (9\% or 1.6\,K lower than the 18\,K derived with OH5).
The two extreme beta models result in $\approx 10$\% higher temperatures (w.r.t. OH5) for $\beta=1.5$\ 
or $14-18$\% lower temperatures for $\beta=2.5$, respectively.
Hence, we can safely assume that the systematic uncertainty of the dust temperatures listed in 
Table\,\ref{tab-res-prof} introduced by the uncertainty in the dust opacity model is $\Delta T \approx \pm1$\,K.
Since the assumption of ISM-like dust (e.g., OH1), which is probably more appropriate for the envelopes than OH5, 
leads to a reduction of the derived temperature by only a few per cent (w.r.t. OH5), the LOS-averaged temperature contrast 
between cores and envelopes could be systematically overestimated by $\approx 0.5$\,K, i.e., less than the absolute 
temperature uncertainty.

The LoS-averaged dust temperature maps presented in this paper provide
a robust estimate of the actual dust temperature only for the envelopes in the projected outer regions, 
where the emission is optically thin at all wavelengths and LoS temperature gradients are negligible. 
Toward the core centers, where cooling and shielding or embedded heating sources can produce significant 
LoS temperature gradients and the observed SEDs are therefore broader than single-temperature SEDs, 
the local central dust temperature is overestimated (in the case of a positive gradient in cold sources) or 
underestimated (negative gradient in internally heated sources). 
E.g., for B\,68, where we already derived a model that accounts for LoS gradients, 
we derive a local central dust temperature of $8.2_{-0.7}^{+2.1}$\,K \citep{nielbock12}, 
which is $3.7\pm1$\,K lower than the LoS-averaged temperature of $11.9\pm1$\,K we derive in this 
survey overview paper. For CB\,17, where the outer dust temperature is lower than in B\,68 
(Table\,\ref{tab-res-prof}), we derive a local central dust temperature that is only $\approx 1$\,K 
lower than the LoS-averaged dust temperature maps derived in this paper \citep{schmalzl13}.
Dust opacity model uncertainties cancel out to first order if we compare 
only results obtained with the same opacity curves.
I.e., in this paper we systematically overestimate the minimal inner dust temperatures by $1-4$\,K,
underestimate the outer envelope dust temperature by $<0.5$\,K,
and underestimate the temperature contrast between cold core and envelope by $1-4$\,K.

The effect of systematically overestimating the actual minimum inner dust temperatures 
on the column density and mass estimates is also not negligible and can be quite significant 
in individual cases. Figure\,\ref{fig-b68-profiles2} compares the column density profiles
of B\,68 derived with an isothermal assumption, the LoS-averaged SED fitting used in this paper, 
and the full RT approach used in \citet{nielbock12}. 
In this rather extreme case (one of the highest core\,--\,envelope temperature contrasts in our sample),
the LoS-averaged SED fitting underestimates the peak column density by a factor of $\approx 1.7$\ 
and the total mass by $\approx 15$\%. In sources with less-pronounced temperature contrast between core and envelope, 
like CB\,17 \citep{schmalzl13}, these systematic effects are smaller.
Still, Fig.\,\ref{fig-b68-profiles2} also illustrates that the column density profiles recovered 
by our SED fitting approach are a much better approximation to the actual underlying source structure 
than those derived with an isothermal assumption.
 
The local temperature minimum and column density peak values are also affected by the angular resolution 
of our analysis. The 36\asp4 beam size we use for all data may smooth out the central parts of the more 
centrally peaked profiles in some sources. Although we do not consider this effect being very significant 
for the starless cores, since the derived flat core radii are all significantly larger than the beam 
(see Sect.\,\ref{ssec-res-seds}), it can only be verified quantitatively by using all data in their original 
resolution and employing full forward modeling. 
Beam-smoothing does affect the temperature and column density estimates much more at the position 
of embedded protostars where temperature and density gradients are far more pronounced and definitely not resolved.
Here, the local temperatures at the position of embedded protostars are usually strongly underestimated and the 
temperature and column density maps are often corrupted by even very small residual pointing mismatches.

Hence we can summarize that the uncertainties introduced by calibration and data reduction issues 
amount to $\approx 3$\% for the temperature and $\approx 15$\% for the column density.
The uncertainties related to the dust opacity models are significantly larger and amount to 
$\pm 1$\,K ($\approx 10$\%) for the temperature and a factor of up to 3 for the column density and mass.
Ignoring LoS temperature gradients leads to an overestimation of the central temperature minima
in starless cores of $\approx 1-3$\,K and an underestimation of the central column density by up to 50\%.
The temperatures and column densities of the outer envelopes are much less affected by this LoS averaging.

The largest (systematic) uncertainties of the derived temperatures and (column) densities are clearly
related to our uncertain knowledge of the underlying dust mass absorption coefficient and its spectral dependence.
Therefore, our goal in future work will be to exploit the \herschel\ data presented here along with 
additional complementary data to derive specific observational constraints on the properties of the dust 
in these globules (see also end of Sect.\,\ref{sec-sum}).


\subsection{Comparison with earlier observations and models of dense cores}  \label{ssec-dis-comp}

In this section we discuss some examples of previous work based on
resolved observations of starless low-mass cores.  For comparison with
our new \herschel\ results, we specifically discuss published
measurements of density and temperature profiles, as well as different
attempts to model said temperature and density profiles.

The observed nonisothermality of dense clouds, derived from
\herschel\ data, was already reported by other authors
\citep[e.g.,][]{stutz10,peretto2010,battersby2011,wilcock2012,pitann12}.
While the \citet{stutz10} paper was the pilot study for the globule
survey presented here, \citet{peretto2010}, \citet{battersby2011},
\citet{wilcock2012}, and \citet{pitann12} found that the LoS-averaged
dust temperatures in IRDCs also systematically
drop toward the highest column
densities. \citet[][Fig.\,5]{wilcock2012} found temperature
profiles of IRDCs that are qualitatively very similar to the $T$\ profiles of
the starless globules we report in this paper
(Fig.\,\ref{fig-profiles-all}), although their outer temperatures were
on average $\sim 5$\,K higher than those found here (20\,K
vs. 15\,K). This is not surprising because IRDCs are generally located
in distant regions of high-mass star formation with strong MIR
background and may on average be exposed to a stronger ISRF than the
nearby isolated globules in the local spiral arm.  
In more nearby sources, low central temperatures and positive
temperature gradients were also inferred from observations and
modeling of molecular lines in prestellar cores
\citep[e.g.,][]{pagani07,crapsi07}.  

\citet{dwt02} already found observational evidence for temperature gradients and colder 
interriors of starless cores using flux ratio maps of 170 and 200\,$\mu$m observed with ISOPHOT.
\citet{Bacmann2000} observed 24 dense starless cores in absorption
with ISOCAM at 6\arcsec\ angular resolution. From modeling of the
7\,$\mu$m absorption against the MIR background, they derived core
structures characterized by a central flat region of $R_{\rm flat}\approx 4000-8000$\,AU, 
central column densities of a few $10^{22}$\,cm$^{-2}$, 
and outer radial column density profiles close
to $N_{\rm H} \propto r^{-1}$.  While the \citet{Bacmann2000} mean
peak column densities agree well with the values we derive, our
analysis results in somewhat larger flat core radii and significantly
steeper column density profiles (Table\,\ref{tab-res-prof}),
which might at least partially be attributable to the larger beam sizes 
of our maps as compared to the Bacmann work.

\citet{schnee05} used 450 and 850\,$\mu$m SCUBA observations to derive
the dust temperature profile of the nearby starless core TMC-1C from
the $450/850\,\mu$m flux ratio (assuming optically thin emission and a
fixed $\beta=1.5$). They found a positive temperature gradient with a
central dust temperature of 6\,K and an outer temperature of 12\,K at
the edge of the core (at 0.08\,pc).  They derived similar column
densities as we find for the globule sources and modeled the density
profile with a broken power-law with a shallow density profile inside
$\sim 4000$\,AU and $\propto r^{-1.8}$\ outside.  Since the break
radius of a broken power-law model of the density is not the same as
the parameter $r_1$ of our column density profile fits
(Eq.\,\ref{eq-densprof}) and should be somewhat smaller, their flat
core radius is comparable or slightly smaller (up to a factor of two)
than what we derive for the globule cores.  Their outer density
profile exponent of 1.8 is significantly smaller than what would be
implied by the column density profiles we derive here
(Table\,\ref{tab-res-prof}) and the density profile of B\,68 we derive
in \citet{nielbock12}.  However, both the dependence of the resulting
temperature on the above-mentioned assumptions as well as the
non-negligible calibration uncertainties of the SCUBA data, in
particular at 450\,$\mu$m, make the temperature scale and the (column)
density profile derived from such flux ratios rather uncertain.

\citet{shirley2000} also observed with SCUBA the 450 and 850\,$\mu$m
dust emission of 21 low-mass cores in different evolutionary
stages. Based on these observations, \citet{evans01} modeled the
emission of three starless cores within the context of a
self-consistent radiative transfer analysis assuming a nonisothermal
BES, modified OH5 dust opacities, and a modified \citet{black94} and
\citet{draine78} ISRF.  They found central densities between 10$^5$
and 10$^6$ cm$^{-3}$\ and required only fractions (0.3\,--\,0.6) of
the adopted ISRF to reproduce the $450/850\,\mu$m flux ratios.  They
also derived positive temperature gradients with central dust
temperatures of $7-8$\,K and outer temperatures of $12-13$\,K and
found that the ISRF was the dominant heating source in the absence of
a nearby star \citep[see also][]{zucconi01}.  Given the fact that our
LoS-averaged dust temperatures overestimate the actual central
temperatures by $1-4$\,K, their derived central temperatures are still
$2-3$\,K lower than what we find here.  However, one has to keep in
mind that this is based on very low number statistics and that their
three cores are on average denser and more massive than ours.
Therefore, this small difference in central temperatures might be real
and reflect different source properties. Their outer temperatures,
which are not well-constrained by the observations, are also $2-3$\,K
below the observationally well-constrained values we derive. However,
without comparing the same sources, it is not possible to verify
whether this reflects real differences in the shielding or is a result
of the specific assumptions for the dust properties and the strength of
the ISRF they made (see also discussion in
Sect.\,\ref{ssec-res-isrf}).  However, the same remarks apply as made
above about the uncertainty of such quantities that are based only on
SCUBA flux ratios.

\citet{shirley02} modeled the Class\,0 cores from the
\citet{shirley2000} sample.  In order to reproduce the 450 and
850\,$\mu$m fluxes, they included the internal radiation field
from the protostar in addition to the heating from the ISRF. They found
that the dust temperature exceeded 20\,K only in the inner
$\approx10^3$\,AU, reached $\approx10$\,K at $r\approx5000-8000$\,AU,
and increased again slightly toward the outer boundary of the core at a few $10^4$\,AU.
This is consistent with what we find, including the temperature rise
at $r>10^4$\,AU, taking again into account that we overestimate the
inner minimum temperature by up to 3\,K, and that their models appear to
underestimate the outer temperatures by $2-3$\,K.

\citet{stamatellos2005} calculated temperature profiles in the context
of an evolutionary model, starting with a 5.4\,\msun\ BES and assuming 
OH5 dust opacities and a modified \citet{black94} ISRF.  In the
initial stage, when only the ISRF heats the core, they found a central
temperature of $\sim$7\,K and outer temperatures of up to $\sim$20\,K.
\citet{stamatellos2007} further studied the
influence of external heating and shielding on the temperature profile
of cores in the $\rho$\ Oph cloud, taking into account also the
surrounding molecular cloud extinction.  They found again that the
inner core temperatures (at $r< 10^4$\,AU) were always below 10\,K,
but that they reached rim temperatures (at $r\approx2 \times
10^4$\,AU) of only $\sim$13-15\,K.  They noted that in
Taurus, where the ambient radiation field is weaker, the core
temperatures are found to be somewhat lower with a central value of
$\approx7$\,K and rim temperature of $\approx12$\,K for TMC-1C.  These
temperatures and temperature contrasts are again consistent with what
we now measure from the \herschel\ data in similar cores. In
particular the outer temperatures we measure for the isolated globules
with their thin halos are all inbetween the values they derived for the
unshielded and the well-shielded cores.


\section{Summary and conclusions}          \label{sec-sum}

We have studied the thermal dust emission from 12 nearby isolated globules 
using FIR dust continuum maps at five wavelength bands between 100 and 500\,$\mu$m obtained with the 
\herschel\ satellite and submm continuum maps at up to three wavelengths beween 450\,$\mu$m and 1.2\,mm 
obtained with ground-based telescopes. 
Great care was taken in both the selection of truly isolated and previously well-characterized sources in 
regions of exceptionally low background emission and in restoring and calibrating all maps in a consistent 
manner to enable a combined in-depth analysis of all data. This careful preselection, data treatment, 
and the very deep 100\,$\mu$m maps ($\sim 50\,\mu$Jy/arcsec$^2$\ rms) turned out to be essential 
to enable our analysis method and lead to the results presented here.

We calibrate all data to the same flux scale, subtract all general background levels in a constistent way, 
convolve all maps to the same angular resolution, assign weights to each pixel according to calibration uncertainties and rms noise, 
and extract full SEDs from 100\,$\mu$m to 1.2\,mm for each image pixel. These individual SEDs are independently fit 
(least squares) with single-temperature modified blackbody curves, using OH5 opacities for mildy coagulated dust grains 
with thin ice mantles. Thus we obtain for each image pixel a column density and an LoS-averaged dust temperature.
The resulting dust temperature maps have rms noise levels of only a few tenth of a K and 
systematic uncertainties of less than $\pm 1$\,K. We then analyze the dust temperature 
and column density maps along with the integrated SEDs of all sources, including additional archival data at shorter wavelengths, 
and obtain the following main results:

\begin{enumerate}

\item The globules in our sample have a mean distance from the Sun of $240\pm100$\,pc, mean total gas mass of $7.3\pm6.8$\,\msun,
  mean bolometric luminosity of $\approx 6.7$\,\lsun, and mean outer diameter of $0.45\pm 0.2$\,pc. Typical peak column densities 
  (averaged over the 36\asp4 beam) are $\langle N_{\rm H}\rangle \sim 3\times 10^{22}$\,cm$^{-2}$, which corresponds to peak visual 
  extinctions of about 45\,mag.

\item The dense cores of the globules have a mean total gas mass of $1.6\pm 1.2$\,\msun\ and a mean 
   diameter of $0.12\pm 0.05$\,pc. They constitute on average $25\pm 6$\,\% of the total globule mass.

\item Out of the 12 globules in our sample, four contain one starless core each, six contain a protostellar core (Class\,0 or I), 
  and two globules contain each one starless and one protostellar core.

\item We confirm the earlier classifications of protostars and embedded YSOs, but the better-sampled thermal SEDs 
  do not allow for better evolutionary stage determination from the SEDs alone since projection effects dominate 
  the spectral appearance of even Class\,0 protostars. However, the \herschel\ data allow us now to also include starless cores in 
  the $L_{\rm smm}\,/\,L_{\rm bol}$\ vs. $T_{\rm bol}$\ diagram (Fig.\,\ref{fig-tbol-lratio}). 
  We find that all starless cores have $T_{\rm bol}<25$\,K and 
  \mbox{$10\%<L_{\rm smm}\,/\,L_{\rm bol}<30\%$}. In particular the criterion $T_{\rm bol}<25$\,K can be used to identify 
  isolated and externally heated starless cores, since even an embedded VeLLO will elevate $T_{\rm bol}$\ above 30\,K.

\item Most globules are surrounded by a thin ($A_V\sim 1-3$\,mag) and warm ($T_{\rm d} \approx$\,15-20\,K) outer halo, 
  the full extent of which we could not properly recover in most cases. These halos are evident in the dust emission-based 
  column density maps, NIR extinction maps, as well as in many cases on optical images as cloudshine.

\item The thermal structure of all globules is dominated by external heating through the ISRF, 
    with warm outer envelopes ($15.5\pm 1.8$\,K) and colder interiors ($<13\pm1$\,K).
   We find a very tight anti-correlation between the minimal central temperatures in starless cores 
    and the total column density toward the outer edge that provides shielding against the ISRF (Fig.\,\ref{fig-tin-nh}).

\item The outer dust temperature of the globules is to first order determined by external heating from an approximately uniform 
  ISRF and shielding by the thin extended halos. We find a clear anti-correlation between the outer dust temperatures and the 
  mean column density / $A_V$\ of the halos (Fig.\,\ref{fig-isrf}).
  Only one of the globules is exposed to a significantly higher than average local ISRF (BHR\,12).

\item The embedded (low-mass and low-luminosity) protostars raise the local temperature of the dense cores only within radii 
 out to about 5000\,AU, but do not significantly affect the overall thermal balance of the globules.

\item Our LoS-averaged dust temperatures are accurate to about $\pm 1$\,K at the outer boundaries of the globules 
 (accounting for both statistical and systematic errors), but systematically overestimate the local central temperature minima 
 by 1\,-\,4\,K, depending on the temperature contrast to the outer rim. 
  I.e., these globules have central temperatures in the range 8\,-\,12\,K and temperature contrasts to the outer rim 
  of 3\,-\,8\,K.

\item Out of the six starless cores in our sample, five cores are found to be thermally stabilized against gravitational collapse.
  Of these, four cores (CB\,17\,-\,SMM, CB\,26\,-\,SMM2, CB\,27, and B\,68) have central densities that are very 
  close to the maximum stable density of a pressure-supported, self-gravitating cloud, i.e., we cannot distinguish wheter they 
  are starless stable or prestellar.
  CB\,4 is thermally sub-critical and must be a purely pressure-confined starless core. 
  The starless core in CB\,244\,-\,SMM2 is clearly super-critical and should very likely be a prestellar core at the verge of collapse.

\item The radius of the single-core globules is very similar to or only slightly larger than their Jeans length ($\sim 0.13$\,pc).
  The projected core separation in the two double-core globules agrees well with the Jeans lengths in these clouds, indicating 
  that large globules tend to fragment gravitationally on the Jeans scale.

\item Many of the globules have one side that is signifcantly warmer at the outer rim compared to the other side.
  Although in some cases the warm side faces the direction toward the Galactic plane, we could not identify 
  a consistent relation between warm side and potential external illuminating sources.

\end{enumerate}

As a next step, we are currently analyzing individual starless cores by employing a ray-tracing algorithm 
that is able to restore LoS temperature gradients and derive the actual volume desnity structure of the cores 
by starting from a 1-D model and allowing for local deviations from spherical symmetry  \citep{nielbock12,schmalzl13}.
This way we hope to better and most robustly constrain the dust temperature structure of these cores without 
introducing too many model-dependent assumptions into the analysis. 
We are also working toward better constraining the dust opacity models. 
For this purpose, we have already obtained deep NIR extinction maps that recover the same angular 
resolution as the \herschel\ data and are currently obtaining additional deep submm and mm continuum maps.
We are planning to employ and compare different approaches, including SED fitting and ray-tracing like 
in this paper and in \citet{nielbock12}, self-consistent radiative transfer modeling like in \citet{shirley11}, 
as well as Bayesian approaches like in \citet{kelly2012}.
Furthermore, we are currently also using complementary molecular line data to derive 
constraints on the distribution of turbulence and molecular gas-phase freeze-out onto dust grains 
in the sources studied in this paper (Lippok et al., in prep.).


\begin{acknowledgements}
  The authors gratefully acknowledge the valuable help from
  H\'{e}l\`{e}ne Roussel in the production of {\it Scanamorphos} PACS
  and SPIRE maps as well as Gonzalo Aniano's assistance
  and careful work with the {\it Herschel} convolution kernels used
  in this work. We greatly benefitted from discussions with Yaroslav Pavlyuchenkov 
  (Moscow) on the data analysis.
  We also thank the anonymous referee for comments and suggestions 
  that helped to improve the clarity and completeness of the paper.
  Part of this work was supported by the German
  \emph{Deut\-sche For\-schungs\-ge\-mein\-schaft, DFG\/} project
  number Ts~17/2--1. This publication makes use of data products from
  the Two Micron All Sky Survey, which is a joint project of the
  University of Massachusetts and the Infrared Processing and Analysis
  Center/California Institute of Technology, funded by the National
  Aeronautics and Space Administration and the National Science
  Foundation. 
  The work of AMS, SR, and JK was supported by the Deutsche Forschungsgemeinschaft
  priority program 1573 ("Physics of the Interstellar Medium"). 
  HL, MN, and ZB are supported by Deutsches Zentrum f\"ur Luft- und Raumfahrt (DLR).
\end{acknowledgements}



\clearpage

\begin{figure}
 \centering
 \includegraphics[width=9cm]{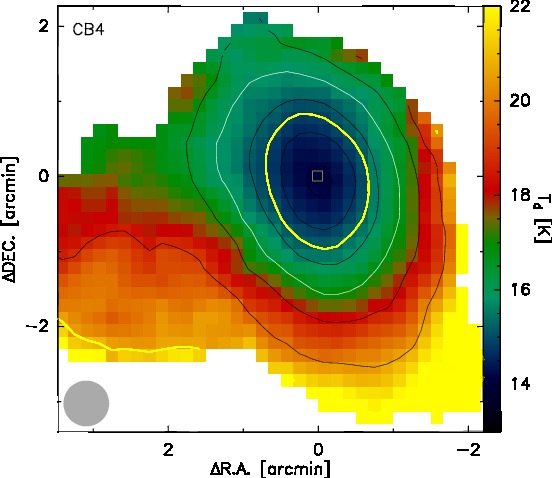}
 \caption{\label{fig-tmap-cb4}
  Dust temperature map of CB\,4 (color scale), overlaid with contours of the 
  hydrogen column density (white: $10^{21}$\,cm$^{-2}$, black: 0.2, 0.5, 2, and 4\,$10^{21}$\,cm$^{-2}$).
  Yellow contours mark $N_{\rm out}$\ (Eq.\,\ref{eq-densprof}) and $N_{1^{\prime}}$\ (Eq.\,\ref{eq-dens1a}).
  The yellow square marks the column density peak and the reference position listed in Table\,\ref{tab-res-prof}.}
\end{figure}

\begin{figure}
 \centering
 \includegraphics[width=9cm]{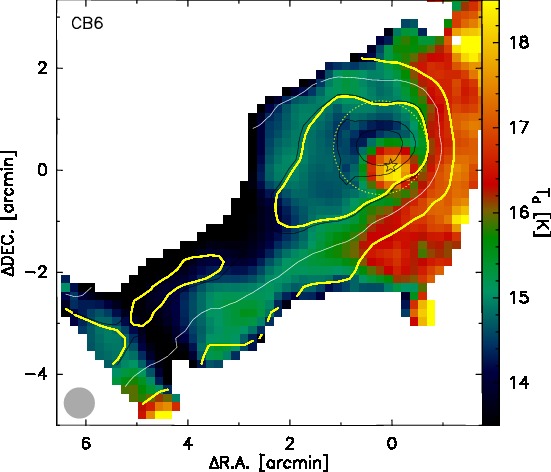}
 \caption{\label{fig-tmap-cb6}
  Dust temperature map of CB\,6 (color scale), overlaid with contours of the 
  hydrogen column density (white: $10^{21}$\,cm$^{-2}$, black: 2, 4, and 6\,$10^{21}$\,cm$^{-2}$).
  Yellow contours mark $N_{\rm out}$\ (Eq.\,\ref{eq-densprof}) and $N_{1^{\prime}}$\ (Eq.\,\ref{eq-dens1a}).
  The dotted yellow ellipse marks the $N_{1^{\prime}}$\ approximation described in Sect.\,\ref{ssec-res-seds}.
  The asterisk marks the position of the embedded YSO.}
\end{figure}

\begin{figure}
 \centering
 \includegraphics[width=9cm]{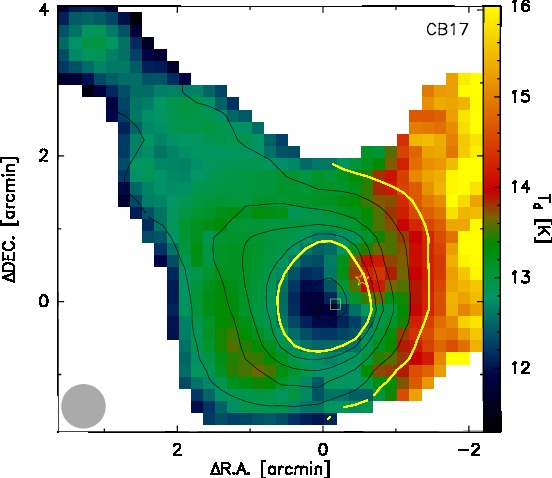}
 \caption{\label{fig-tmap-cb17}
  Dust temperature map of CB\,17 (color scale), overlaid with contours of the 
  hydrogen column density (white: $10^{21}$\ - masked by the yellow $N_{\rm out}$\ contour - and $10^{22}$\,cm$^{-2}$, 
  black: 2, 4, 6, 8, and 20\,$10^{21}$\,cm$^{-2}$).
  Yellow contours mark $N_{\rm out}$\ (Eq.\,\ref{eq-densprof}) and $N_{1^{\prime}}$\ (Eq.\,\ref{eq-dens1a}).
   The yellow square marks the column density peak and position of CB\,17-SMM.
   The asterisk marks the position of CB\,17-IRS \citep[Table\,\ref{tab-res-prof}; see also][]{lau10}.}
\end{figure}

\begin{figure}
 \centering
 \includegraphics[width=9cm]{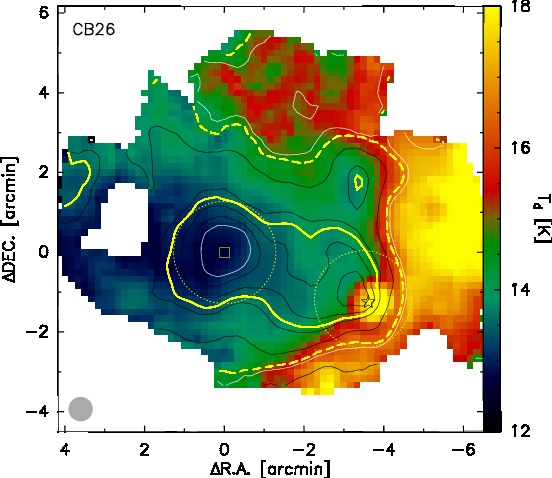}
 \caption{\label{fig-tmap-cb26}
  Dust temperature map of CB\,26 (color scale), overlaid with contours of the 
  hydrogen column density (white: $10^{21}$\ and $10^{22}$\,cm$^{-2}$, black: 0.5, 2, 4, 6, and 8\,$10^{21}$\,cm$^{-2}$). 
  Yellow contours mark $N_{\rm out}$\ (Eq.\,\ref{eq-densprof}) and $N_{1^{\prime}}$\ (Eq.\,\ref{eq-dens1a}, 
  referring to the starless core SMM2).
  Dotted yellow ellipses mark the $N_{1^{\prime}}$\ approximations for the two subcores described in Sect.\,\ref{ssec-res-seds}.
  The yellow square marks the column density peak and position of CB\,26-SMM2; 
  the asterisk marks the position of CB\,26-SMM1 (Table\,\ref{tab-res-prof}).}
\end{figure}

\begin{figure}
 \centering
 \includegraphics[width=9cm]{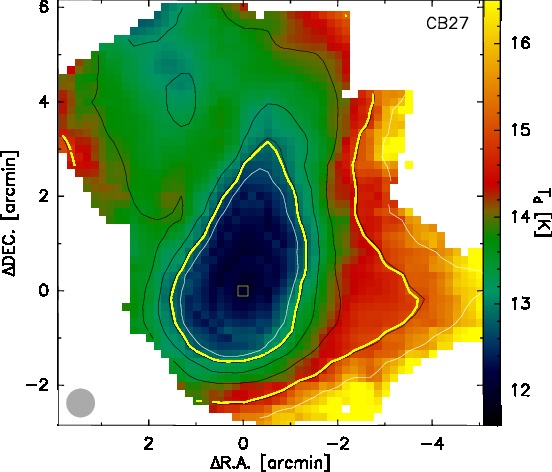}
 \caption{\label{fig-tmap-cb27}
  Dust temperature map of CB\,27 (color scale), overlaid with contours of the 
  hydrogen column density (white: $10^{21}$\ and $10^{22}$\,cm$^{-2}$, black: 2, 4, 6, and 8\,$10^{21}$\,cm$^{-2}$).
  Yellow contours mark $N_{\rm out}$\ (Eq.\,\ref{eq-densprof}) and $N_{1^{\prime}}$\ (Eq.\,\ref{eq-dens1a}).
  The yellow square marks the column density peak and the reference position listed in Table\,\ref{tab-res-prof}.}
\end{figure}

\begin{figure}
 \centering
 \includegraphics[width=9cm]{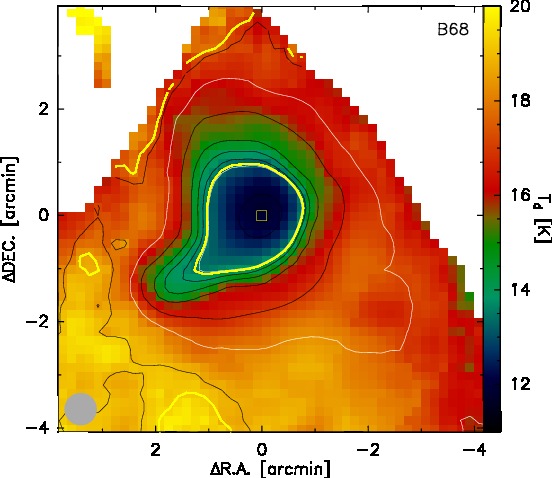}
 \caption{\label{fig-tmap-b68}
  Dust temperature map of B\,68 (color scale), overlaid with contours of the 
  hydrogen column density (white: $10^{21}$\ and $10^{22}$\,cm$^{-2}$, black: 0.5, 2, 4, 6, 8 and 20\,$10^{21}$\,cm$^{-2}$).
  Yellow contours mark $N_{\rm out}$\ (Eq.\,\ref{eq-densprof}) and $N_{1^{\prime}}$\ (Eq.\,\ref{eq-dens1a}).
  The yellow square marks the column density peak and the reference position listed in Table\,\ref{tab-res-prof}.}
\end{figure}

\begin{figure}
 \centering
 \includegraphics[width=9cm]{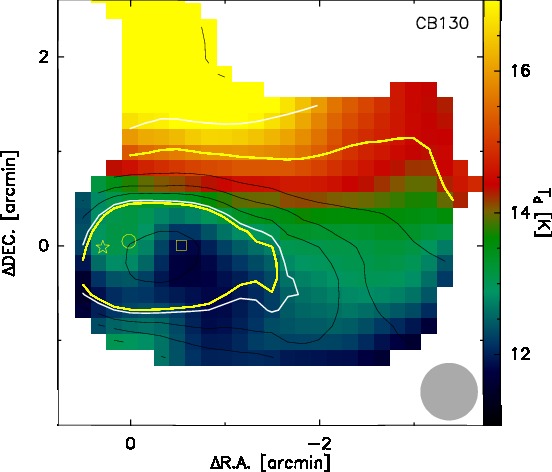}
 \caption{\label{fig-tmap-cb130}
  Dust temperature map of CB\,130 (color scale), overlaid with contours of the 
  hydrogen column density (white: $10^{21}$\ and $10^{22}$\,cm$^{-2}$, black: 0.5, 2 - masked by yellow $N_{\rm out}$\ contour - 
  4, 6, 8, and 20\,$10^{21}$\,cm$^{-2}$).
  Yellow contours mark $N_{\rm out}$\ (Eq.\,\ref{eq-densprof}) and $N_{1^{\prime}}$\ (Eq.\,\ref{eq-dens1a}).
  The yellow circle marks the position of the protostellar core CB\,130-SMM1 with the embedded Class\,0 VeLLO.
  The yellow square marks the position of the secondary starless core CB\,130-SMM2.
  The asterisk marks the position of the Class\,I-II YSO CB\,130-IRS, which most likely associated with CB\,130, 
  but not embedded in its dense core \citep[see][]{lau10}.}
\end{figure}

\begin{figure}
 \centering
 \includegraphics[width=9cm]{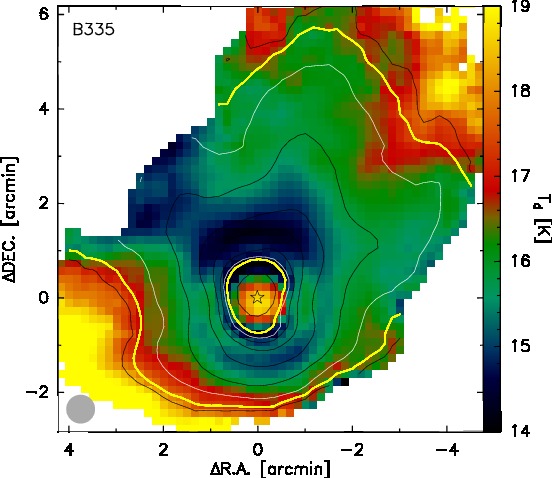}
 \caption{\label{fig-tmap-b335}
  Dust temperature map of B\,335 (color scale), overlaid with contours of the 
  hydrogen column density (white: $10^{21}$\ and $10^{22}$\,cm$^{-2}$, black: 0.5, 2, 4, 6, 8, and 20\,$10^{21}$\,cm$^{-2}$).
  Yellow contours mark $N_{\rm out}$\ (Eq.\,\ref{eq-densprof}) and $N_{1^{\prime}}$\ (Eq.\,\ref{eq-dens1a}).
  The asterisk marks the position of the protostar and the reference position listed in Table\,\ref{tab-res-prof}.}
\end{figure}

\begin{figure}
 \centering
 \includegraphics[width=9cm]{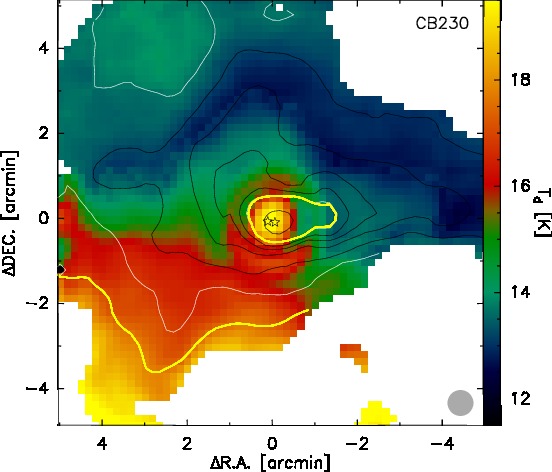}
 \caption{\label{fig-tmap-cb230}
  Dust temperature map of CB\,230 (color scale), overlaid with contours of the 
  hydrogen column density (white: $10^{21}$\ and $10^{22}$\,cm$^{-2}$, black: 2, 4, 6, 8, and 20\,$10^{21}$\,cm$^{-2}$).
  Yellow contours mark $N_{\rm out}$\ (Eq.\,\ref{eq-densprof}) and $N_{1^{\prime}}$\ (Eq.\,\ref{eq-dens1a}).
  The two asterisks mark the positions of the embedded YSOs \citep[see][]{lau10}.}
\end{figure}

\begin{figure}
 \centering
 \includegraphics[width=9cm]{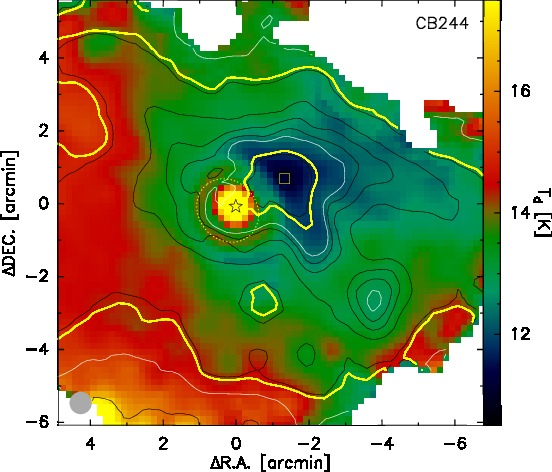}
 \caption{\label{fig-tmap-cb244}
  Dust temperature map of CB\,244 (color scale), overlaid with contours of the 
  hydrogen column density (white: $10^{21}$\ and $10^{22}$\,cm$^{-2}$, black: 0.5, 2, 4, 6, 8, and 20\,$10^{21}$\,cm$^{-2}$).
  Yellow contours mark $N_{\rm out}$\ (Eq.\,\ref{eq-densprof}) and $N_{1^{\prime}}$\ (Eq.\,\ref{eq-dens1a}) for 
  the starless core CB\,244-SMM2.
  The dotted yellow ellipse marks the $N_{1^{\prime}}$\ approximation for CB244-SMM1 described in Sect.\,\ref{ssec-res-seds}.
  The yellow square marks the column density peak of CB\,244-SMM2 and the reference position listed in Table\,\ref{tab-res-prof}.
  The asterisk marks the position of of the protostar CB244-SMM1.}
\end{figure}


\Online

\begin{appendix} 
\section{Images on large-scale morphology}
\label{sec-figs-morph}

These figures are discussed in Sect.\,\ref{ssec-res-env}.

\begin{figure}[htb]
 \centering
 \includegraphics[width=9cm]{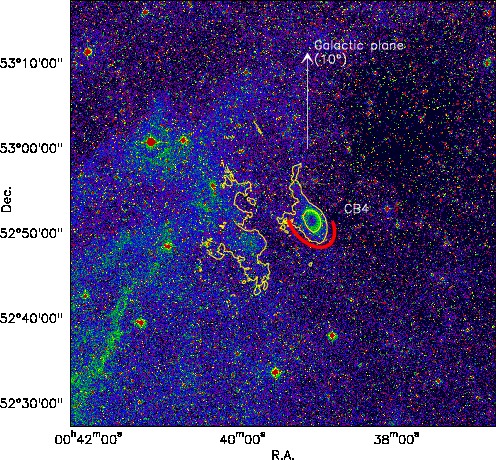}
 \caption{\label{fig-cb4-morph}
  CB\,4: Visual (red) DSS2 image with SPIRE\,500\,$\mu$m continuum contours overlaid. 
  The red arc marks the warm rim of CB\,4.
}
\end{figure}

\begin{figure}[htb]
 \centering
 \includegraphics[width=9cm]{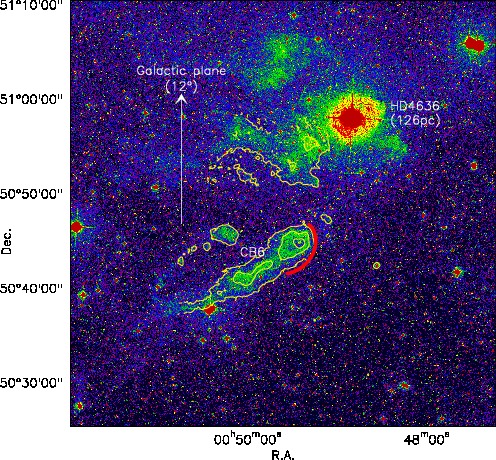}
 \caption{\label{fig-cb6-morph}
  CB\,6: Visual (red) DSS2 image with SPIRE\,500\,$\mu$m continuum contours overlaid. 
  The red arc marks the warm rim of CB\,6.
}
\end{figure}

\begin{figure}[htb]
 \centering
 \includegraphics[width=9cm]{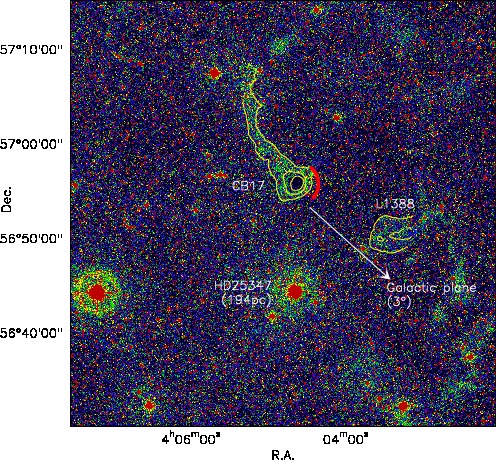}
 \caption{\label{fig-cb17-morph}
  CB\,17: Visual (red) DSS2 image with SPIRE\,500\,$\mu$m continuum contours overlaid. 
  The red arc marks the warm rim of CB\,17.
}
\end{figure}

\begin{figure}[htb]
 \centering
 \includegraphics[width=9cm]{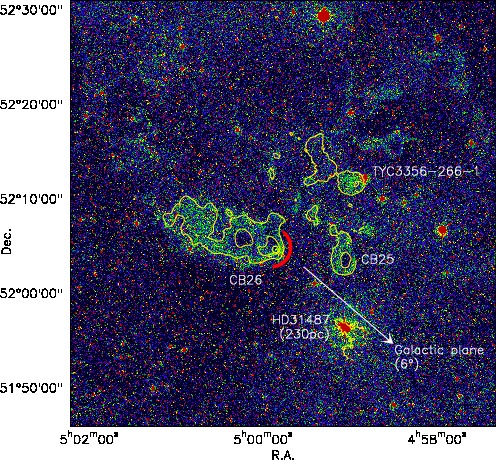}
 \caption{\label{fig-cb26-morph}
  CB\,26: Visual (red) DSS2 image with SPIRE\,500\,$\mu$m continuum contours overlaid. 
  The red arc marks the warm rim of CB\,26.
}
\end{figure}

\begin{figure}[htb]
 \centering
 \includegraphics[width=9cm]{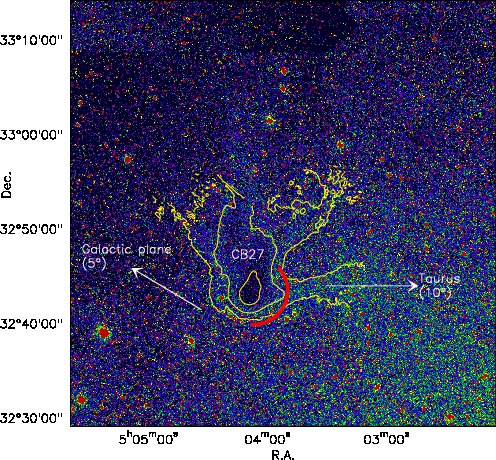}
 \caption{\label{fig-cb27-morph}
  CB\,27: Visual (red) DSS1 image with SPIRE\,500\,$\mu$m continuum contours overlaid. 
  The red arc marks the warm rim of CB\,27.
}
\end{figure}

\begin{figure}[htb]
 \centering
 \includegraphics[width=9cm]{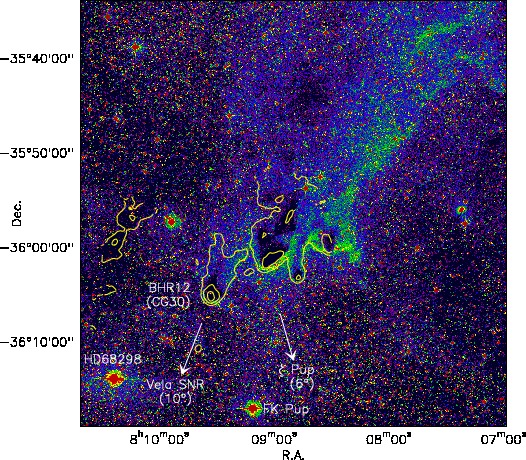}
 \caption{\label{fig-bhr12-morph}
  BHR\,12: Visual (red) DSS2 image with SPIRE\,500\,$\mu$m continuum contours overlaid. 
  The tail side is warmer than the head side, but there is nor clear warm rim side.
}
\end{figure}

\begin{figure}[htb]
 \centering
 \includegraphics[width=9cm]{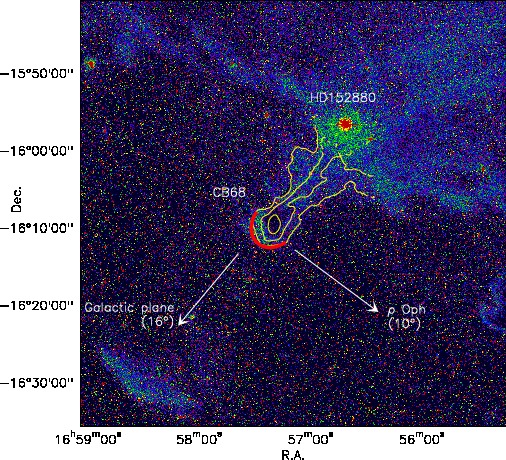}
 \caption{\label{fig-cb68-morph}
  CB\,68: Visual (red) DSS2 image with SPIRE\,500\,$\mu$m continuum contours overlaid. 
  The red arc marks the warm rim of CB\,68.
}
\end{figure}

\begin{figure}[htb]
 \centering
 \includegraphics[width=9cm]{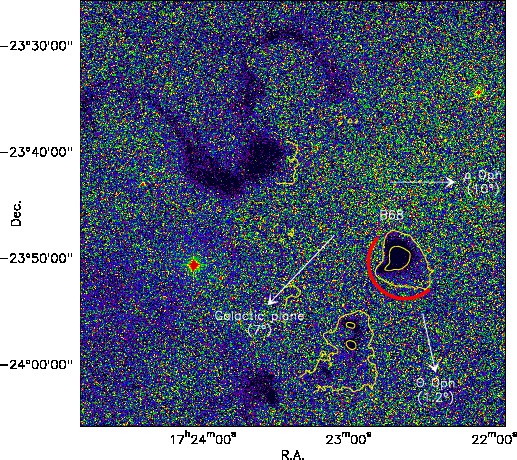}
 \caption{\label{fig-b68-morph}
  B\,68: Visual (red) DSS1 image with SPIRE\,500\,$\mu$m continuum contours overlaid. 
  The red arc marks the warm rim of B\,68. The B2\,IV star $\Theta$\,Oph is located 
  at a distance from the Sun of 134\,pc, i.e., its physical distance from B\,68 is only $\approx 2.8$\,pc.
}
\end{figure}

\begin{figure}[htb]
 \centering
 \includegraphics[width=9cm]{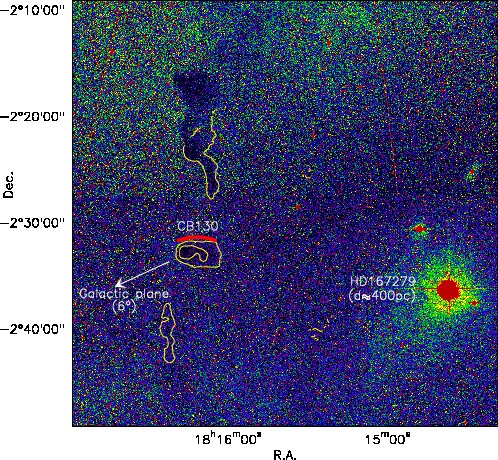}
 \caption{\label{fig-cb130-morph}
  CB\,130: Visual (red) DSS2 image with SPIRE\,500\,$\mu$m continuum contours overlaid. 
  The red arc marks the warm rim of CB\,130.
}
\end{figure}

\begin{figure}[htb]
 \centering
 \includegraphics[width=9cm]{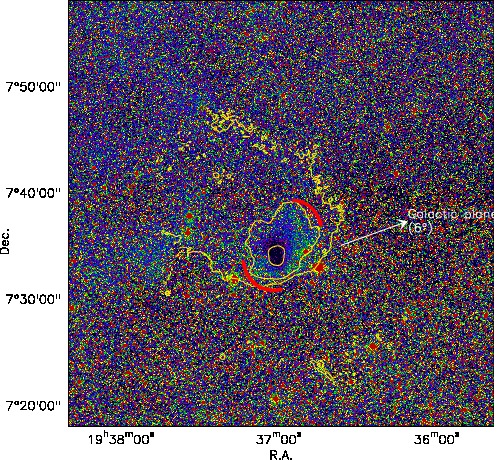}
 \caption{\label{fig-b335-morph}
  B\,335: Visual (red) DSS2 image with SPIRE\,500\,$\mu$m continuum contours overlaid. 
  The red arc marks the warm rim of B\,335.
}
\end{figure}

\begin{figure}[htb]
 \centering
 \includegraphics[width=9cm]{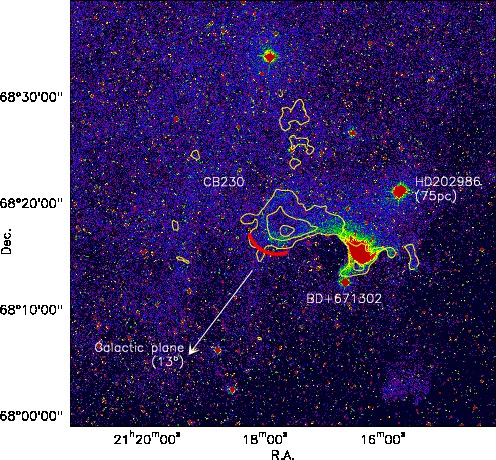}
 \caption{\label{fig-cb230-morph}
  CB\,230: Visual (red) DSS2 image with SPIRE\,500\,$\mu$m continuum contours overlaid. 
  The red arc marks the warm rim of CB\,230.
}
\end{figure}

\begin{figure}[htb]
 \centering
 \includegraphics[width=9cm]{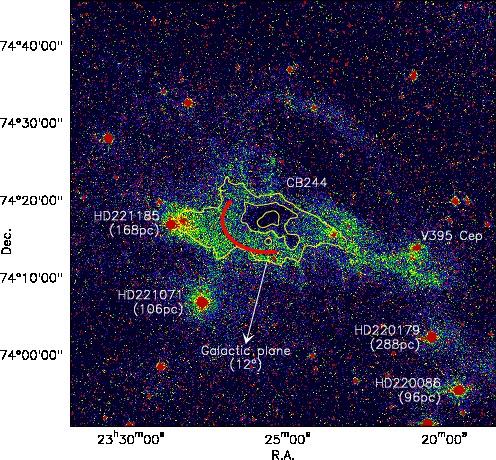}
 \caption{\label{fig-cb244-morph}
  CB\,244: Visual (red) DSS2 image with SPIRE\,500\,$\mu$m continuum contours overlaid.  
  The red arc marks the warm rim of CB\,244.
}
\end{figure}

\end{appendix}


\end{document}